\documentclass[preprint,12pt]{elsarticle}
\usepackage{natbib}
\usepackage{url} 
\usepackage{amsmath}
\usepackage{natbib}
\bibpunct{(}{)}{;}{a}{}{,}
\usepackage{subcaption}
\usepackage[colorlinks = true,urlcolor  = green,
            citecolor = blue]{hyperref}
\usepackage{url}
\usepackage{graphicx}
\usepackage{amsmath, amssymb}
\usepackage{bbold}
\usepackage{mathbbol}
\usepackage{algorithm}
\usepackage{algorithmic}
\usepackage{enumitem}
\usepackage{float}
\usepackage{xcolor}
\usepackage{bm} 
\usepackage{setspace}
\usepackage{multirow}
\usepackage{lscape}
\usepackage{rotating}
\usepackage{adjustbox}
\usepackage{booktabs}
\usepackage{ragged2e}

\usepackage{float}
\def\*#1{\mathbf{#1}}
\def\^#1{\amsmathbb{#1}}
\def\##1{\mathbb{#1}}
\DeclareSymbolFontAlphabet{\amsmathbb}{AMSb}
\usepackage[symbol]{footmisc}
\usepackage{amsthm}

\DeclareMathOperator*{\argmin}{argmin}

\DeclareSymbolFontAlphabet{\amsmathbb}{AMSb}%

\usepackage[a4paper,margin=1in]{geometry}

\begin{document}

\def\spacingset#1{\renewcommand{\baselinestretch}%
{#1}\small\normalsize} \spacingset{1}
\begin{frontmatter}


\title{Variable Selection for Fixed and Random Effects in Multilevel Functional Mixed Effects Models}
 \author{Rahul Ghosal$^{1,\ast}$, Marcos Matabuena$^{2}$,
Enakshi Saha$^{1}$ \\[4pt]
\textit{$^{1}$ Department of Epidemiology and Biostatistics, University of South Carolina \\
$^{2}$ Department of Biostatistics, Harvard University}
\\[2pt]
{rghosal@mailbox.sc.edu}}

\begin{abstract}
We develop a new method for simultaneously selecting fixed and random effects in a multilevel functional regression model. The proposed method is motivated by accelerometer-derived physical activity data from the 2011-12 cohort of the National Health and Nutrition Examination Survey (NHANES), with the aim of identifying age and race-specific heterogeneity in covariate effects on the diurnal pattern of physical activity across the lifespan. Existing methods for variable selection in function-on-scalar regression have primarily been designed for fixed effect selection and for single-level functional data. In high-dimensional multilevel functional regression, the presence of cluster-specific heterogeneity in covariate effects could be detected through sparsity in fixed and random effects, and for this purpose, we propose a multilevel functional mixed effects selection (MuFuMES) method. The fixed and random functional effects are modelled using splines, with spike-and-slab group lasso (SSGL) priors on the unknown parameters of interest, and a computationally efficient MAP estimation approach is employed for mixed effect selection through an Expectation Conditional Maximization (ECM) algorithm. Numerical analysis using simulation study illustrates the satisfactory selection accuracy of the variable selection method in having a negligible false-positive and false-negative rate. 
The proposed method is applied to the NHANES 2011-12 accelerometer data, where it effectively identifies age and race-specific heterogeneity in covariate effects on the diurnal pattern of physical activity, recovering biologically meaningful insights.
\end{abstract}
 
\vfill
\end{frontmatter}
\newpage
\spacingset{1.5} 
\section{Introduction}
\label{intro}

Functional regression models \citep{Ramsay05functionaldata, crainiceanu2024functional} are widely used for modelling dynamic effects of scalar and functional covariates on a functional response of interest, varying over some continuous index such as time, space, age, or other similar domain. Functional regression models have diverse applications in various disciplines ranging from agriculture \citep{montesinos2018bayesian,park2023crop}, growth curve modelling \citep{tang2008pairwise,leroux2018dynamic}, economics \citep{kosiorowski2014functional}, biological sciences \citep{goldsmith2016assessing,xu2017functional}, imaging \citep{morris2011automated,zipunnikov2011functional,xiao2016fast}, physical activity research \citep{goldsmith2016new,kowal2020bayesian,cui2021additive}, and many other areas. When each observational unit has a unique functional observation associated with them, the data structure is referred to as single-level functional data. Multiple methods have been developed over the last two decades for smooth estimation and prediction \citep{reiss2010fast,bauer2018introduction,ivanescu2015penalized,ghosal2023shape} for such single-level functional regression models. See \cite{wang2016functional}  and the references therein for existing modelling approaches in single-level functional data.

With the rapid development of technology in biomedical sciences, multiple functional curves can now be measured for each observational unit of interest, for example, functional observations over multiple longitudinal visits \citep{di2009multilevel}, hierarchical structure or nested functional data \citep{shou2015structured} with multiple replications for each unit of interest, leading to a {\it multilevel} functional data structure. Multiple regression and modelling approaches \citep{zipunnikov2014longitudinal,goldsmith2015generalized,park2015longitudinal,li2021multilevel,cui2022fast,li2022fixed,cui2023fast,koner2023second,koner2024profit} have been developed for such ``second-generation" and multilevel functional data, accounting for the within-cluster correlation.

Functional regression models are increasingly becoming high-dimensional due to the advancement of sensors and medical devices \citep{yan2018real}, which can enable the collection of multiple functional observations from different modalities along with a large number of scalar covariates such as demographic, clinical, and genetic information of subjects or units. Several variable selection methods have been developed in single-level functional regression models using both frequentist \citep{chen2016variable,barber2017function,parodi2018simultaneous,ghosal2020variable,ghosal2021variable,ghosal2024variable} and Bayesian approaches \citep{kowal2020bayesian,mehrotra2022simultaneous,sousa2023bayesian,bai2023scalable}, which are aimed at estimating the dynamic association of the influential predictors on the functional response of interest. These methods enable the selection of the functional fixed effects, thus enhancing interpretability. 
While multiple regression frameworks have been developed for functional mixed effects models \citep{scheipl2015functional} and multilevel functional regression models \citep{goldsmith2015generalized,park2015longitudinal,cui2022fast,sergazinov2023case,sun2025ultra}, high dimensionality in such hierarchical or clustered functional data can pose significant computation challenges. Understanding the cluster-specific heterogeneity in covariate effects on the functional outcome requires accurate identification of the fixed and random functional effects in such multilevel functional mixed effects models. However, due to its computational complexity, variable selection methods for fixed and random effects have been less explored for such models.


\subsection{Motivating Application}

Our motivating application comes from the accelerometer-derived physical activity data in the 2011-2012 cohort of the National Health and Nutrition Examination Survey (NHANES). We are interested in identifying and understanding whether there exists any age and race-specific heterogeneity in the effects of key demographic (e.g., gender), clinical (e.g., BMI), lifestyle (e.g., diet), and socioeconomic covariates (e.g., income) on the diurnal patterns of physical activity (PA) across the lifespan. 
\begin{figure}[ht]
\centering
\includegraphics[width=0.8\linewidth , height=1\linewidth]{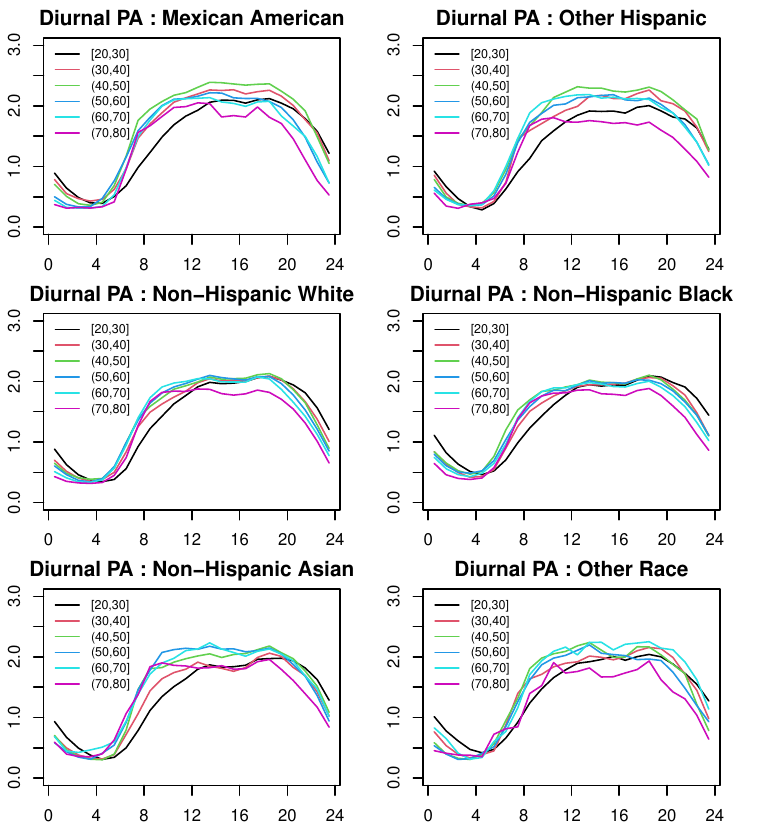}
\caption{The average diurnal patterns of PA (log-transformed) for the six age and ethnicity groups in NHANES 2011-12.}
\label{fig:fig1}
\end{figure}
Previous research in the NHANES 2003-06 cohorts has demonstrated considerable heterogeneity in the effect of aging and race/ethnicity on summary-level PA metrics and diurnal PA patterns \citep{hawkins2009objectively,xiao2015quantifying,varma2017re,cui2022fast,ghosal2023variable}. However, the key drivers of the diurnal PA patterns, and the age and race-specific heterogeneity in their effects have been less explored. NHANES 2011-2012 reports individuals’ acceleration in Monitor Independent Movement Summary (MIMS) unit \citep{john2019open}. Figure \ref{fig:fig1} displays the observed diurnal pattern in PA among adults across six age groups ($20-30,30-40,\ldots,70-80$) and six races (ethnicities). It can be observed that there exists considerable heterogeneity in the diurnal PA patterns among the age-groups and across races. Our objective is to develop a method that can simultaneously (i) identify key drivers of diurnal patterns in PA, (ii) detect covariates with age or race-specific heterogeneity in their effects, and (iii) estimate their dynamic associations with PA.

\subsection{Contributions}
In this article, we propose a new method for simultaneously selecting fixed and random effects
in a multilevel functional regression model. 
We model the fixed and random smooth functional effects using splines and impose spike-and-slab group lasso (SSGL) priors on both the fixed effect basis coefficients \citep{bai2022spike,bai2023scalable} and random effect covariance parameters, represented by a Cholesky decomposition of the covariance matrix \citep{bondell2010joint,ibrahim2011fixed}. We develop a computationally efficient approach for mixed effect selection through an expectation conditional maximization (ECM) algorithm which derives maximum a posteriori (MAP) estimates \citep{rovckova2014emvs,bai2023scalable,mclain2025efficient}. The SSGL prior has been shown to perform adaptive shrinkage \citep{bai2022spike,bai2023scalable} compared to other group-based penalties. The resulting MAP estimator enjoys exact sparsity for the fixed and random functional effects and under a non-separable beta-Bernoulli prior on the mixing proportions, enables sharing of information across the fixed and random functional effects. The proposed method provides a flexible framework for selecting the important fixed and random functional effects, while also simultaneously estimating these dynamic effects. While several methods have been developed for variable selection for fixed and random effects in linear mixed-effects models \citep{chen2003random,bondell2010joint,ibrahim2011fixed,lin2013fixed,hui2017joint,rashid2020modeling} using frequentist and Bayesian approaches, to the best of our knowledge, this is the first work exploring functional fixed and random effect selection (and hence a first group selection problem of fixed and random effects in mixed effects models) in multilevel functional regression models.

The key methodological contributions of this article include i) A joint variable selection approach of fixed and random functional effects in multilevel functional mixed models using a semiparametric spline-based formulation. ii) Employing a computationally efficient MAP estimation strategy using an Expectation Conditional Maximization (ECM) algorithm under a spike-and-slab group lasso (SSGL) prior on the fixed effect coefficients and random effect covariance parameters of interest corresponding to a Cholesky decomposition of the random effect covariance matrix.  iii) An automated data-driven procedure for tuning parameter selection for fixed and random effects, which are allowed to be different for these two components, thus accounting for the scale difference in the two kinds of parameters. The proposed method also directly provides a way of doing group-level selection of fixed and random effects in linear mixed effects models. Note that the proposed method for the multilevel functional mixed effects selection (MuFuMES) is flexible and can accommodate both scalar or functional covariates, which may or may not vary within the repeated observations of the clusters.

Numerical analysis using simulations illustrates the satisfactory selection and estimation accuracy of the proposed method. Finally, the proposed MuFuMES method is applied to accelerometer data from the 2011-2012 cohort of the National Health and Nutrition Examination
Survey (NHANES) to identify and understand the age and race-specific heterogeneity
in covariate effects on the diurnal patterns of physical activity across the lifespan. The rest of this article is organized as follows. We introduce the modeling framework and the multilevel functional mixed effect model in Section \ref{sec:method1} and illustrate the proposed variable selection method MuFuMES. The finite sample performance of the proposed method is investigated via simulations in Section \ref{sim}. Real data application of the proposed  MuFuMES method is demonstrated on the NHANES 2011-12 accelerometer data in Section \ref{case study}. We conclude with a discussion on the strengths and limitations of the proposed method 
and some possible extensions of this work in Section \ref{Disc}.

\section{Methodology}
\label{sec:method1}
\subsection{Modeling Framework}
\label{mf1}
We denote the functional response for the $j$-th replication within the $i$-th cluster as $Y_{ij}(s), s\in \mathcal{S}$ ($i=1,\ldots,n$, $j=1,\ldots,J_i$), where $\mathcal{S}$ is a compact domain. In practice, the functional response is often observed on a finite set of points. In this article, we assume the functions are observed on a dense and regular grid $\mathcal{T}_{m}= \{s_{1},s_{2},\ldots,s_{m} \} \subset \mathcal{S}=[0,1]$, without loss of generality, although this can be relaxed and easily extended to accommodate more general scenarios, e.g., sparse and irregular domain. For example, in our motivating application, $Y_{ij}(s)$ denotes the diurnal PA pattern of the $j$-th participant within the $i$-th cluster (age-by-race cluster) at time-of-day $\mathcal{S}$. The fixed effect predictors of interest are denoted by $\*X_{ij}=(X_{ij1},\ldots,X_{ijp})^T\in \^R^p$ (e.g., BMI, Gender, Income, etc.) and the random effect predictors of interest are denoted by $\*Z_{ij}=(Z_{ij1},\ldots,Z_{ijq})^T\in \^R^q$. Often we set $\*Z_{ij}=\*X_{ij}$ but this is not necessary \citep{bondell2010joint,cui2022fast}. We allow for both fixed and random intercepts, by absorbing them in the $X_{ij1},Z_{ij1}$, respectively. We posit the following multilevel functional mixed effects model,

\begin{equation}
    Y_{ij}(s)=\sum_{k=1}^pX_{ijk}\beta_k(s)+\sum_{r=1}^qZ_{ijr}u_{ir}(s)+v_{ij}(s)+\epsilon_{ij}(s). \label{mfmem}
\end{equation}
The functional fixed effects (including intercepts) are denoted by $\beta_k(\cdot), k=1,\ldots,p$ and the functional random effects (including intercepts) are denoted by $u_{ir}(\cdot), r=1,\ldots,q$, which are generally assumed to be smooth \citep{parodi2018simultaneous,ghosal2021variable}. Here $v_{ij}(s)$ denotes the cluster-subject specific deviation (for subject $j$ within cluster $i$) which is assumed to be a zero mean stochastic process with unknown covariance structure and independent of cluster-specific effects $u_{ir}(s)$. The measurement error $\epsilon_{ij}(s)$ are assumed to be i.i.d and distributed as $N(0,\sigma^2)$. The within-curve correlations are introduced through the functional random effects $u_{ir}(\cdot)$ and $v_{ij}(\cdot)$. Note that, for each fixed $s\in \mathcal{S}$, the model (\ref{mfmem}) corresponds to the classical linear mixed effects model \citep{bondell2010joint,ibrahim2011fixed}. Hence the proposed model serves as a generalization of the linear mixed effects model in functional domains. In this article, our primary objective is to simultaneously select the important fixed and random predictors among  $\*X_{ij}, \*Z_{ij}$ and to also estimate their smooth effects.  We illustrate the multilevel functional mixed effects
selection (MuFuMES) approach in the following section.

\subsection{Multilevel Functional Mixed Effects
Selection}
We model the fixed functional effects using cubic B-spline basis expansions as $\beta_{k}(s)=\sum_{u=1}^{d} \gamma_{ku}B_{ku}(s)$ ($k=1,\ldots,p$), here $\bm\gamma_k=(\gamma_{k1},\ldots,\gamma_{kd})^T$ are unknown fixed basis coefficients. Similarly, we model the cluster-level functional random effects using cubic B-spline basis expansions as $u_{ir}(s)=\sum_{v=1}^{d'} \eta_{irv}B_{rv}(s)$ ($r=1,\ldots,q$), where $\bm\eta_{ir}=(\eta_{ir1},\ldots,\eta_{ird})^T$ are random coefficients. Finally, the cluster-subject level functional random error  $V_{ij}(s)$ is also modelled using a cubic B-spline basis expansion as $v_{ij}(s)=\sum_{l=1}^{L} \zeta_{ijl}B_{l}(s)$. In this article, we have used B-spline basis functions for both these effects; however, in general, any basis functions can be used. Plugging in these basis expansions in model (\ref{mfmem}), we get
\begin{eqnarray}
    Y_{ij}(s) &=\sum_{k=1}^pX_{ijk}\sum_{u=1}^{d} \gamma_{ku}B_{ku}(s)+\sum_{r=1}^qZ_{ijr}\sum_{v=1}^{d'} \eta_{irv}B_{rv}(s)+\sum_{l=1}^{L} \zeta_{ijl}B_{l}(s)+\epsilon_{ij}(s) \notag \\
    &=\sum_{k=1}^p\sum_{u=1}^{d} \gamma_{ku} X_{ijk}B_{ku}(s)+\sum_{r=1}^q\sum_{v=1}^{d'} \eta_{irv}Z_{ijr}B_{rv}(s)+\sum_{l=1}^{L} \zeta_{ijl}B_{l}(s)+\epsilon_{ij}(s) \notag\\
    &= \tilde{X}_{ij}(s)^T\bm\gamma+ \tilde{Z}_{ij}(s)^T\bm\eta_i+\tilde{W}_{ij}(s)^T\bm\zeta_{ij} +\epsilon_{ij}(s).
\end{eqnarray}
Here $\tilde{X}_{ij}(s)^T$ denotes the stacked row vector of $\{X_{ijk}B_{ku}(s)\}_{k=1,u=1}^{p,d}$ with length $dp$, $\tilde{Z}_{ij}(s)^T$ denotes the stacked row vector of $\{Z_{ijr}B_{rv}(s)\}_{r=1,v=1}^{q,d'}$ with length $d'q$ and $\tilde{W}_{ij}(s)^T=(B_{1}(s),B_{2}(s),\ldots,B_{L}(s))$. Similarly $\bm\gamma=(\gamma_{11},\gamma_{12},\ldots,\gamma_{pd})^T$ is the stacked fixed effect basis coefficients, $\bm\eta_i=(\eta_{i11},\eta_{i12},\ldots,\eta_{iqd'})^T$ is the stacked random effect basis coefficients for cluster $i$, and $\bm\zeta_{ij}=(\zeta_{ij1},\zeta_{ij2},\ldots,\zeta_{ijL})^T$. Now stacking the terms $Y_{ij}(s),e_{ij}(s)$ and $\tilde{X}_{ij}(s)^T,\tilde{Z}_{ij}(s)^T$, $\tilde{W}_{ij}(s)^T$ first across the observed functional domain $\mathcal{S}= \{s_{1},s_{2},\ldots,s_{m} \}$ and then across replications $j=1,\ldots,J_i$ we can reformulate (2) as,
\begin{eqnarray}
    \*Y_{i}= \^X_{i}\bm\gamma + \^Z_{i}\bm\eta_i+\^W_{i}\bm\zeta_i+\bm\epsilon_i. \label{vecmod}
\end{eqnarray}
Here $\*Y_{i}=(Y_{i1}(s_1),\ldots,Y_{i1}(s_m),Y_{i2}(s_1),\ldots,Y_{i2}(s_m),\ldots,Y_{iJ_i}(s_1),\ldots,Y_{iJ_i}(s_m))^T$ is the stacked response vector and $\bm\epsilon_i$ is defined analogously. $\^X_{i}=(\tilde{X}_{i1}(s_1),\ldots,\tilde{X}_{i1}(s_m),\ldots,\tilde{X}_{iJ_i}(s_1),\ldots,\tilde{X}_{iJ_i}(s_m))^T$ is the stacked known basis matrix of dimension $mJ_i\times dp$. We denote by the stacked matrix $\^Z_{i}=(\tilde{Z}_{i1}(s_1),\ldots,\tilde{Z}_{i1}(s_m),\ldots,\tilde{Z}_{iJ_i}(s_1),\ldots,\tilde{Z}_{iJ_i}(s_m))^T$, a known basis matrix of dimension $mJ_i\times d'q$ and similarly define $\^W_{i}$. We assume in model (\ref{vecmod}), $\bm\eta_i \sim N(0,\^D)$, $\bm\zeta_i \sim N(0,\*\Omega)$ and stacked error vector $\bm\epsilon_i \sim N(0, \sigma^2\^I)$, where $\^I$ is the identity matrix of dimension $mJ_i\times mJ_i$, and they are independent and identically distributed across $i$.

For the purpose of selecting the important fixed and random functional predictors, it is crucial to impose sparsity in the fixed effect coefficients $\bm\gamma$ and eliminate appropriate variance components in $\^D$ \citep{chen2003random,bondell2010joint,ibrahim2011fixed} of the cluster-level random effects. For this purpose, we use a Cholesky decomposition technique \citep{ibrahim2011fixed} of the covariance matrix $\^D=\^L\^L^T$. Such Cholesky decomposition or modified Cholesky decompositions have been used in the linear mixed models for enforcing sparsity in the cluster-level random effect components \citep{chen2003random,bondell2010joint,ibrahim2011fixed}. Sparsity can be enforced in the fixed functional effects by enforcing group sparsity in the fixed effect coefficients $\bm\gamma$ \citep{ghosal2020variable,bai2023scalable}. This ensures the functional effect is zero only if all the basis coefficients corresponding to the coefficient function are zero, thus giving rise to a natural group selection problem. For enforcing sparsity in the random functional effects, we will illustrate how a group-based shrinkage on the rows of $\^L$ can be used. Writing the random coefficient  $\bm\eta_i =\^L\*b_i$, where $\*b_i\sim N(0,\^I_{d'q})$, we reformulate model (\ref{vecmod}) as,
\begin{eqnarray}
    \*Y_{i}= \^X_{i}\bm\gamma + \^Z_{i}\^L\* b_i+\^W_{i}\bm\zeta_i+\bm\epsilon_i. \label{vecmodfinal}
\end{eqnarray}
We also assume that $\^D$ is positive semidefinite, which allows certain components of $\bm\eta_i=\^L\*bi$ to be zero, with probability 1.
Next, we endow the unknown parameters $\bm\gamma$ and $\^L$, $\*\Omega$ with appropriate priors and illustrate a MAP estimation strategy that maximizes the posterior distribution using an Expectation Conditional Maximization (ECM) algorithm \citep{bai2023scalable,mclain2025efficient}. 

\subsection{Prior Specification for Fixed and Random Effect Parameters}



We use a spike-and-slab group lasso (SSGL) prior \citep{bai2022spike,bai2023scalable} on the vector of basis coefficients $\bm\gamma = (\bm\gamma_1^T, \dots, \bm\gamma_p^T)^T$ to enforce fixed effect selection, where the grouping corresponds to each functional fixed effect $\beta_k(\cdot)$. The SSGL prior is given by,
\begin{align}
\pi(\bm\gamma \mid \theta) = \prod_{k=1}^p [(1 - \theta) \Psi(\bm\gamma_k \mid \lambda_0) + \theta \Psi(\bm\gamma_k \mid \lambda_1)]. \label{SSGL pr} 
\end{align}
Here $\theta \in (0, 1)$ denotes a mixing proportion, capturing the expected proportion of nonzero $\gamma_k$' coefficient vectors. We denote by $\Psi(\cdot \mid \lambda)$ a multivariate Laplace density with hyperparameter $\lambda$,
\begin{align}
\Psi(\bm\gamma_k \mid \lambda) = \frac{\lambda^d e^{-\lambda ||\bm\gamma_k||_2}}{2^d \pi^{(d-1)/2} \Gamma((d+1)/2)}, \quad k = 1, \dots, p.
\end{align}
Under the above SSGL prior denoted as $SSGL(\lambda_0, \lambda_1, \theta)$, the posterior mode for $\bm\gamma$ becomes exactly sparse \citep{bai2022spike} and hence can used for simultaneous estimation and variable selection. Now, to enforce sparsity in the random effects, we use a group-based shrinkage on the rows of $\^L$ in (4). In particular, let us denote a partition of the $d'q\times d'q$ lower triangular matrix $\^L$ as $\^L=\left[
\^L_1^T,\ldots,\^L_q^T
\right]^T$, where each $\^L_r$ ($r=1,\ldots q$) has $d'$ rows and $d'q$ columns. Now if an entire block $\^L_r$ is zero, the variance-covariance components in $\^D$ corresponding to the random effects $\bm\eta_{ir}=(\eta_{ir1},\ldots,\eta_{ird'})$ and it's covariance with $\bm\eta_{is}$ ($s\neq r$) is zero, which ensures that the functional random effects $u_{ir}(\cdot)$ is zero for all $i$. We further denote the stacked (by row) nonzero elements of the matrix $\^L_r$ as $\tilde{\*L}_{r}$ and define $\tilde{\*L}=(\tilde{\*L}_{1}^T,\ldots,\tilde{\*L}_{q}^T)^T$. We endow on $\tilde{\*L}$ a SSGL prior $SSGL(\nu_0, \nu_1, \theta^{*})$, given by
\begin{align}
\pi(\tilde{\*L} \mid \theta^*) = \prod_{r=1}^q [(1 - \theta^*) \Psi(\tilde{\*L}_{r} \mid \nu_0) + \theta^* \Psi(\tilde{\*L}_{r} \mid \nu_1)]. \label{SSGL pr2} 
\end{align}
Here $\theta^* \in (0, 1)$ is a mixing proportion capturing the expected proportion of nonzero $\tilde{\*L}_{r}$ and the multivariate Laplace density $\Psi(\tilde{\*L}_{r} \mid \nu )$ is given by,
\begin{align}
\Psi(\tilde{\*L}_{r} \mid \nu) = \frac{\nu^{N_{r}} e^{-\nu ||\tilde{\*L}_{r}||_2}}{2^{N_{r}} \pi^{(N_{r}-1)/2} \Gamma((N_{r}+1)/2)}, \quad r = 1, \dots, q.
\end{align}
Here $N_{r}$ denotes the length of the vector $\tilde{\*L}_{r}$ and hence the  $SSGL(\nu_0, \nu_1, \theta^{*})$ prior adopts to the varying group size in $\tilde{\*L}$, ensuring that the overall amount of shrinkage remains comparable. In the SSGL priors (5,7) we set $\lambda_0>>\lambda_1$ (and similarly $\nu_0>>\nu_1$) so the spike component $\Psi(\bm\gamma_k \mid \lambda_0)$ (or $\Psi(\tilde{\*L}_{r} \mid \nu_0)$) is heavily concentrated around the zero vector. The slab component $\Psi(\bm\gamma_k \mid \lambda_1)$ (or $\Psi(\tilde{\*L}_{r} \mid \nu_1)$) prevents shrinking parameters with large magnitude. This is an advantage of SSGL over other group penalties such as group lasso \citep{yuan2006model} or its nonconvex extensions such as the group minimax concave
penalty (MCP), since SSGL can perform adaptive shrinkage due to the two tuning parameters (spike and slab) controlling the sparsity as opposed to one.
Hence, the groups with larger coefficients can be identified due to the minimal shrinkage imposed by the slab component. Another key innovation of the proposed SSGL prior for mixed effect selection is that we enhance flexibility by allowing the fixed and random components to have different degrees of sparsity, accounting for their scale difference via different sets of tuning parameters $(\lambda_0, \lambda_1)$ and $(\nu_0, \nu_1)$.

The unknown mixing proportions $\theta$ and $\theta^*$  are endowed with beta priors,
\begin{equation}
    \theta\sim \mathcal{B}(a_0,b_0), \theta^*\sim \mathcal{B}(a_1,b_1).
\end{equation}
Here, $a_0,b_0,a_1,b_1> 0$ are positive constant hyperparameters. These priors on $\theta$ and $\theta^*$ render our Bayesian penalty in (5, 7) non-separable, making the groups $\bm\gamma_k$ and $\^L_r$ a priori dependent \citep{bai2022spike,bai2023scalable}, and enabling sharing of information across groups. For particular choices of hyperparameters such as $a_0=1,b_0=p$ and $a_1=1,b_1=q$, the SSGL prior has been shown to perform automatic multiplicity adjustment and favor parsimonious models in higher dimensions \citep{scott2010bayes}. For the covariance parameter $\*\Omega$ we use an Inverse-Wishart prior which is conditionally conjugate,
\begin{equation}
    \*\Omega \sim {IW}(\nu, \*\Delta), \nu>L-1, \textit{ $\*\Delta$ is positive definite.} 
\end{equation}

Finally, for the error variance $\sigma^2$, we use the Inverse-Gamma prior,
\begin{equation}
    \sigma^2 \sim IG(c_0/2,d_0/2),
\end{equation}
where $c_0,d_0$ are hyperparameters. We have used  $c_0=1,d_0=1$ throughout this article, which results in a weakly informative prior.
Next, based on the model formulation in (4) and the above priors, we illustrate a computationally efficient and scalable Expectation Conditional Maximization (ECM) algorithm that maximizes the posterior distribution, i.e., performs a MAP estimation.
\subsection{ECM Algorithm for MAP Estimation}
Let us denote the set of unknown parameters in model (4) and in priors (5, 7, 9, 10, 11) as $\bm\Phi=\{\bm\gamma,\tilde{\*L},\*b,\bm\zeta,\theta,\theta^*,\*\Omega,\sigma^2\}$. Here $\*b=(\*b_1^T,\ldots, \*b_n^T)^T$ and $\bm\zeta=(\bm\zeta_1^T,\ldots, \bm\zeta_n^T)^T$ . Note that the matrix $\^L$ in model (4) is directly related to $\tilde{\*L}$ as $vec(\^L)=\^J\tilde{\*L}$, for a non-singular $(d'q)^2 \times \frac{d'q(d'q+1)}{2}$ dimensional matrix $\^J$ \citep{ibrahim2011fixed} which transforms $\tilde{\*L}$ to $vec(\^L)$ (vectorization of the matrix $\^L$). Hence, we can rewrite the model (4) as,
$
\*Y_{i}= \^X_{i}\bm\gamma +  (b_i^T \otimes \^Z_{i})\^J\tilde{\*L} +\^W_{i}\bm\zeta_i+\bm\epsilon_i. \label{vecmodfinal2}
$
For obtaining the MAP estimator, the log-posterior density (up to an additive constant) is given by,
\begin{eqnarray}
    log \{\pi (\bm\Phi|\*Y,
    \^X,\^Z,\^W)\} = -\frac{N}{2}log(\sigma^2)-\sum_{i=1}^n \frac{|| \*Y_{i}- \^X_{i}\bm\gamma - (b_i^T \otimes \^Z_{i})\^J\tilde{\*L}-\^W_{i}\bm\zeta_i ||_2^2}{2\sigma^2} \notag\\
   -\frac{1}{2}\sum_{i=1}^n\*b_i^T\*b_i+\sum_{k=1}^{p} \log \{ (1 - \theta) \lambda_0^d e^{-\lambda_0 \|\bm\gamma_k\|_2} + \theta \lambda_1^d e^{-\lambda_1 \|\bm\gamma_k\|_2} \}\notag\\+
    \sum_{r=1}^{q} \log \{ (1 - \theta^*) \nu_0^d e^{-\nu_0 \|\tilde{\*L}_{r}\|_2} + \theta^* \nu_1^d e^{-\nu_1 \|\tilde{\*L}_{r}\|_2} \} + (a_0 - 1) \log \theta + (b_0 - 1) \log (1 - \theta)\notag\\+ (a_1 - 1) \log \theta^* + (b_1 - 1) \log (1 - \theta^*) -\left( \frac{c_0 + 2}{2} \right) \log \sigma^2 - \frac{d_0}{2 \sigma^2}\notag\\
    +\frac{n}{2} log (det(\*\Omega^{-1})) -\frac{1}{2}\sum_{i=1}^n \bm\zeta_i^T\bm\zeta_i+ \frac{\nu+L+1}{2} log (det(\*\Omega^{-1}))  -\frac{1}{2} tr(\*\Delta\*\Omega^{-1}).  \hspace{1 mm}
\end{eqnarray}
Now to perform variable selection for the fixed and random effects, we further introduce latent binary indicators $\tau_k \in\{0,1\}, k=1,\ldots,p$ and $\tau_r^* \in \{0,1\},r=1,\ldots,q$, which indicates if the effect is coming from the slab component or the spike component. In particular, we posit the following hierarchical formulation for the SSGL priors $SSGL(\lambda_0, \lambda_1, \theta)$ and $SSGL(\nu_0, \nu_1, \theta^*)$ as the marginal priors under beta-Bernoulli priors $\pi(\bm\tau \mid \theta)$ and $\pi(\bm\tau^* \mid \theta^*)$,
\begin{eqnarray}
    \pi(\bm\gamma \mid \bm\tau) = \prod_{k=1}^p [(1 - \tau_k) \Psi(\bm\gamma_k \mid \lambda_0) + \tau_k \Psi(\bm\gamma_k \mid \lambda_1)], \notag \\
     \pi(\bm\tau \mid \theta)=\prod_{k=1}^p\theta^{\tau_k} (1-\theta)^{1-\tau_k},
\end{eqnarray}
and similarly,
\begin{eqnarray}
   \pi(\tilde{\*L} \mid \tau^*) = \prod_{r=1}^q [(1 - \theta^*) \Psi(\tilde{\*L}_{r} \mid \nu_0) + \theta^* \Psi(\tilde{\*L}_{r} \mid \nu_1)]. , \notag \\
     \pi(\bm\tau^* \mid \theta^*)=\prod_{r=1}^q(\theta^*)^{\tau_r^*} (1-\theta^*)^{1-\tau_r^*},
\end{eqnarray}
where $\bm\tau=(\tau_1,\ldots,\tau_p)^T$ and $\bm\tau^*=(\tau_1^*,\ldots,\tau_q^*)^T$, are unknown. The augmented log-posterior $log \{\pi (\Phi,\bm\tau,\bm\tau^*|\*Y,
    \^X,\^Z,\^W)\}$ is given by:
  \begin{eqnarray}
    log \{\pi (\bm\Phi,\bm\tau,\bm\tau^*|\*Y,
    \^X,\^Z,\^W)\} = -\frac{N}{2}log(\sigma^2)-\sum_{i=1}^n \frac{|| \*Y_{i}- \^X_{i}\bm\gamma - (b_i^T \otimes \^Z_{i})\^J\tilde{\*L}-\^W_{i}\bm\zeta_i  ||_2^2}{2\sigma^2} \notag \\-\frac{1}{2}\sum_{i=1}^n\*b_i^T\*b_i +\sum_{k=1}^{p} \log \{ (1 - \tau_k) \lambda_0^d e^{-\lambda_0 \|\bm\gamma_k\|_2} + \tau_k \lambda_1^d e^{-\lambda_1 \|\bm\gamma_k\|_2} \}\notag\\+
    \sum_{r=1}^{q} \log \{ (1 - \tau_r^*) \nu_0^d e^{-\nu_0 \|\tilde{\*L}_{r}\|_2} + \tau_r^* \nu_1^d e^{-\nu_1 \|\tilde{\*L}_{r}\|_2} \} \notag\\+ (a_0 - 1+\sum_{k=1}^p \tau_k) \log \theta + (b_0 - 1+p-\sum_{k=1}^p \tau_k) \log (1 - \theta)+ (a_1 - 1+\sum_{r=1}^q \tau_r^*) \log \theta^* + \notag\\ (b_1 - 1+q-\sum_{r=1}^q \tau_r^*) \log (1 - \theta^*) - \left( \frac{c_0 + 2}{2} \right) \log \sigma^2 - \frac{d_0}{2 \sigma^2}\notag\\
     +\frac{n}{2} log (det(\*\Omega^{-1})) -\frac{1}{2}\sum_{i=1}^n \bm\zeta_i^T\bm\zeta_i+ \frac{\nu+L+1}{2} log (det(\*\Omega^{-1}))  -\frac{1}{2} tr(\*\Delta\*\Omega^{-1}).  \hspace{4 mm}
\end{eqnarray}
Treating the indicators $\bm\tau,\bm\tau^*$ as missing data, we follow an ECM algorithm \citep{meng1993maximum} to calculate the expected log-posterior $Q(\bm\Phi \mid \bm\Phi^{(t-1)} )=E_{\bm\tau,\bm\tau^*}( log \{\pi (\bm\Phi,\bm\tau,\bm\tau^*|\*Y,
    \^X,\^Z,\^W)\} \mid \bm\Phi^{(t-1)})$ given the current parameter estimates at $(t-1)$-th stage in the first step (E-step) and then iteratively maximize the expected log-posterior $Q(\bm\Phi \mid \bm\Phi^{(t-1)})$ iteratively with respect to the parameters of interest $\bm\Phi$ (CM step). The E and CM steps are iterated until convergence. For brevity, we present the explicit details of the E and the CM steps in Appendix A of the supplementary material. We summarize the key steps below as an algorithm in Algorithm \ref{algorithm1} (refer to the supplemental equations for further details).

\subsection{Choice of Tuning Parameters}
The proposed MuFuMES method illustrated above involves several tuning parameters: $\lambda_0>>\lambda_1$ and  $\nu_0>>\nu_1$, which are the spike and slab parameters corresponding to the fixed effect and random effect variance components, respectively. We fix the slab parameters in the SSGL priors
$SSGL(\lambda_0, \lambda_1, \theta)$ and $SSGL(\nu_0, \nu_1, \theta^*)$ to be $\lambda_1=1,\nu_1=1$. The spike parameters are chosen in a data-driven way, based on a two-dimensional grid search, from a grid of decreasing $\lambda_0$ values and $\nu_0$ values \citep{bai2023scalable}. To accommodate the varying group sizes in $\tilde{\*L}_r$ we additionally scale up $\nu_0$ by $\sqrt{N_r}$ for each group while implementing the group-LASSO optimization \citep{bai2022spike} within the CM step. The optimal $(\lambda_0,\nu_0)$ combination is chosen based on a BIC-type criterion \citep{bondell2010joint,bai2023scalable} defined as:
\begin{equation}
BIC(\lambda_0,\nu_0)=-2\sum_{i=1}^{n}\ell(\*Y_i,\hat{\Phi})+ log(N)\times df_{\lambda_0,\nu_0}.
\end{equation}
Here $\hat{\Phi}$ denotes the MAP estimator, $\ell$ is the marginal log-likelihood of $\*Y_i$ based on model (4) with marginal covariance $\*\Sigma_i=\^Z_i\^D\^Z_i^T+\^W_i\*\Omega\^W_i^T+\sigma^2\^I$. Here $df_{\lambda_0,\nu_0}$ denotes the total number of nonzero elements in the MAP estimator $\hat{\bm\gamma}$ and $\hat{\tilde{\*L}}$. The hyperparameters are set to $a_0=1,b_0=p$ and $a_1=1,b_1=q$ favouring parsimonious models. We also set the hyperparameters $\nu=L+2, \Delta=\^I$, and $c_0=d_0=1$, resulting in weakly informative priors. Finally, we have used a moderate number of basis functions to control the smoothness of the fixed and random functional effects using a truncated basis approach \citep{fan2015functional}. This can also be chosen in a data-driven way based on the BIC criterion (16) and is illustrated in our application.

\section{Simulation Study}
\label{sim}
We investigate the performance of the proposed MuFuMES method using numerical simulations. 
We generate observations following a multilevel functional mixed effects model (denoted as Scenario A) given by,
\begin{equation}
    Y_{ij}(s)=\sum_{k=1}^{8}X_{ijk}\beta_k(s)+\sum_{r=1}^8Z_{ijr}u_{ir}(s)+v_{ij}(s)+\epsilon_{ij}(s), s\in [0,1] \label{mfmem: sim1},
\end{equation}
for clusters $i=1,\ldots,n$ and replications $ j=1,\ldots,J_i$. In this model, we consider $p=8$ covariates for functional fixed effects (including a fixed intercept) and $q=8$ covariates for functional random effects (including a random intercept). The fixed effects coefficient 

\begin{algorithm}[H]
	\begin{flushleft}
		\textbf{Input:} Initial values $\bm{\gamma}^{(0)}$, $\tilde{\*L}^{(0)}$,$\*b^{(0)}$,$\bm\zeta^{(0)}$,$\theta^{(0)}$,$\theta^{*^{(0)}}$, $\*\Omega^{(0)}$,$\sigma^{2(0)}$, $t=0$, and fixed $\lambda_0,\nu_0,\lambda_1=1,\nu_1=1$, grouping vectors $\*G_1,\*G_2$ for $\bm{\gamma}$, $\tilde{\*L}$.  \\
		\textbf{Output:} Selected covariates with fixed (among $X_{ijk}$, $k = 1, \ldots, p$) and random (among $Z_{ijr}$, $r = 1, \ldots, q$) functional effects, and estimates of fixed functional effects $\widehat{\beta}_k(\bm{s}),~k = 1, \ldots, p$, and MAP estimates $\hat{\tilde{\*L}}$ (or $\^L$),$\hat{\*b}$.
		\vspace{.2cm}
		
		\textbf{while} $\text{diff}_1 > \epsilon_1$ or $\text{diff}_2 > \epsilon_2$ \textbf{do}
		\begin{enumerate}
			\item Increment $t$.
			\item
			\textbf{E-step}

			\begin{enumerate}
				\item For $k = 1, \ldots, p$, compute $p_k(\bm\gamma_k^{(t-1)},\theta^{(t-1)})=\frac{\theta^{(t-1)}\Psi(\bm\gamma_k^{(t-1)}\mid \lambda_1)}{\theta^{(t-1)}\Psi(\bm\gamma_k^{(t-1)}\mid \lambda_1)+(1-\theta^{(t-1)})\Psi(\bm\gamma_k^{(t-1)}\mid \lambda_0)}$ .
                \item For $r = 1, \ldots, q$, compute $p_r(\tilde{\*L}_{r}^{(t-1)},\theta^{*^{(t-1)}})=\frac{\theta^{*^{(t-1)}}\Psi(\tilde{\*L}_{r}^{(t-1)}\mid \nu_1)}{\theta^{*^{(t-1)}}\Psi(\tilde{\*L}_{r}^{(t-1)}\mid \nu_1)+(1-\theta^{*^{(t-1)}})\Psi(\tilde{\*L}_{r}^{(t-1)}\mid \nu_0)}$ .
				\item Calculate the expected log-posterior $Q(\bm\Phi \mid \bm\Phi^{(t-1)})$.
			\end{enumerate}
			\item
			\textbf{CM-step}
			\begin{enumerate}
				\item Update $\theta^{(t)}=\frac{a_0-1+\sum_{k=1}^p p_k}{a_0+b_0+p-2}$ and Update $ \theta^{*^{(t)}}=\frac{a_1-1+\sum_{r=1}^q p_{r^*}}{a_1+b_1+q-2}$ .
				\item For $i = 1, \ldots, n$, update $ \*b_i^{(t)}=\{(\^Z_{i}\^L^{(t-1)})^T(\^Z_{i}\^L^{(t-1)})+\sigma^{2^{(t-1)}}\^I\}^{-1}(\^Z_{i}\^L^{(t-1)})^T(\*Y_i-\^X_i\bm\gamma^{(t-1)}-\^W_i\bm\zeta_i^{(t-1)})$ and $\bm\zeta_i^{(t)}=\{\^W_{i}^T\^W_{i}+\sigma^{2^{(t-1)}}\{\*\Omega^{(t-1)}\}^{-1}\}^{-1}(\^W_{i})^T(\*Y_i-\^X_i\bm\gamma^{(t-1)}-\^Z_i\^L^{(t-1)} \*b_i^{(t)})$.
                
				\item Update $\bm{\gamma}^{(t)}=\underset{(\bm\gamma)}{\operatorname{\argmin}} \sum_{i=1}^n || \*Y_{i}- \^X_{i}\bm\gamma - (\*b_i^{{(t)}^T} \otimes \^Z_{i})\^J\tilde{\*L}^{(t-1)} -\^W_i\bm\zeta_i^{(t-1)}||_2^2+\sum_{k=1}^{p}2\lambda_k^*\sigma^{2^{(t-1)}}\|\bm\gamma_k\|_2 $.
                \item Update  $\tilde{\*L}^{(t)}=\underset{(\tilde{\*L})}{\operatorname{\argmin}}\sum_{i=1}^n || \tilde{\*Y}_{i}^2- (\*b_i^{{(t)}^T} \otimes \^Z_{i})\^J\tilde{\*L} ||_2^2+\sum_{r=1}^{q}2\nu_r^*\sigma^{2^{(t-1)}}\|\tilde{\*L}_{r}\|_2, \tilde{\*Y}_{i}^2=\*Y_{i}-\^X_{i}\bm\gamma^{(t)}-\^W_i\bm\zeta_i^{(t)}$.
				\item Update $\*\Omega^{(t)}=\frac{1}{n+\nu+L+1} (\*\Delta+\sum_{i=1}^n\zeta_i^{(t)}(\zeta_i^{(t)})^T)$ and $\sigma^{2(t)}=\sum_{i=1}^n \frac{|| \*Y_{i}- \^X_{i}\bm\gamma^{(t)} - (\*b_i^{{(t)}^T} \otimes \^Z_{i})\^J\tilde{\*L}^{(t)}-\^W_i\bm\zeta_i^{(t)} ||_2^2+d_0}{N+c_0+2}.$
			\end{enumerate}
			\item Set $\text{diff}_1 = \lVert \bm{\gamma}^{(t)} - \bm{\gamma}^{(t-1)} \rVert_2^2 / \lVert \bm{\gamma}^{(t-1)} \rVert_2^2$ and $\text{diff}_2 = \lVert \tilde{\*L}^{(t)}- \tilde{\*L}^{(t-1)} \rVert_2^2 / \lVert \tilde{\*L}^{(t-1)} \rVert_2^2$. 
		\end{enumerate}
		\item 
		\textbf{return} Indices of selected covariates with fixed (among $X_{ijk}$, $k = 1, \ldots, p$) and random (among $Z_{ijr}$, $r = 1, \ldots, q$) functional effects. $\widehat{\beta}_k(\bm{s}) = \sum_{u=1}^{d} \widehat{\gamma}_{ku} B_{ku}(s)$, $k=1, \ldots, p$, $\hat{\tilde{\*L}}$,$\hat{\*b}$. 
	\end{flushleft}
	\caption{ECM algorithm for MAP estimation under MuFuMES} \label{algorithm1}
\end{algorithm}
functions are given by $\beta_1(s)=8sin(2\pi s)$ (intercept), $\beta_2(s)=2\bm\phi(s,0.6,0.15^2)$, $\beta_3(s)=2.5\bm\phi(s,0.6,0.15^2)$, $\beta_4(s)=3cos(2\pi s)$, $\beta_5(s)=5sin(2\pi s)+5cos(2\pi s)$ and $\beta_k(s)=0$ for $k=6,7,\ldots,11$. Here, we denote by $\bm\phi(s,a,b^2)$ the density at $s$ for Normal distribution with mean $a$ and variance $b^2$.  So, only the first 5 fixed effect covariates are relevant. The fixed effect covariates $X_{ijk}$ are independently generated from a $\mathcal{N}(0,2^2)$ distribution for $k=2,\ldots,8$ and $X_{ij1}=1$ (for all $i,j$) corresponds to the intercept. 
The random effect covariates $Z_{ijr}$ are exactly the same as the fixed effect covariates $X_{ijk}$ \citep{bondell2010joint} in this scenario.
The functional random effects are given by $u_{i1}(s)=c_{i1}sin(2\pi s)+d_{i1}cos(2\pi s)$ (random intercept), where $c_{i1}\sim \mathcal{N}(0,3^2\sigma^2_B),d_{i1}\sim \mathcal{N}(0,1.5^2\sigma^2_B)$. Similarly, $u_{i4}(s)=c_{i4}sin(2\pi s)+d_{i4}cos(2\pi s)+e_{i4}sin(\pi s)+f_{i4}cos(\pi s)$, where $c_{i4}\sim \mathcal{N}(0,1.5^2\sigma^2_B),d_{i4}\sim \mathcal{N}(0,0.75^2\sigma^2_B),e_{i4}\sim \mathcal{N}(0,0.5^2\sigma^2_B),f_{i4}\sim \mathcal{N}(0,0.25^2\sigma^2_B)$. 
The rest of the functional random effects $u_{ik}(s)$ are considered to be zero. Hence only the first and the fourth random effects of the covariates are important. We set $\sigma^2_B$ based on $SNR_B=0.5$, where $SNR_B$ is the standard deviation of the fixed effects functions divided by the standard deviation of the random effects \citep{cui2022fast,scheipl2015functional}. The cluster-subject-level functional random effect $v_{ij}(s)$ is given by $v_{ij}(s)=v_{ij1}sin(\pi s)+v_{ij2}cos(\pi s)$, where $v_{ij1}\sim \mathcal{N}(0,0.8^2\sigma^2_S),v_{ij2}\sim \mathcal{N}(0,0.4^2\sigma^2_S)$. We set $\sigma^2_S$ based on $SNR_S=2$, where $SNR_S$ is the standard deviation of the fixed effects functions divided by the standard deviation of the cluster-visit level random effects. 
The random errors $\epsilon_{ij}(s)\sim \mathcal{N}(0,\sigma_\epsilon^2)$, where $\sigma_\epsilon$ is chosen based on
a signal to noise ratio of $SNR_{\epsilon}=4$, which represents the 
standard deviation of the linear predictors (fixed and random predictors combined) divided by the standard deviation of the noise $\sigma_\epsilon$. The functional response $Y_{ij}(s)$ is observed on a grid of $m = 10$ equidistant time points in $S=[0,1]$. Cluster size $n\in \{25,50,100\}$ is considered for this scenario, and $J_i=J=10$ replications are considered within each cluster.
We generate 100 replicated data sets to assess the performance of the proposed variable selection method. We also explore an additional simulation scenario (Scenario B), where the fixed and random effect covariates differ. The details of this scenario are presented in Appendix B of the supplementary material.

\subsection{Simulation Results}
\hspace*{- 1 mm}
\textbf{Scenario A}:\\
We evaluate the performance of the proposed MuFuMES method in terms of selection accuracy and estimation accuracy. We also compare the performance of MuFuMES to existing group selection methods that can handle the multilevel functional mixed effects model (1) or group-selection in its longitudinal mixed effects model representation in (3). Among existing candidates with available implementation for comparison, we identified I) The \texttt{glmmLasso} R package \citep{groll2014variable} which can perform fixed effect selection in linear mixed models with an $L_1$ penalization. However, this method cannot handle this scenario since it can only do fixed effect selection and does not impose group penalization, which is essential for selection of functional effects. II) The \texttt{grpreg} R package \citep{bre2015}, which can do group selection of the fixed effects using group LASSO \citep{yuan2006model}, but is unable to handle random effects or account for clustering in the data. We use the longitudinal mixed effects model representation in (3) for the selection of the functional fixed effects and compare its performance to the proposed MuFuMES for fixed effect selection. III) The \texttt{glmmPen} R package \citep{rashid2020modeling} which uses a Monte Carlo Expectation Conditional Minimization (MCECM) algorithm for fixed and random effect selection, but this is again not directly applicable or comparable to our case as this does not handle group selection of fixed effects (or structured group selection of random effects). Moreover, with $80$-dimensional fixed effect parameters ($\bm\gamma$) and a $80$-dimensional random effect parameter $\bm\eta_i$ coming from model (3) with the total number of observations $N \in {2500,5000,10000}$, we found a very high computational cost of this method (e.g., the method did not converge for a single replication after running 24 hours), which is expected as this method was not developed for functional mixed effect selection. Hence, after reviewing, we compare the selection and estimation performance of the proposed method to its closest competitor II) the fixed effect selection performance from group LASSO implemented using \texttt{grpreg}. Table \ref{tab1sel2} reports the selection performance of MuFuMES for the selection of the fixed effects and random effects in terms of true positive and false positive rates. The performance of group LASSO is also reported for the fixed effects as a competing method. The tuning parameters of the MuFuMES method are chosen based on the BIC criterion (16) and the tuning parameters of group LASSO are chosen based on a 10-fold cross-validation \citep{bre2015}.

\begin{figure}[H]
\centering
\includegraphics[width=0.8\linewidth , height=0.6\linewidth]{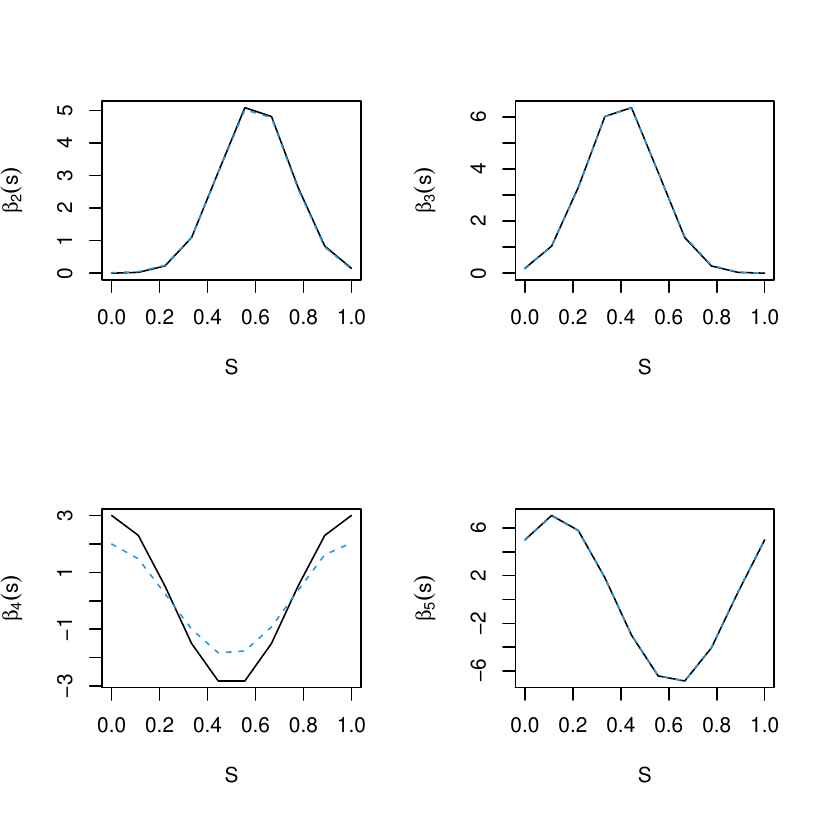}
\caption{Displayed are the true (solid) and M.C mean (dashed) of estimated functional slopes $\beta_k(s)$ (k=2,\ldots,5) from MuFuMES, $n=100$.}
\label{fig:fig2}
\end{figure}

\begin{table*}[h]
\centering
\caption{Average true positive rate (TP), false positive rate (FP) for fixed (TPF, FPF) and random (TPR, FPR)  effects of the MuFuMES method, Scenario B. The performance of group LASSO is reported in parentheses for the fixed (TPF, FPF) effects. }
\label{tab1sel2}
\begin{tabular}{ccccc}
\hline
Sample Size               & TPF & FPF    & TPR  & FPR  \\ \hline
\multirow{1}{*}{n=25} & 0.93 (0.998)  & 0.003 (0.64)& 0.98 & 0       \\ \cline{2-5} 
\multirow{1}{*}{n=50}  & 0.99 (0.996)  & 0(0.76) & 1 & 0.002       \\ \cline{2-5} 
                       
\multirow{1}{*}{n=100}  & 0.99 (1)  & 0 (0.73) & 0.99   & 0.003    \\ \cline{2-5} 
\hline                   
\end{tabular}
\end{table*}
The proposed MuFuMES method is again seen to have a negligible false positive rate ($<5\%$ in all cases) and a high true positive rate for both fixed and random functional effects across all the sample sizes. However, the naive group LASSO can be seen to have a very high false positive rate for the fixed effects, reported previously in \citep{ghosal2020variable}. To assess the estimation performance, we again display the Monte Carlo (MC) mean estimates (averaged estimated coefficient function over 100 replications) of the 
functional fixed effect slopes  $\beta_k(s)$ ($k=2,3,4,5$) in Figure \ref{fig:fig2} for sample size $n=100$ for this scenario. The estimated coefficient functions closely capture the true effects in most cases. Interestingly, in this scenario, we observe the estimated fixed effect $\beta_4(s)$ to be more biased than the other estimates, which might be due to the covariate $X_{ij4}$ having both a fixed and a random slope.


We report the mean integrated squared error (MISE) of the functional fixed effect slope estimates from the MuFuMES method in Table \ref{tab:my-table4}. We compare these performances with the estimates from the group LASSO-based approach, which only selects fixed effects and does not account for random effects and hence clustering in the data.

\begin{table}[ht]
\centering
\caption{Mean integrated squared error (MISE) of the functional fixed effect slope estimates from the MuFuMES method, scenario B. The performance of group LASSO is reported in parentheses.}
\label{tab:my-table4}
\begin{tabular}{lllll}
\hline
Sample Size & MISE $\beta_2(\cdot)$   & MISE $\beta_3(\cdot)$   & MISE $\beta_4(\cdot)$ & MISE $\beta_5(\cdot)$ \\ \hline
n=25       & 0.19 (1.25)  & 0.18 (1.05)  & 4.40 (4.46)  & 0.20 (1.37)  \\ \hline
n=50       & 0.08 (0.526)  & 0.07 (0.540)  & 2.08 (3.24)  & 0.09 (0.55)\\ \hline
n=100       & 0.04 (0.29)  & 0.04 (0.30) & 1.61 (1.79)  & 0.04 (0.28) \\ \hline
\end{tabular}
\end{table}
We can observe that properly accounting for the random effects and clustering in our MuFuMES method dramatically improves the estimation performance of the fixed effect parameters in most cases, as compared to naive group LASSO. Interestingly, even for the covariate $X_{ij4}$, which has both a fixed and a random slope, the estimation performance of the MuFuMES estimator $\hat{\beta_4}(\cdot)$ is comparable or better than the group LASSO estimator, particularly for smaller sample sizes.   

Overall, our simulation results illustrate that the proposed MuFuMES method is able to select the true fixed and random functional effects with high accuracy and also provides an impressive estimation performance due to accounting for clustering in the data. The results from the additional simulation scenario B are presented in Appendix B of the supplementary material.

\section{NHANES 2011-12 Case Study}
\label{case study}
We consider identifying and investigating the age and race-specific heterogeneity in the effect of key demographic, lifestyle, and socioeconomic covariates, on the diurnal functional patterns of physical activity (PA) among adults based on the accelerometer data from the 2011-2012 cohort of NHANES.  The NHANES is a nationally representative sample of the non-institutionalized US population and provides a broad range of descriptive summaries related to health and nutrition.  In addition, in NHANES 2011-2012, accelerometer data were collected using the wrist-worn ActiGraph GT3X+ accelerometer (ActiGraph of Pensacola, FL). Participants were asked to wear the physical activity monitor continually for seven full days and remove it on the morning of the 9th day. We focus on the minute-level accelerometer data, which reports individuals’ acceleration in Monitor Independent Movement Summary (MIMS) unit, an open-source, device-independent universal summary metric  \citep{john2019open}.
Particularly, we consider the MIMS triaxial value (sum of X,Y,Z axis MIMS) reported per minute as the measure of physical activity. Our final sample for analysis consists of $3402$ adults (aged more than 20 years) with available physical activity data (physical
activity monitoring available at least ten hours per day for
at least four days) and covariate information (reported in Supplementary Table S3). We considered gender,  body mass index (BMI), healthy eating index (HEI) score \citep{krebs2018update} as a measure of diet quality (ranging from 0-100, higher indicating better), combined grip strength (potential indicator of frailty, \cite{blodgett2015frailty}), and income quantified by the ratio of family income to poverty (INDFMPIR) (a measure of income relative to the poverty threshold, with higher value indicating higher income relative to the poverty line) as potential drivers of physical activity across the lifespan, which also might show race and age-specific heterogeneity. Table S1 in the Supplementary Material also presents the descriptive statistics of this sample.

As illustrated in Figure \ref{fig:fig1} 
there exists considerable heterogeneity in the diurnal PA patterns among the six age-groups ($20-30,30-40,\ldots,70+$) and the six reported races/ethnicities. In this article, we are interested in identifying the key drivers of diurnal patterns of PA and also identifying the covariates having  cluster-specific (age or race-specific) heterogeneity in their effects on diurnal PA patterns. We define our outcome as $Y_{ij} (s)$ as the daily MIMS at minute s for the $j$ th subject within cluster $i =1,2,\ldots,36$ (age-by-race groups). This is  calculated based on averaging all the available MIMS at time $s$ across days for the subject $j$ within cluster $i$, i.e., averaging $\{Y_{ijd}(s)\}_{d=1}^{7}$ to represent the typical daily pattern of PA.

Since, the raw minute-level MIMS profile can be noisy, to extract and understand smooth diurnal patterns of PA we pre-smooth the data \citep{ghosal2021variable} using 1 hour windows, resulting in hourly observations $\{Y_{ij}(s_1),Y_{ij}(s_2),\ldots,Y_{ij}(s_m)\}_{i=1,j=1}^{36,J_i}$, $m=24$ and $\mathcal{T}_{m} =\{s_{1},s_{2},\ldots,s_{m} \}$ are equi-spaced grid points between 12.30 am (midpoint of the first window) to 11.30 pm (midpoint of last window). This also helps in reducing the computational complexity of handling minute-level data in functional models (see \cite{cui2022fast} for a fast multi-level estimation approach) and the additional complexity coming from our proposed SSGL-based MuFuMES method with increasing $m$. We consider the following multilevel functional mixed effects model for the diurnal PA $Y_{ij} (s)$:
\begin{eqnarray}
    Y_{ij}(s)=\sum_{k=1}^{11}X_{ijk}\beta_k(s)+\sum_{r=1}^{8}Z_{ijr}u_{ir}(s)+v_{ij}(s)+\epsilon_{ij}(s).  \label{eq:app} 
\end{eqnarray}

\begin{table}[ht]
\centering
\caption{Selection of fixed and random functional effects in the NHANES application from the MuFuMES method. Variables being selected are indicated by selection=1 and 0 otherwise.}
\label{tab:my-tableapp}
\begin{tabular}{llll}
\hline
Fixed             & Selection & Random           & Selection \\ \hline
Intercept         & 1         & Intercept        & 1         \\ \hline
BMI               & 1         & BMI              & 1         \\ \hline
Gender            & 1         & Gender           & 1         \\ \hline
INDFMPIR          & 1         & INDFMPIR         & 1         \\ \hline
MGDCGSZ           & 1         & MGDCGSZ          & 1         \\ \hline
HEI               & 0         & HEI              & 1          \\ \hline
$X_{7}$ (pseudo)  & 0         & $Z_{7}$ (pseudo) & 0          \\ \hline
$X_{8}$ (pseudo)  & 0         & $Z_{8}$ (pseudo) & 0          \\ \hline
$X_{9}$ (pseudo)  & 0         &                  &           \\ \hline
$X_{10}$ (pseudo) & 0         &                  &           \\ \hline
$X_{11}$ (pseudo) & 0         &                  &           \\ \hline
\end{tabular}
\end{table}
\hspace*{- 6 mm}
  Here $X_{ij1}$ is the column corresponding to the fixed functional intercept, $(X_{ij2},X_{ij3},X_{ij4},X_{ij5},X_{ij6})$ corresponds to the covariates BMI, Gender (female), INDFMPIR, MGDCGSZ (grip strength), HEI respectively for the $j$ th replication in the $i$ th cluster (age-by-race). Additionally, the covariates $X_{ijk}\sim \mathcal{N}(0,1), k=7,8,\ldots,11$ are generated as pseudo-covariates and added as fixed effect covariates to assess the performance of the proposed variable selection method \citep{wu2007controlling,ghosal2022variable}. Similarly, $Z_{ij1}$ is the column corresponding to the random functional intercept, $(Z_{ij2},Z_{ij3},Z_{ij4},Z_{ij5},Z_{ij6})$ corresponds to the covariates BMI, Gender (female), INDFMPIR, MGDCGSZ (grip strength), HEI respectively for the $j$ th replication in the $i$ th cluster. We also add pseudo-covariates $Z_{ij7}=X_{ij7},Z_{ij8}=X_{ij8}$ to further assess the selection performance. We apply the proposed MuFuMES method to select the important fixed and random functional effects. All the continuous covariates are standardized before applying MuFuMES. The optimal number of basis functions, and the spike parameters $(\lambda_0,\nu_0)$ were chosen using a grid-search based on the proposed BIC criterion in Section 2.5. The optimally selected values of the spike parameters were $\lambda_0=350,\nu_0=30$, and we used $7$ cubic B-spline basis functions to model the functional fixed intercept and $5$ cubic B-spline basis functions to model the functional fixed and random slope parameters (as well as the random functional intercept). The selection result from MuFuMES is reported in Table \ref{tab:my-tableapp}.
\begin{figure}[ht]
\centering
\includegraphics[width=0.8\linewidth , height=0.8\linewidth]{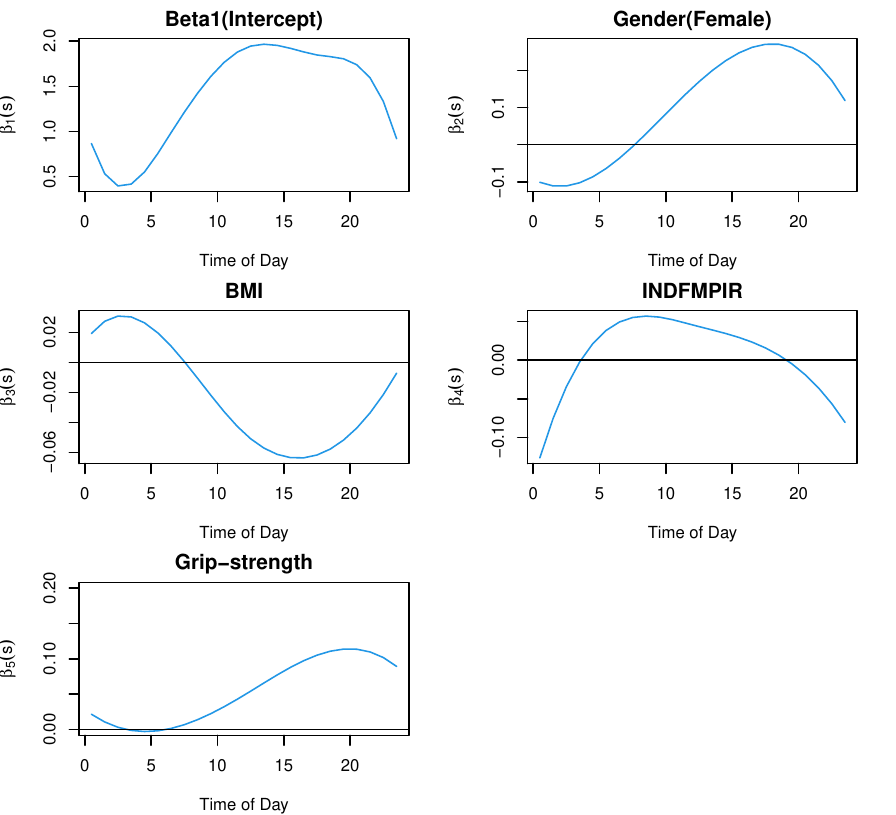}
\caption{Estimated functional fixed effects in the NHANES application from the MuFuMES method.}
\label{fig:fig4}
\end{figure}
We observe that the variables BMI, Gender, INDFMPIR (ratio of family income to poverty), and MGDCGSZ (grip strength) are selected as important functional fixed effects (along with the fixed effect functional intercept), highlighting these variables influence the diurnal PA patterns across the lifespan and all races. Among the functional random effects, first, the intercept is selected, underlining that there exists age and race-specific heterogeneity in the diurnal PA patterns. Furthermore, we observe that BMI, Gender, INDFMPIR (ratio of family income to poverty), MGDCGSZ (grip strength), and HEI (healthy eating index) exhibit age and race-specific heterogeneity in their effects on the diurnal PA patterns. All the pseudo-variables (fixed and random) are discarded by the proposed MuFuMES method, highlighting its robust selection performance. We display the estimated functional fixed effects for the selected covariates in Figure \ref{fig:fig4}. 

We observe that the fixed effect intercept $\beta_1(s)$ captures the overall diurnal pattern of PA. Females can be seen to have a higher diurnal PA during the daytime (positive $\beta_2(s)$ after 8 a.m.) and lower PA during the night compared to the males \citep{xiao2015quantifying,crainiceanu2024functional}. A higher BMI is observed to be associated (negative $\beta_3(s)$) with a lower diurnal PA \citep{cardenas2018association} during the day between 8 a.m. - 8 p.m. and a higher PA during night (disrupted sleep) across all age-groups and races. A higher income is observed to be associated (positive $\beta_4(s)$) with a higher diurnal PA during the day between 5 a.m. - 8 p.m. and a lower PA during the night (better sleep) across all age-groups and races \citep{armstrong2018association,kakinami2018association}. A higher grip strength is found to be associated \citep{kim2017adiposity} with a higher diurnal PA during the daytime across all age groups and races. Thus MuFuMES can identify the key drivers of diurnal PA and estimate their dynamic effects, while also identifying the key factors exhibiting age and race-specific heterogeneity in their effects on the diurnal PA. It is also possible with MuFuMES to quantify and understand these age and race-specific heterogeneous effects.

For illustration, we display the predicted diurnal PA trajectories from the MuFuMES across the 36 age-by-race groups arising from the interaction of six age groups ($20-30,30-40,\ldots,70+$) and six races (MA=1, OH=2, NHW=3, NHB=4, NHA=6, OR=7, see supplementary Table S1 for description) in Supplementary figures S2-S37. For each of the age-by-race clusters, the predicted trajectories for male (reference group) are given by $\hat{\beta}_1(s)$ for the fixed effect prediction and $\hat{\beta}_1(s)+\hat{\beta}_{1i}(s)$ for the mixed effects prediction (since continuous variables were standardized and hence centered), with the other covariates being held at their average values across the sample. We observe considerable age and race-specific heterogeneity in the predicted diurnal PA patterns among males.
For example, among the Mexican Americans (MA), the age groups $20-30,30-40,40-50$ have a higher diurnal PA compared to the average diurnal PA, the age group $50-60$ has a dip in the evening (Figure S4), and the age groups $60-70,70+$ have considerably lower late afternoon-evening PA compared to the average population (Figures S5, S6). Similar heterogeneity but possibly different patterns can also be seen among the other races. For example, among the non-Hispanic white population, the age groups $20-30,30-40,40-50$ have a lower diurnal PA (Figures S13-S15) compared to the average population in the morning-afternoon (4 p.m.) but have a higher diurnal PA in the evening (after 4 p.m.). Interestingly, this trend begins to get reversed in the older ages, and the older white males (age 70+) can be seen to consistently have a lower predicted PA compared to the average population, particularly later in the day (after 12 p.m.). These observations also match with observed diurnal PA trajectories in Figure \ref{fig:fig1}.

Thus, with the MuFuMES, we can successfully identify and understand the key drivers of diurnal PA across the lifespan among all races, and also identify and understand the age and race-specific heterogeneities in the effect of these drivers on the diurnal patterns of physical activity.

\section{Discussion}
\label{Disc}
In this article, we have proposed a new method (MuFuMES) for variable selection of fixed and random effects in multilevel functional mixed effects models. To the best of our knowledge, this is the first work exploring variable selection of fixed and random effects in multilevel functional mixed effects models. With the use of spike-and-slab group
lasso (SSGL) priors on the B-spline basis coefficients, a computationally efficient MAP estimation approach is employed for mixed effect selection and estimation through an
Expectation Conditional Maximization (ECM) algorithm. Numerical analysis using
simulation have shown a satisfactory selection accuracy of the variable selection method for both fixed and random functional effects. The method is applied on the NHANES 2011-12 accelerometer
data to identify gender, BMI, poverty income ratio and grip strength as the key drivers of diurnal PA across the lifespan among all races. The estimated diurnal functional effects highlight the dynamic association of PA with these factors and could be useful for designing time-of-day specific PA interventions \citep{cho2024exploring}. Additionally, the method also identifies BMI, gender, poverty income ratio,  grip strength, and healthy eating index
(HEI) score as the key factors exhibiting age and race-specific heterogeneity in their effects on
the diurnal PA.

Multiple research directions could be explored based on our current work. In this article, we have focused on the linear effects of the multilevel predictors. This could be extended to accommodate nonlinear effects of the predictors  \citep{ghosal2021variable} or interaction effects through single index type models \citep{ghosal2024variable}. In this article, we have primarily focused on selection and estimation of the model components in a multilevel functional mixed effects model. The proposed method can be computationally intensive when working with ultra-dense (e.g., minute-level activity counts) functional data. If interested in fixed effects inference and estimation, fast and scalable methods e.g., \cite{cui2022fast,cui2023fast} based on univariate mixed effects models as building blocks would serve as suitable candidates. For scalable inference on random effects (without selection), recent inferential methods for prediction based on functional random effects \citep{zhou2025prediction} would be appealing.
Future work would therefore benefit from exploration of uncertainty quantification and inference while doing simultaneous selection in the MuFuMES based on efficient Markov chain Monte Carlo (MCMC) based approaches \citep{sun2025ultra}. Post-selection inference-based approaches could also serve as viable alternatives \citep{lee2016exact,taylor2018post} in this scenario and remain areas for future research.

\section*{Supplementary Material}
Appendix A-D, along with the Supplementary Tables and Supplementary Figures referenced in this article, are available online as Supplementary Material. Software implementation via R \citep{Rsoft} and illustration of the proposed framework are included with this article and will be made available on GitHub.


\onehalfspacing
\bibliographystyle{asa}
\bibliography{refs}
\end{document}


\def\spacingset#1{\renewcommand{\baselinestretch}%
{#1}\small\normalsize} \spacingset{1}


\if0\blind
{
  \title{\bf Supplementary Material for Variable Selection for Fixed and Random Effects in Multilevel Functional Mixed Effects Models}
  \maketitle
} \fi

\if1\blind
{
  \bigskip
  \bigskip
  \bigskip
  \begin{center}
    {\LARGE\bf Title}
\end{center}
  \medskip
} \fi

\bigskip

\vfill

\newpage
\spacingset{1.5} 
\section{Appendix A: ECM Algorithm for MuFuMES}
\subsubsection*{E Step}    
For calculating the expected log-posterior $Q(\bm\Phi \mid \bm\Phi^{(t-1)})$ based on equation (15) of the paper, we first obtain $E(\tau_k \mid \*Y,
    \^X,\^Z,\^W, \bm\Phi^{(t-1)})$ and $E(\tau_r^* \mid \*Y,
    \^X,\^Z,\^W,\bm\Phi^{(t-1)})$. 
 \begin{eqnarray}
     E(\tau_k \mid \*Y,
    \^X,\^Z,\^W, \bm\Phi^{(t-1)})=p_k(\bm\gamma_k^{(t-1)},\theta^{(t-1)}),\\
    p_k(\bm\gamma_k,\theta)= \frac{\theta\Psi(\bm\gamma_k\mid \lambda_1)}{\theta\Psi(\bm\gamma_k\mid \lambda_1)+(1-\theta)\Psi(\bm\gamma_k\mid \lambda_0)}, k=1,\ldots,p.
 \end{eqnarray}   
Here $p_k(\bm\gamma_k,\theta)$ is the conditional posterior probability of $\bm\gamma_k$ coming from the slab distribution rather than the spike component. For notational simplicity, we denote $p_k(\bm\gamma_k^{(t-1)},\theta^{(t-1)})$ computed given the current ($(t-1)$-th stage) estimates as $p_k$. Similarly, we obtain 
\begin{eqnarray}
     E(\tau_r^* \mid \*Y,
    \^X,\^Z,\^W, \bm\Phi^{(t-1)})=p_r(\tilde{\*L}_{r}^{(t-1)},\theta^{*^{(t-1)}}),\\
    p_r(\tilde{\*L}_{r},\theta^*)= \frac{\theta^*\Psi(\tilde{\*L}_{r}\mid \nu_1)}{\theta^*\Psi(\tilde{\*L}_{r}\mid \nu_1)+(1-\theta^*)\Psi(\tilde{\*L}_{r}\mid \nu_0)}, r=1,\ldots,q.
 \end{eqnarray} 
 We denote $p_r(\tilde{\*L}_{r}^{(t-1)},\theta^{*^{(t-1)}})$ computed given the current ($(t-1)$-th stage) estimates as $p_r^*$. Next, we define $\lambda_k^*=\lambda_0(1-p_k)+\lambda_1p_k$ for $k=1,\ldots,p$ and $\nu_r^*=\nu_0(1-p_r^*)+\nu_1p_r^*$ for $r=1,\ldots,q$. It can be seen that $E_{\tau} (\log \{ (1 - \tau_k) \lambda_0^d e^{-\lambda_0 \|\bm\gamma_k\|_2} + \tau_k \lambda_1^d e^{-\lambda_1 \|\bm\gamma_k\|_2} \}) = -\lambda_k^*\|\bm\gamma_k\|_2$ up to an additive constant which does on depend on the parameters in $\bm\Phi$. Similarly, we have $E_{\tau^*} (\log \{ (1 - \theta^*) \nu_0^d e^{-\nu_0 \|\tilde{\*L}_{r}\|_2} + \theta^* \nu_1^d e^{-\nu_1 \|\tilde{\*L}_{r}\|_2} ) = -\nu_r^*\|\tilde{\*L}_{r}\|_2$ up to an additive constant which does not depend on the parameters in $\bm\Phi$. Plugging in these conditional expectations into $Q(\bm\Phi \mid \bm\Phi^{(t-1)} )=E_{\bm\tau,\bm\tau^*}( log \{\pi (\bm\Phi,\bm\tau,\bm\tau^*|\*Y,
    \^X,\^Z,\^W)\} \mid \bm\Phi^{(t-1)})$ we have,
\begin{eqnarray}
        Q(\bm\Phi \mid \bm\Phi^{(t-1)} )=E_{\bm\tau,\bm\tau^*}( log \{\pi (\bm\Phi,\bm\tau,\bm\tau^*|\*Y,
    \^X,\^Z,\^W)\} \mid \bm\Phi^{(t-1)})= C-\frac{N}{2}log(\sigma^2) \notag \\-\sum_{i=1}^n \frac{|| \*Y_{i}- \^X_{i}\bm\gamma - (b_i^T \otimes \^Z_{i})\^J\tilde{\*L}-\^W_{i}\bm\zeta_i  ||_2^2}{2\sigma^2}-\frac{1}{2}\sum_{i=1}^n \*b_i^T\*b_i-\sum_{k=1}^{p}\lambda_k^*\|\bm\gamma_k\|_2 -
    \sum_{r=1}^{q} \nu_r^*\|\tilde{\*L}_{r}\|_2 \notag\\+ (a_0 - 1+\sum_{k=1}^p p_k) \log \theta + (b_0 - 1+p-\sum_{k=1}^p p_k) \log (1 - \theta)+ (a_1 - 1+\sum_{r=1}^q p_{r^*}) \log \theta^* + \notag\\ (b_1 - 1+q-\sum_{r=1}^q p_r^*) \log (1 - \theta^*) - \left( \frac{c_0 + 2}{2} \right) \log \sigma^2 - \frac{d_0}{2 \sigma^2},\notag\\
     +\frac{n}{2} log (det(\*\Omega^{-1})) -\frac{1}{2}\sum_{i=1}^n \bm\zeta_i^T\bm\zeta_i+ \frac{\nu+L+1}{2} log (det(\*\Omega^{-1}))  -\frac{1}{2} tr(\*\Delta\*\Omega^{-1}).  \hspace{4 mm}
\end{eqnarray}
where $C$ is a constant not depending on the parameters.

\subsubsection*{CM Step}
In the CM step we maximize $Q(\bm\Phi \mid \bm\Phi^{(t-1)})$ in (5) with respect to the parameters $\bm\Phi=\{\bm\gamma,\tilde{\*L},\*b,\bm\zeta,\theta,\theta^*,\*\Omega,\sigma^2\}$ through two iterative steps.

\hspace*{- 8 mm}
\textbf{Step 1}:\\
In the first CM step we optimize $Q(\bm\Phi \mid \bm\Phi^{(t-1)})$ with respect to $(\theta,\theta^*,\*b,\bm\zeta)$ holding $(\bm\gamma,\tilde{\*L},\*\Omega,\sigma^2)$ fixed at the current value.

$$(\theta^{(t)},\theta^{*{(t)}},\*b^{(t)},\bm\zeta^{(t)})=\underset{(\theta,\theta^*,\*b,\bm\zeta)}{\operatorname{\argmax}} \{Q(\bm\Phi \mid \bm\Phi^{(t-1)})\}|_{\bm\gamma^{(t-1)},\tilde{\*L}^{(t-1)},\*\Omega^{(t-1)},\sigma^{2^{(t-1)}}}.$$

\hspace*{- 8 mm}
\textbf{Step 2}:\\
In the second CM step we optimize $Q(\bm\Phi \mid \bm\Phi^{(t-1)})$ with respect to $(\bm\gamma,\tilde{\*L},\*\Omega,\sigma^2)$ holding $(\theta,\theta^*,\*b,\bm\zeta)$ fixed at the updated current value.

$$(\bm\gamma^{(t)},\tilde{\*L}^{(t)},\*\Omega^{(t)},\sigma^{2^{(t)}})=\underset{(\bm\gamma,\tilde{\*L},\*\Omega,\sigma^2)}{\operatorname{\argmax}} \{Q(\bm\Phi \mid \bm\Phi^{(t-1)})\}|_{\theta^{(t)},\theta^{*{(t)}},\*b^{(t)},\bm\zeta^{(t)}}.$$
Next, we present the computational details for performing step 1 and 2. For step 1, based on $Q(\bm\Phi \mid \bm\Phi^{(t-1)})$ in equation (5), we see that both $\theta^{(t)}$ and $\theta^{*^{(t)}}$ have close form updates: 
\begin{eqnarray}
    \theta^{(t)}=\frac{a_0-1+\sum_{k=1}^p p_k}{a_0+b_0+p-2}, 
\end{eqnarray}
and 
\begin{eqnarray}
    \theta^{*^{(t)}}=\frac{a_1-1+\sum_{r=1}^q p_{r^*}}{a_1+b_1+q-2}.
\end{eqnarray}
The random effect parameter $\*b_i$ (for $i=1,\ldots,n$) based on (5) are updated as,
\begin{eqnarray}
    \*b_i^{(t)}=\{(\^Z_{i}\^L^{(t-1)})^T(\^Z_{i}\^L^{(t-1)})+\sigma^{2^{(t-1)}}\^I\}^{-1}(\^Z_{i}\^L^{(t-1)})^T(\*Y_i-\^X_i\bm\gamma^{(t-1)}-\^W_i\bm\zeta_i^{(t-1)}), \hspace{4 mm} i=1,\ldots,n.
\end{eqnarray}
where $\^L$ is as in model (4) of the paper and we have $vec(\^L^{(t-1)})=\^J\tilde{\*L}^{(t-1)}$. The random effect parameter $\bm\zeta_i$ (for $i=1,\ldots,n$) based on (5) are then updated as, 

\begin{eqnarray}
    \bm\zeta_i^{(t)}=\{\^W_{i}^T\^W_{i}+\sigma^{2^{(t-1)}}\{\*\Omega^{(t-1)}\}^{-1}\}^{-1}(\^W_{i})^T(\*Y_i-\^X_i\bm\gamma^{(t-1)}-\^Z_i\^L^{(t-1)} \*b_i^{(t)}), \hspace{4 mm} i=1,\ldots,n.
\end{eqnarray}

For the step 2 of the CM algorithm, we first update $\bm\gamma$ by maximizing the expected log-posterior (5) with respect to $\bm\gamma$ holding $(\theta^{(t)},\theta^{*{(t)}},\*b^{(t)},\bm\zeta^{(t)})$ and $\tilde{\*L}^{(t-1)},\*\Omega^{(t-1)},\sigma^{2^{(t-1)}}$ fixed. This optimization can be reduced to,
\begin{eqnarray}
\bm\gamma^{(t)}=\underset{(\bm\gamma)}{\operatorname{\argmax}} -\sum_{i=1}^n \frac{|| \*Y_{i}- \^X_{i}\bm\gamma - (\*b_i^{{(t)}^T} \otimes \^Z_{i})\^J\tilde{\*L}^{(t-1)}-\^W_{i}\bm\zeta_i^{(t)} ||_2^2}{2\sigma^{2^{(t-1)}}}-\sum_{k=1}^{p}\lambda_k^*\|\bm\gamma_k\|_2 \notag\\
    =\underset{(\bm\gamma)}{\operatorname{\argmin}} \sum_{i=1}^n || \*Y_{i}- \^X_{i}\bm\gamma - (\*b_i^{{(t)}^T} \otimes \^Z_{i})\^J\tilde{\*L}^{(t-1)} -\^W_{i}\bm\zeta_i^{(t)}||_2^2+\sum_{k=1}^{p}2\lambda_k^*\sigma^{2^{(t-1)}}\|\bm\gamma_k\|_2 \notag\\
    =\underset{(\bm\gamma)}{\operatorname{\argmin}} \sum_{i=1}^n || \tilde{\*Y}_{i}^1- \^X_{i}\bm\gamma ||_2^2+\sum_{k=1}^{p}2\lambda_k^*\sigma^{2^{(t-1)}}\|\bm\gamma_k\|_2 ,
\end{eqnarray}

where $ \tilde{\*Y}_{i}^1= \*Y_{i}-(\*b_i^{{(t)}^T} \otimes \^Z_{i})\^J\tilde{\*L}^{(t-1)}-\^W_{i}\bm\zeta_i^{(t)}$. This can now be identified as a group-LASSO problem \citep{yuan2006model}, with adaptive group-specific weights. We use the coordinate descent algorithm by \cite{bre2015} for the above optimization.  Next, we update $\tilde{\*L}$ by maximizing (5) with respect to $\tilde{\*L}$, holding $(\theta^{(t)},\theta^{*{(t)}},\*b^{(t)},\bm\zeta^{(t)})$ and $\bm\gamma^{(t)},\*\Omega^{(t-1)},\sigma^{2^{(t-1)}}$ fixed. This optimization similarly reduces to the following group-LASSO type problem with varying group size and adaptive weights:
\begin{eqnarray}
\tilde{\*L}^{(t)}=\underset{(\tilde{\*L})}{\operatorname{\argmax}} -\sum_{i=1}^n \frac{|| \*Y_{i}- \^X_{i}\bm\gamma^{(t)} - (\*b_i^{{(t)}^T} \otimes \^Z_{i})\^J\tilde{\*L} -\^W_{i}\bm\zeta_i^{(t)}||_2^2}{2\sigma^{2^{(t-1)}}}- \sum_{r=1}^{q} \nu_r^*\|\tilde{\*L}_{r}\|_2 \notag\\
    =\underset{(\tilde{\*L})}{\operatorname{\argmin}}\sum_{i=1}^n || \tilde{\*Y}_{i}^2- (\*b_i^{{(t)}^T} \otimes \^Z_{i})\^J\tilde{\*L}||_2^2+\sum_{r=1}^{q}2\nu_r^*\sigma^{2^{(t-1)}}\|\tilde{\*L}_{r}\|_2,
\end{eqnarray}
where $\tilde{\*Y}_{i}^2=\*Y_{i}-\^X_{i}\bm\gamma^{(t)}-\^W_{i}\bm\zeta_i^{(t)}$. Finally, $\*\Omega$ and $\sigma^2$ are updated in Step 2, with closed-form updates given by:

\begin{eqnarray}
   \*\Omega^{(t)}=\frac{1}{n+\nu+L+1} (\*\Delta+\sum_{i=1}^n\zeta_i^{(t)}(\zeta_i^{(t)})^T)
\end{eqnarray}

\begin{eqnarray}
   \sigma^{2^{(t)}}=\sum_{i=1}^n \frac{|| \*Y_{i}- \^X_{i}\bm\gamma^{(t)} - (\*b_i^{{(t)}^T} \otimes \^Z_{i})\^J\tilde{\*L}^{(t)}  -\^W_{i}\bm\zeta_i^{(t)}||_2^2+d_0}{N+c_0+2}.
\end{eqnarray}
Here $N=m(\sum_{i=1}^{n} J_i)$ denotes the total number of observations.

\section{Appendix B: Additional Simulations}

\subsection*{Scenario B} 
We generate observations following a multilevel functional mixed effects model given by,
\begin{equation}
    Y_{ij}(s)=\sum_{k=1}^{11}X_{ijk}\beta_k(s)+\sum_{r=1}^8Z_{ijr}u_{ir}(s)+v_{ij}(s)+\epsilon_{ij}(s), s\in [0,1] \label{mfmem: sim1},
\end{equation}
for clusters $i=1,\ldots,n$ and replications $ j=1,\ldots,J_i$. In this model, we consider $p=11$ covariates for functional fixed effects (including a fixed intercept) and $q=8$ covariates for functional random effects (including a cluster-level random intercept). We denote by $\bm\phi(s,a,b^2)$ the density at $s$ for Normal distribution with mean $a$ and variance $b^2$. The fixed effects coefficient functions are given by $\beta_1(s)=8sin(2\pi s)$ (intercept), $\beta_2(s)=2\bm\phi(s,0.6,0.15^2)$, $\beta_3(s)=2.5\bm\phi(s,0.6,0.15^2)$, $\beta_4(s)=3cos(2\pi s)$, $\beta_5(s)=5sin(2\pi s)+5cos(2\pi s)$ and $\beta_k(s)=0$ for $k=6,7,\ldots,11$. So, only the first 5 fixed effect covariates are relevant. The fixed effect covariates $X_{ijk}$ are independently generated from a $\mathcal{N}(0,2^2)$ distribution for $k=2,\ldots,11$ and $X_{ij1}=1$ (for all $i,j$) corresponds to the intercept. The cluster-level functional random effects are given by $u_{i1}(s)=c_{i1}sin(2\pi s)+d_{i1}cos(2\pi s)$ (random intercept), where $c_{i1}\sim \mathcal{N}(0,3^2\sigma^2_B),d_{i1}\sim \mathcal{N}(0,1.5^2\sigma^2_B)$. Similarly, $u_{i4}(s)=c_{i4}sin(2\pi s)+d_{i4}cos(2\pi s)+e_{i4}sin(\pi s)+f_{i4}cos(\pi s)$, where $c_{i4}\sim \mathcal{N}(0,1.5^2\sigma^2_B),d_{i4}\sim \mathcal{N}(0,0.75^2\sigma^2_B),e_{i4}\sim \mathcal{N}(0,0.5^2\sigma^2_B),f_{i4}\sim \mathcal{N}(0,0.25^2\sigma^2_B)$. 
The rest of the functional random effects $u_{ik}(s)$ are considered to be zero. Hence only the first and the fourth random effect of the covariates are important. In this scenario, $Z_{ijr}$ are independently generated from a $\mathcal{N}(0,2^2)$ distribution for $r=2,\ldots,8$ and $Z_{ij1}=1$ (for all $i,j$) corresponds to the random intercept term. We set $\sigma^2_B$ based on $SNR_B=0.5$, where $SNR_B$ is the standard deviation of the fixed effects functions divided by the standard deviation of the random effects \citep{cui2022fast,scheipl2015functional}. The cluster-subject-level functional random effect $v_{ij}(s)$ is given by $v_{ij}(s)=v_{ij1}sin(\pi s)+v_{ij2}cos(\pi s)$, where $v_{ij1}\sim \mathcal{N}(0,0.8^2\sigma^2_S),v_{ij2}\sim \mathcal{N}(0,0.4^2\sigma^2_S)$. We set $\sigma^2_S$ based on $SNR_S=2$, where $SNR_S$ is the standard deviation of the fixed effects functions divided by the standard deviation of the cluster-visit level random effects. The random errors $\epsilon_{ij}(s)\sim \mathcal{N}(0,\sigma_\epsilon^2)$, where $\sigma_\epsilon$ is chosen based on
a signal to noise ratio of $SNR_{\epsilon}=4$, which represents the 
standard deviation of the linear predictors (fixed and random predictors combined) divided by the standard deviation of the noise $\sigma_\epsilon$. The functional response $Y_{ij}(s)$ is observed on a grid of $m = 10$ equidistant time points in $S=[0,1]$. Cluster size $n\in \{25,50,100\}$ are considered for this scenario, and $J_i=J=10$ replications are considered within each cluster.

\subsection*{Simulation Results: Scenario B}

The performance of the proposed variable selection (MuFuMES) method is evaluated in terms of selection accuracy and estimation accuracy. The tuning parameters of the MuFuMES method are chosen based on the proposed BIC criterion (26). 
Table \ref{tab1sel1} reports the selection performance of MuFuMES for the fixed effects and random effects in terms of true positive and false positive rates for both the fixed and random functional effects.

\begin{table*}[h]
\centering
\caption{Average true positive rate (TP), false positive rate (FP) for fixed (TPF, FPF) and random (TPR, FPR)  effects, Scenario A.}
\label{tab1sel1}
\begin{tabular}{cccccc}
\hline
Sample Size            & Method   & TPF & FPF    & TPR  & FPR  \\ \hline
\multirow{1}{*}{n=25} & MuFuMES & 0.95  & 0.02 & 1 & 0.015       \\ \cline{2-6} 
\multirow{1}{*}{n=50} & MuFuMES & 0.99  & 0.003 & 1 & 0.003       \\ \cline{2-6} 
                       
\multirow{1}{*}{n=100} & MuFuMES & 1  & 0.005 & 1   & 0.01    \\ \cline{2-6} 
\hline
                       
\end{tabular}
\end{table*}

\begin{figure}[H]
\centering
\includegraphics[width=1\linewidth , height=0.8\linewidth]{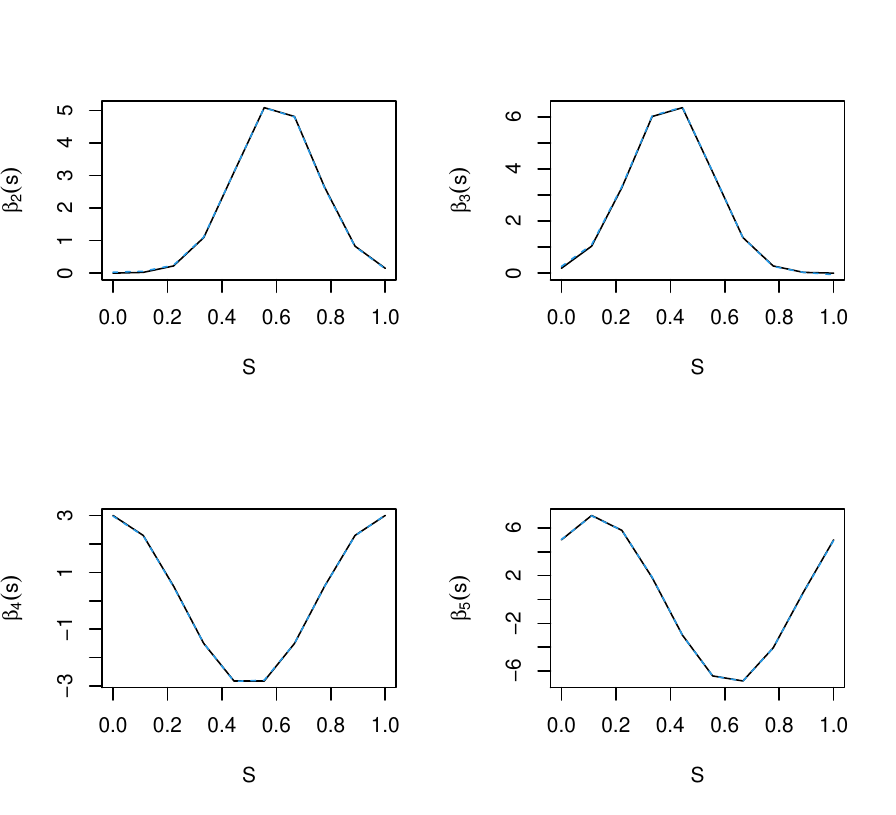}
\caption{Displayed are the true (solid) and M.C mean (dashed) of estimated functional slopes $\beta_k(s)$ (k=2,\ldots,5) from MuFuMES, $n=100$.}
\label{fig:figsim1}
\end{figure}

We observe that the proposed MuFuMES method has a negligible false positive rate and a high true positive rate for both fixed and random effects across all the sample sizes, illustrating parsimonious and accurate model selection. The performance can also be seen to improve with an increasing sample size. Next, we illustrate the estimation performance of the proposed MuFuMES method. We display the Monte Carlo (MC) mean estimates (averaged estimated coefficient function over 100 replications) of the 
functional fixed effect slopes  $\beta_k(s)$ ($k=2,3,4,5$) in Figure \ref{fig:figsim1} for sample size $n=100$. 
The estimated coefficient functions are superimposed on the true curves and can be seen to capture the true effects closely, indicating a satisfactory performance of the proposed method in terms of estimation. 
We also report the mean integrated squared error (MISE) of the functional fixed effect slope estimates in Table \ref{tab:my-table2} across the three sample sizes. The MISE of the functional estimate \(\hat{\beta_k}(s)\) is defined as:  
\[
MISE_k = \frac{1}{B}\sum_{b=1}^{B} \left( \int_0^1 \left( \hat{\beta}_{kb}(s) - \beta_k(s) \right)^2 \, ds \right),
\]
where \(\hat{\beta}_{kb}(s)\) represents the estimate of \(\beta_k(s)\) for the \(b\)-th Monte-Carlo (M.C.) replication, and \(B\) is the total number of replications.

\begin{table}[H]
\centering
\caption{Mean integrated squared error (MISE) of the functional fixed effect slope estimates from the MuFuMES method, scenario A.}
\label{tab:my-table2}
\begin{tabular}{lllll}
\hline
Sample Size & MISE $\beta_2(\cdot)$   & MISE $\beta_3(\cdot)$   & MISE $\beta_4(\cdot)$ & MISE $\beta_5(\cdot)$ \\ \hline
n=25       & 0.160  & 0.158  & 0.154  & 0.162  \\ \hline
n=50       & 0.071  & 0.071  & 0.071  & 0.075\\ \hline
n=100       & 0.037  & 0.038 & 0.038  & 0.036 \\ \hline
\end{tabular}
\end{table}

The proposed MuFuMES method can be seen to estimate the true functional parameters accurately, with MISE decreasing with an increasing sample size, indicating the consistency of the estimators.

\section{Appendix C: Supplementary Tables}

\begin{table}[H]
\small
\caption{List of Predictors and their descriptive statistics (mean and standard deviation for continuous variables and percentages for the categorical variables) in the NHANES 2011-12 study (N=3402). The races MA, OH, NHW, NHB, NHA, OR refer to the races Mexican American, Other Hispanic, Non-Hispanic White, Non-Hispanic Black, Non-Hispanic Asian, and Other Races, respectively. }
\label{tab:my-table}
\begin{tabular}{lll}
\hline
Predictor                                                                  & Description                                                                                                                                                              & mean (sd) / $\%$             \\ \hline
Gender (female)                                                            & Categorical                                                                                                                                                       & $50.8$                         \\ \hline
Age                                                                        & Continuous                                                                                                                                                        & 48.7 (17.6)                \\ \hline
BMI                                                                        & Body Mass Index (Kg/$m^2$)                                                                                                                                                        &  29.16 (7)                 \\ \hline
Race (categorical)                                                                       &MA, OH, NHW, NHB, NHA, OR                                                      & 9.3, 9, 41.2, 25.9, 11.6, 2.9   \\ \hline
INDFMPIR (Continuous)                                                             & Ratio of family income to poverty                                                                                                                                                     & 2.5 (1.7)                         \\ \hline
MGDCGSZ (Continuous)                                                                  & Combined grip strength (Kg)                                                                                                                                                       & 71.0 (22.4)                          \\ \hline
HEI (continuous)                                                                  & Healthy Eating Index                                                                                                                                                      & 55.2 (13.3)                          \\ \hline

\end{tabular}
\end{table}

\section{Appendix D: Supplementary Figures}
We display the predicted trajectories of the 36 clusters from the MuFuMES method for the male group, with the other covariates held at the average value. The interaction of six age groups ($20-30,30-40,\ldots,70-80$) and six races (MA=1, OH=2, NHW=3, NHB=4, NHA=6, OR=7)  leading to the 36 age-by-race groups are indicated in the header of each of the following figures.

\begin{figure}[H]
\includegraphics[page=1,scale=.95]{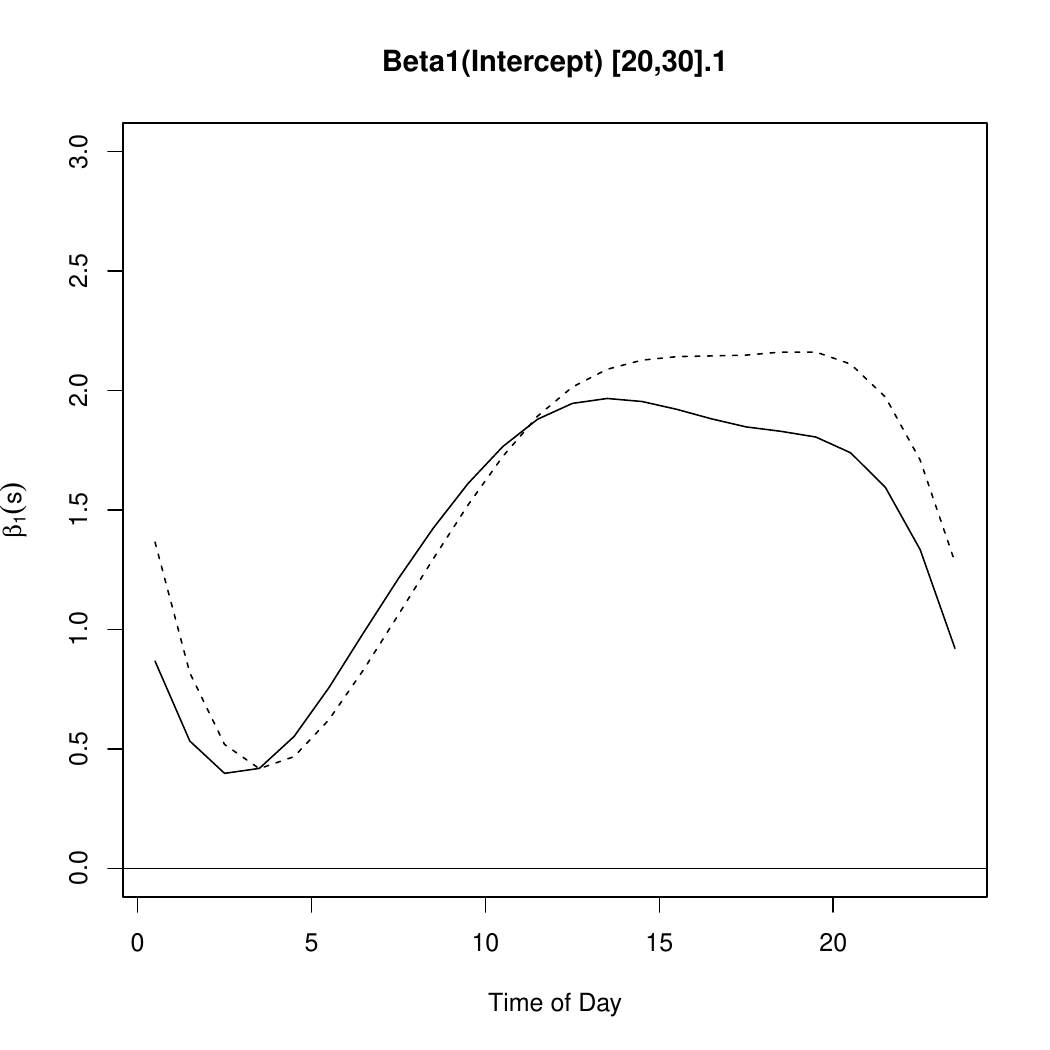}
  \caption{Predicted functional trajectories ($\hat{\beta}_1(s)+\hat{\beta}_{1i}(s)$) of diurnal MIMS for males from the MuFuMES method in the NHANES application. The other continuous covariates were held at their average values. The solid line corresponds to the fixed effect prediction $\hat{\beta}_1(s)$ and the dashed line corresponds to the mixed effects prediction $\hat{\beta}_1(s)+\hat{\beta}_{1i}(s)$. The age-by-race groups are indicated in the header.}
  \end{figure}
\begin{figure}[H]
\includegraphics[page=2,scale=.95]{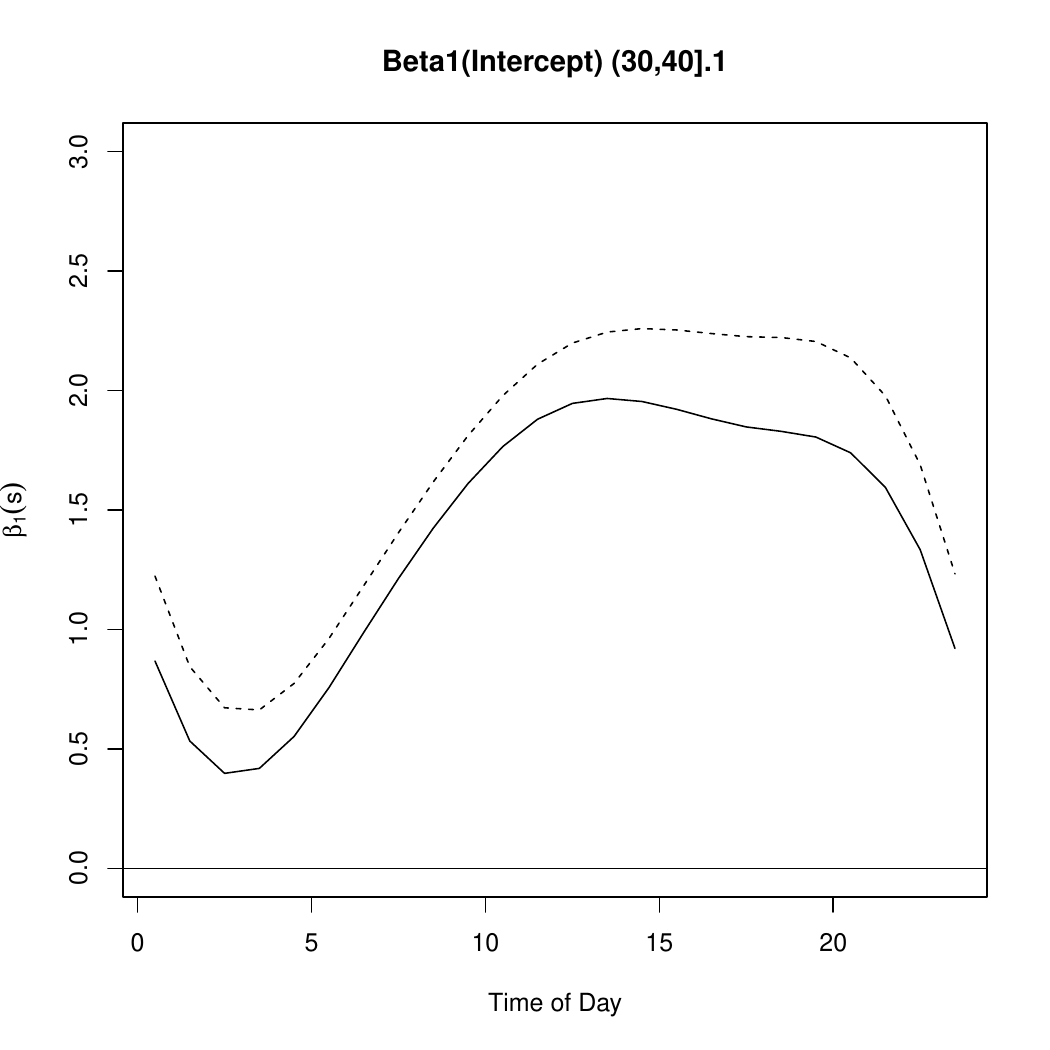}
  \caption{Predicted functional trajectories ($\hat{\beta}_1(s)+\hat{\beta}_{1i}(s)$) of diurnal MIMS for males from the MuFuMES method in the NHANES application. The other continuous covariates were held at their average values. The solid line corresponds to the fixed effect prediction $\hat{\beta}_1(s)$ and the dashed line corresponds to the mixed effects prediction $\hat{\beta}_1(s)+\hat{\beta}_{1i}(s)$. The age-by-race groups are indicated in the header.}
  \end{figure}
  \begin{figure}[H]
\includegraphics[page=3,scale=.95]{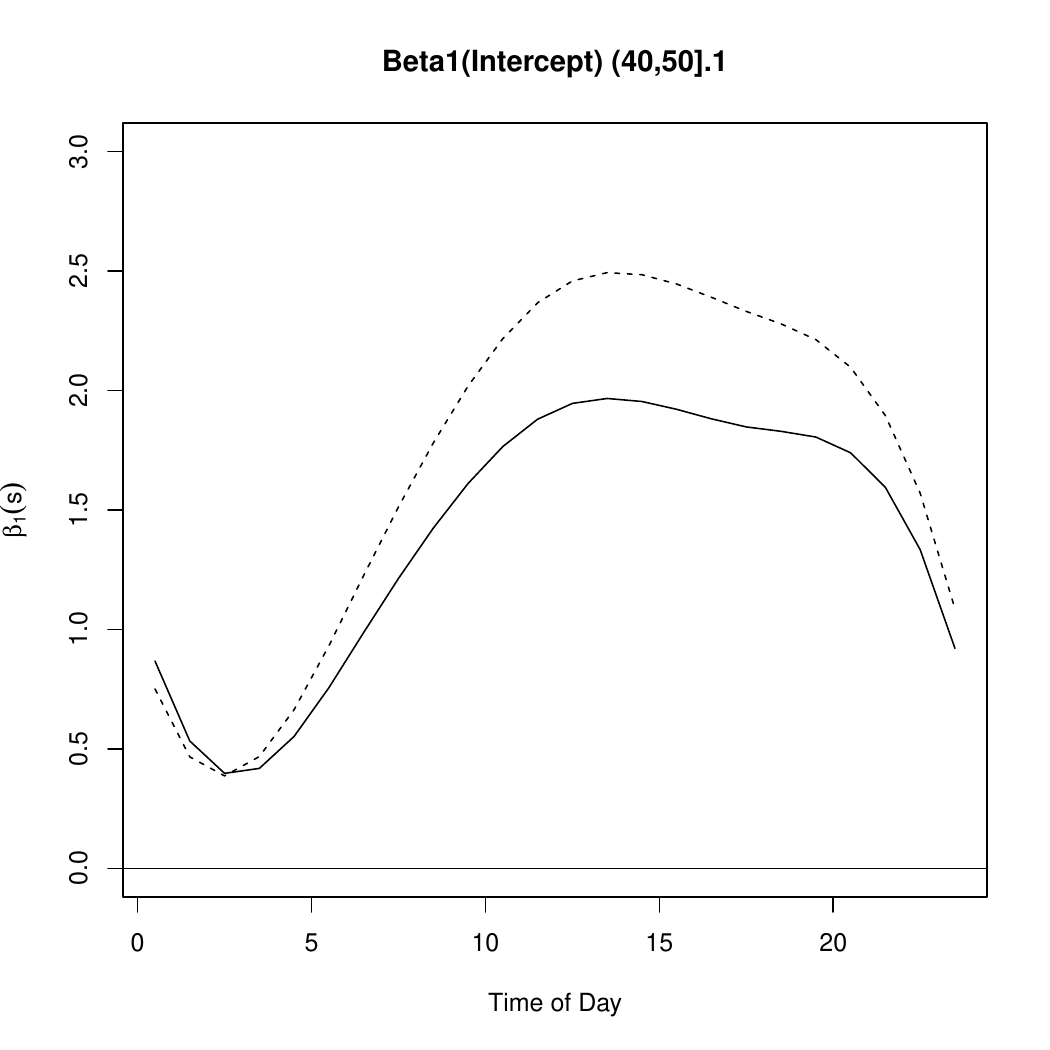}
  \caption{Predicted functional trajectories ($\hat{\beta}_1(s)+\hat{\beta}_{1i}(s)$) of diurnal MIMS for males from the MuFuMES method in the NHANES application. The other continuous covariates were held at their average values. The solid line corresponds to the fixed effect prediction $\hat{\beta}_1(s)$ and the dashed line corresponds to the mixed effects prediction $\hat{\beta}_1(s)+\hat{\beta}_{1i}(s)$. The age-by-race groups are indicated in the header.}
  \end{figure}
  \begin{figure}[H]
\includegraphics[page=4,scale=.95]{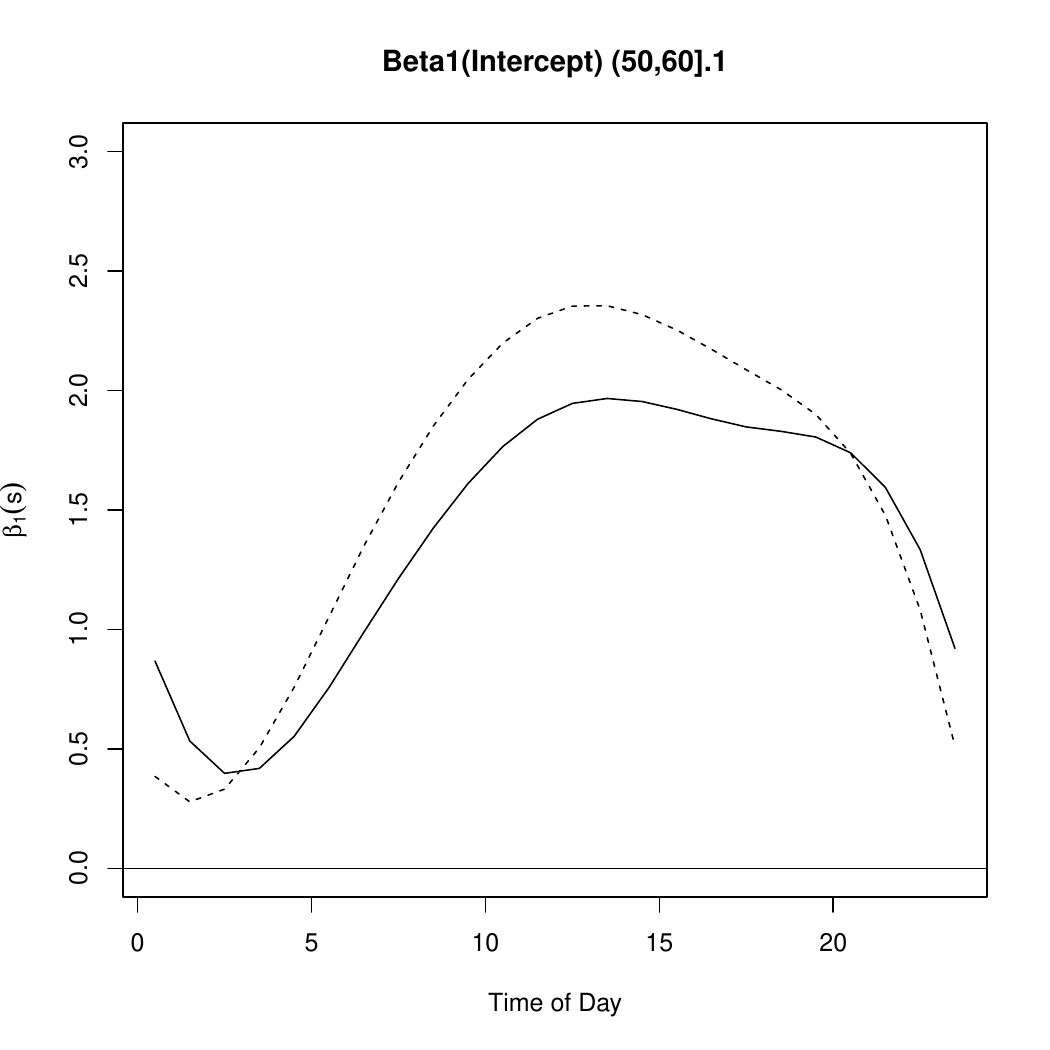}
  \caption{Predicted functional trajectories ($\hat{\beta}_1(s)+\hat{\beta}_{1i}(s)$) of diurnal MIMS for males from the MuFuMES method in the NHANES application. The other continuous covariates were held at their average values. The solid line corresponds to the fixed effect prediction $\hat{\beta}_1(s)$ and the dashed line corresponds to the mixed effects prediction $\hat{\beta}_1(s)+\hat{\beta}_{1i}(s)$. The age-by-race groups are indicated in the header.}
  \end{figure}
  \begin{figure}[H]
\includegraphics[page=5,scale=.95]{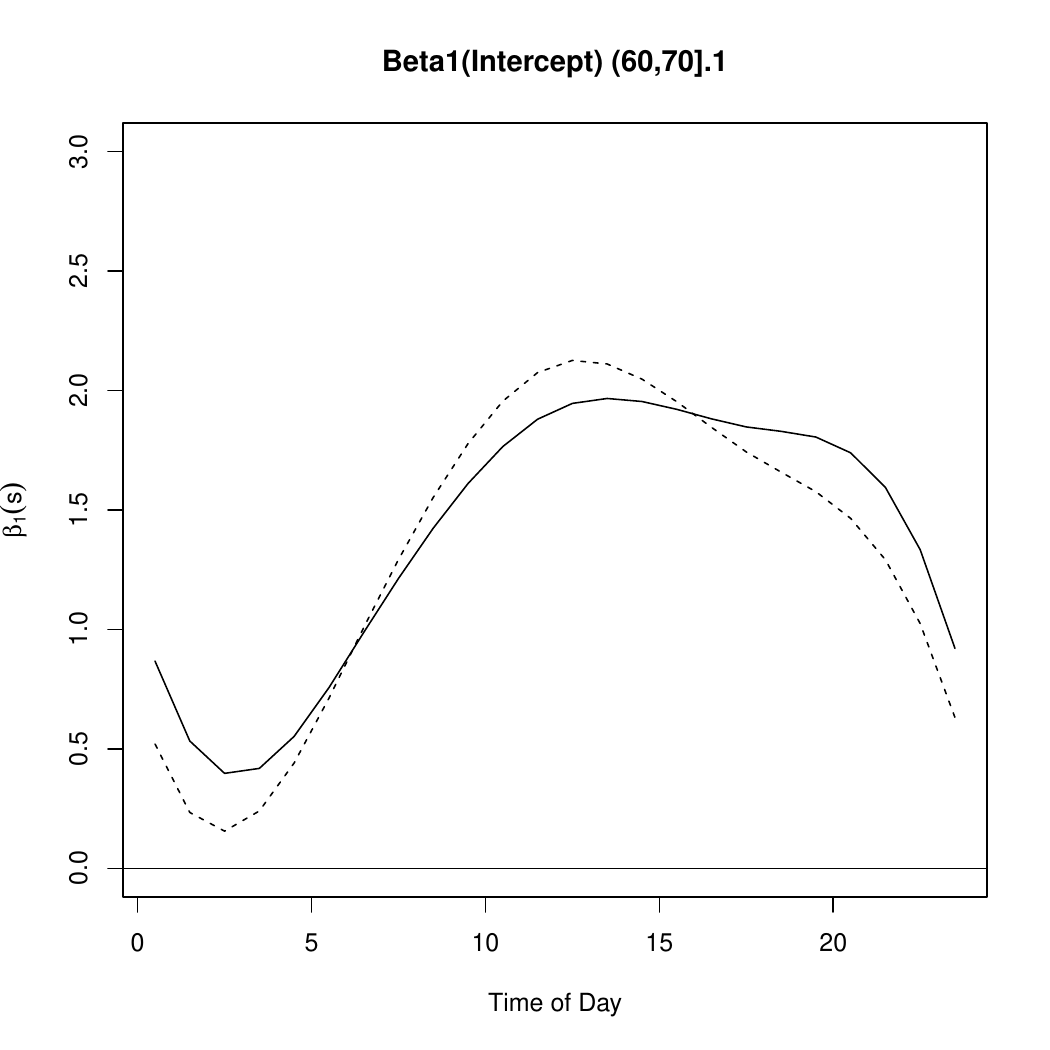}
  \caption{Predicted functional trajectories ($\hat{\beta}_1(s)+\hat{\beta}_{1i}(s)$) of diurnal MIMS for males from the MuFuMES method in the NHANES application. The other continuous covariates were held at their average values. The solid line corresponds to the fixed effect prediction $\hat{\beta}_1(s)$ and the dashed line corresponds to the mixed effects prediction $\hat{\beta}_1(s)+\hat{\beta}_{1i}(s)$. The age-by-race groups are indicated in the header.}
  \end{figure}
  \begin{figure}[H]
\includegraphics[page=6,scale=.95]{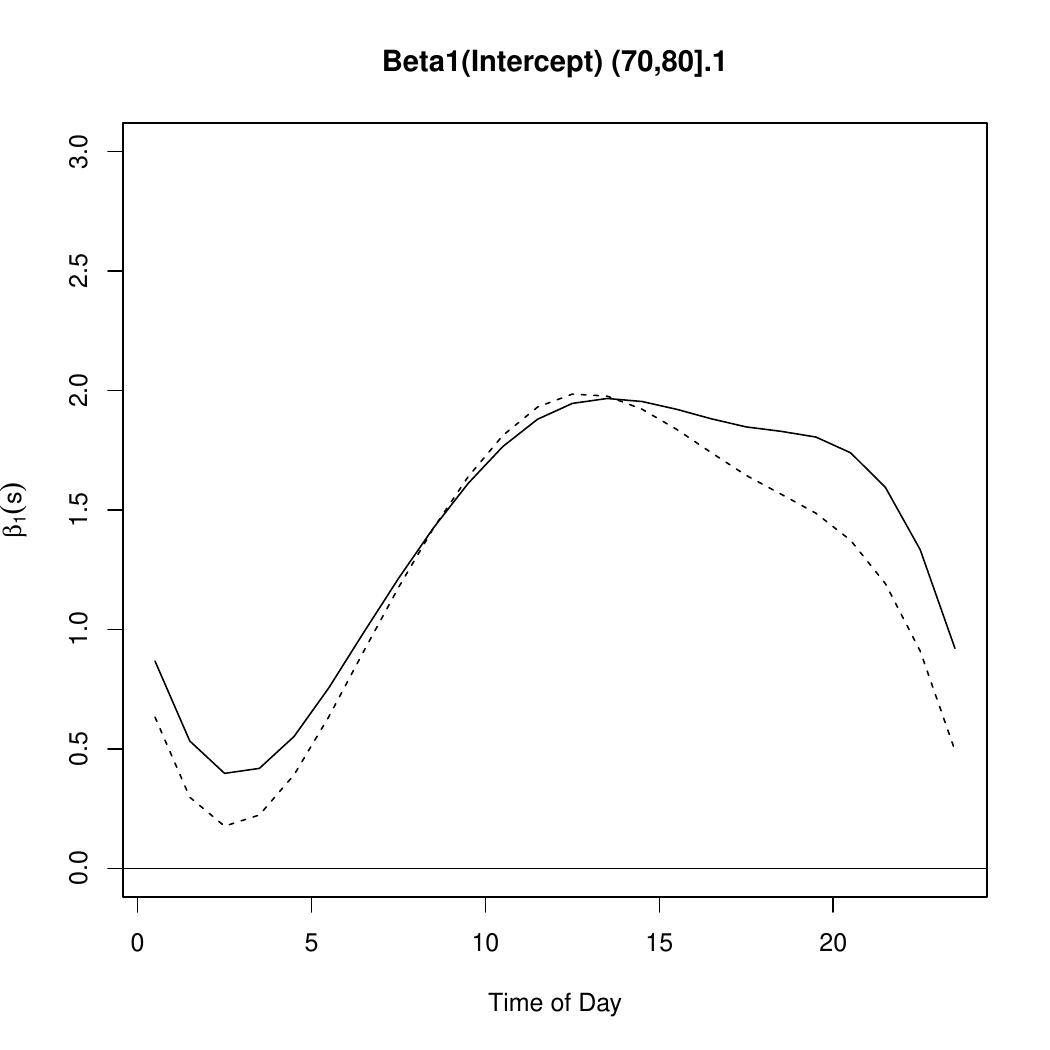}
  \caption{Predicted functional trajectories ($\hat{\beta}_1(s)+\hat{\beta}_{1i}(s)$) of diurnal MIMS for males from the MuFuMES method in the NHANES application. The other continuous covariates were held at their average values. The solid line corresponds to the fixed effect prediction $\hat{\beta}_1(s)$ and the dashed line corresponds to the mixed effects prediction $\hat{\beta}_1(s)+\hat{\beta}_{1i}(s)$. The age-by-race groups are indicated in the header.}
  \end{figure}
  \begin{figure}[H]
\includegraphics[page=7,scale=.95]{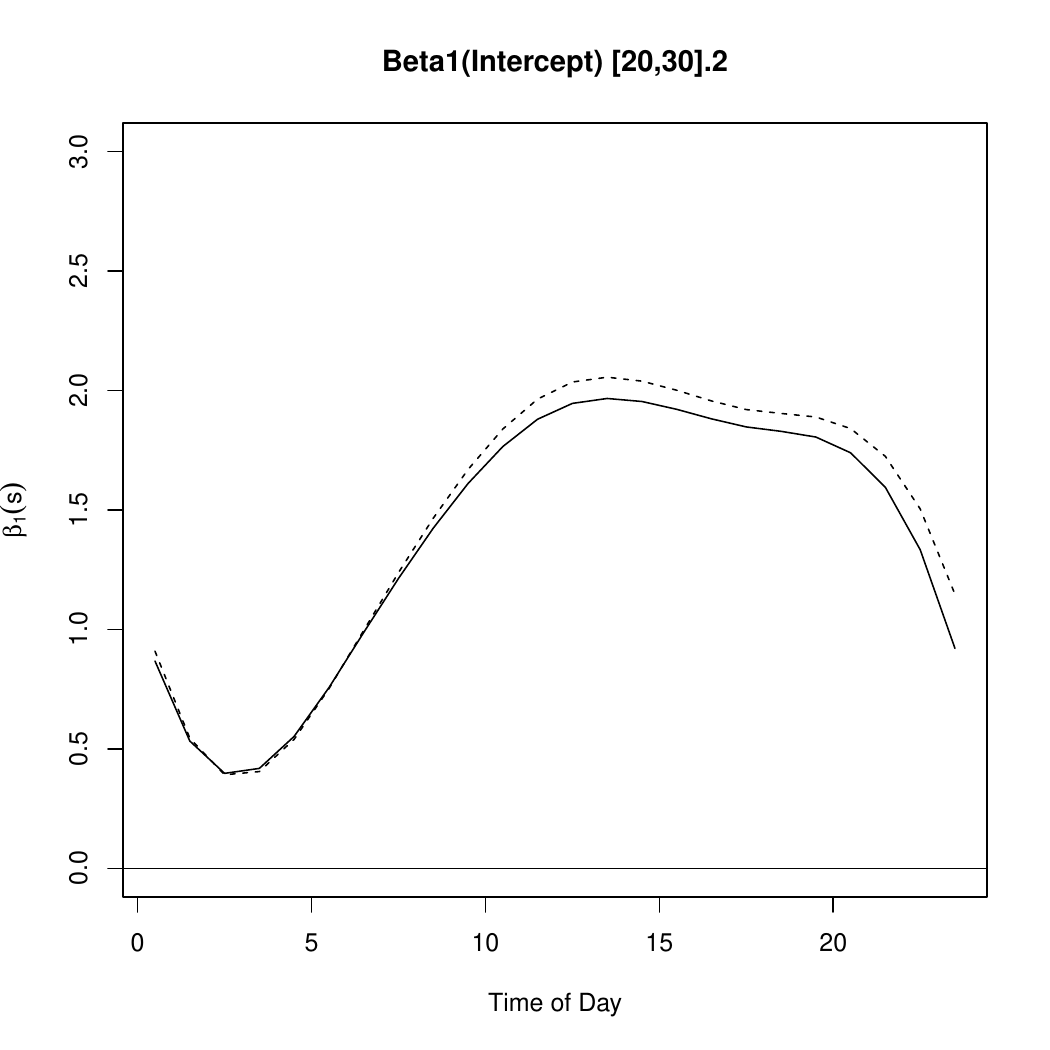}
  \caption{Predicted functional trajectories ($\hat{\beta}_1(s)+\hat{\beta}_{1i}(s)$) of diurnal MIMS for males from the MuFuMES method in the NHANES application. The other continuous covariates were held at their average values. The solid line corresponds to the fixed effect prediction $\hat{\beta}_1(s)$ and the dashed line corresponds to the mixed effects prediction $\hat{\beta}_1(s)+\hat{\beta}_{1i}(s)$. The age-by-race groups are indicated in the header.}
  \end{figure}
  \begin{figure}[H]
\includegraphics[page=8,scale=.95]{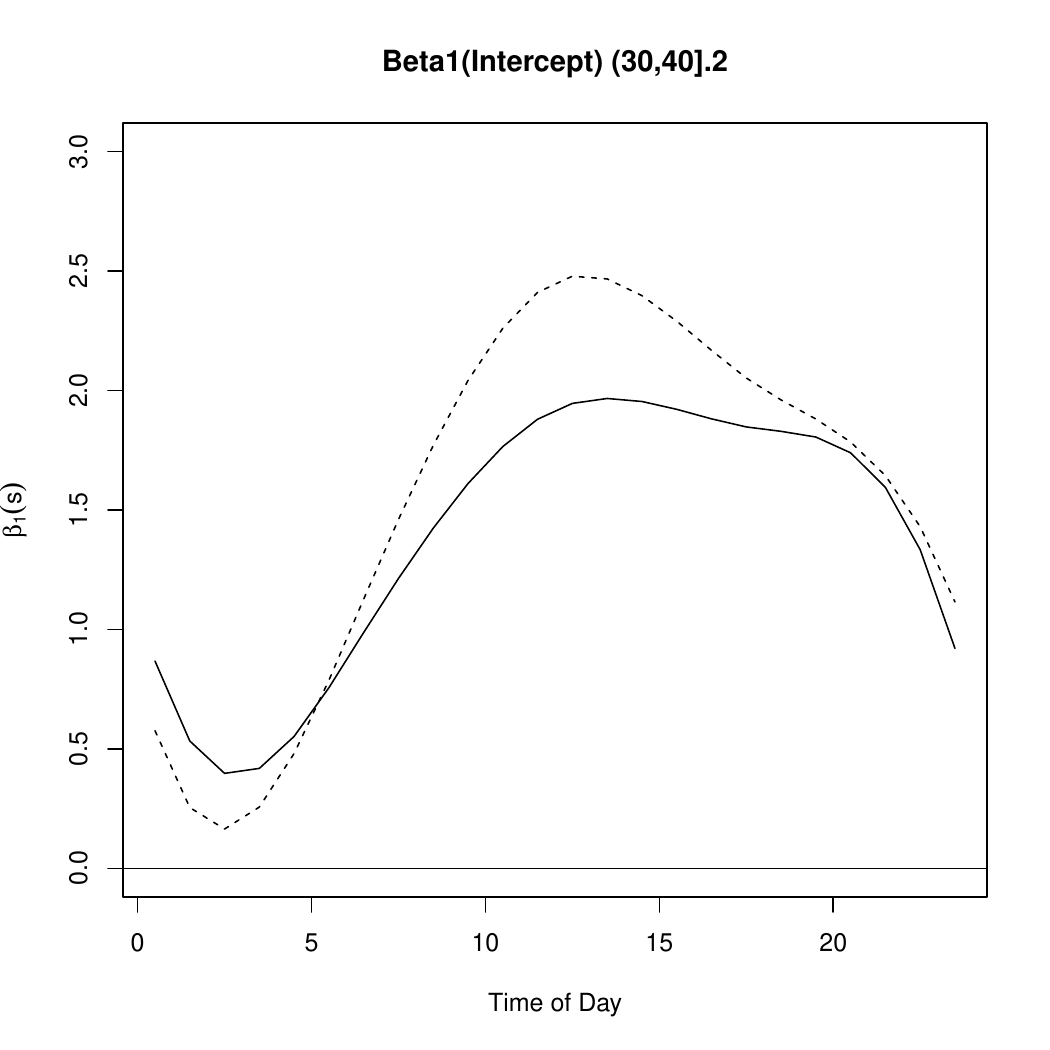}
  \caption{Predicted functional trajectories ($\hat{\beta}_1(s)+\hat{\beta}_{1i}(s)$) of diurnal MIMS for males from the MuFuMES method in the NHANES application. The other continuous covariates were held at their average values. The solid line corresponds to the fixed effect prediction $\hat{\beta}_1(s)$ and the dashed line corresponds to the mixed effects prediction $\hat{\beta}_1(s)+\hat{\beta}_{1i}(s)$. The age-by-race groups are indicated in the header.}
  \end{figure}
  \begin{figure}[H]
\includegraphics[page=9,scale=.95]{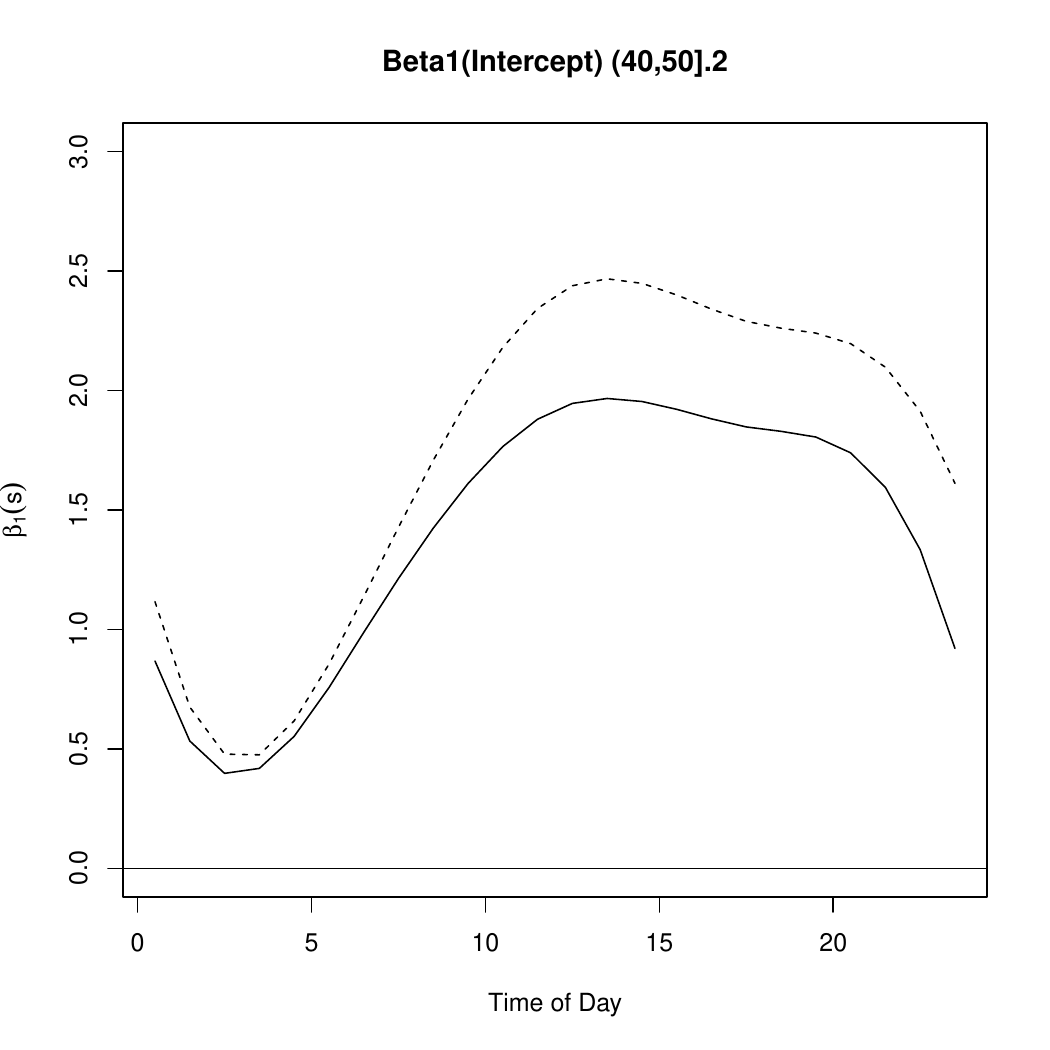}
  \caption{Predicted functional trajectories ($\hat{\beta}_1(s)+\hat{\beta}_{1i}(s)$) of diurnal MIMS for males from the MuFuMES method in the NHANES application. The other continuous covariates were held at their average values. The solid line corresponds to the fixed effect prediction $\hat{\beta}_1(s)$ and the dashed line corresponds to the mixed effects prediction $\hat{\beta}_1(s)+\hat{\beta}_{1i}(s)$. The age-by-race groups are indicated in the header.}
  \end{figure}
  \begin{figure}[H]
\includegraphics[page=10,scale=.95]{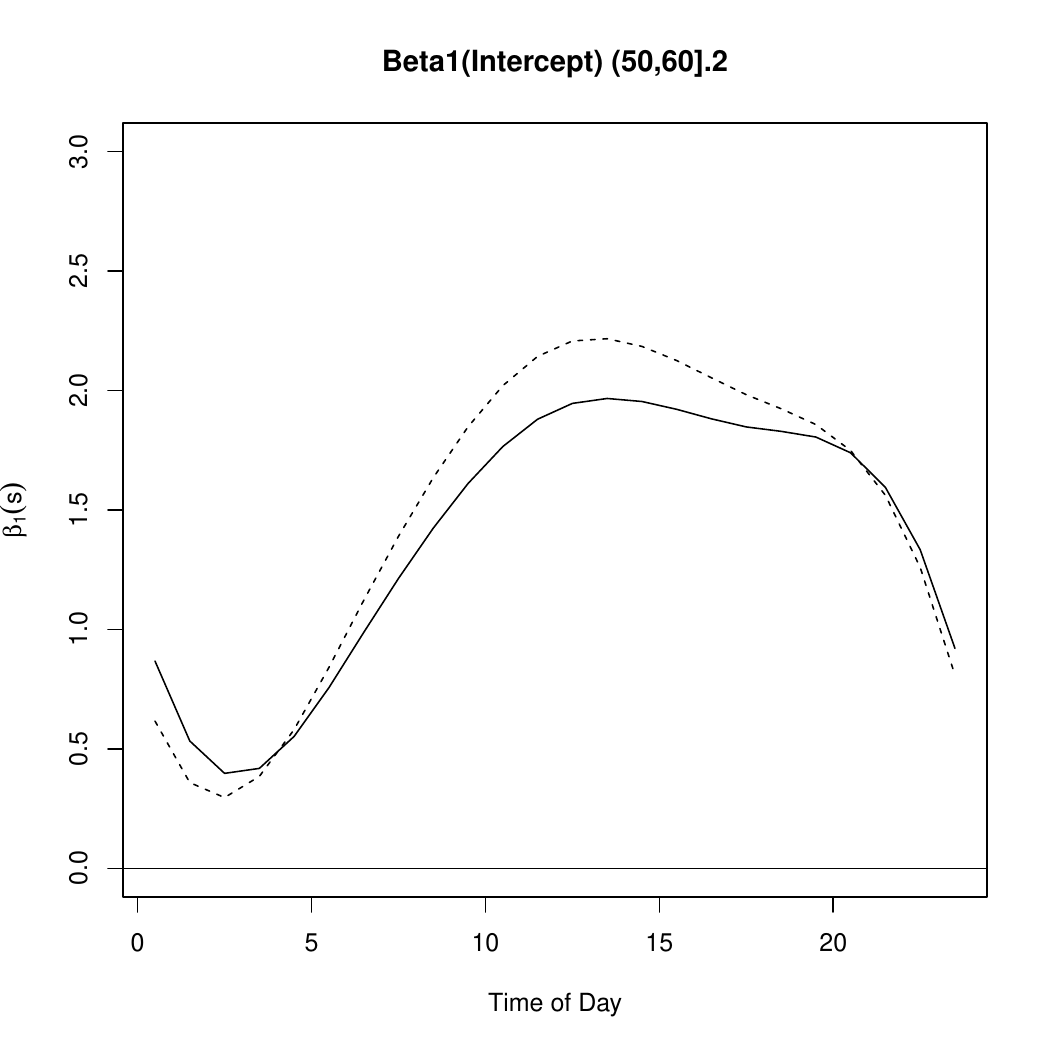}
  \caption{Predicted functional trajectories ($\hat{\beta}_1(s)+\hat{\beta}_{1i}(s)$) of diurnal MIMS for males from the MuFuMES method in the NHANES application. The other continuous covariates were held at their average values. The solid line corresponds to the fixed effect prediction $\hat{\beta}_1(s)$ and the dashed line corresponds to the mixed effects prediction $\hat{\beta}_1(s)+\hat{\beta}_{1i}(s)$. The age-by-race groups are indicated in the header.}
  \end{figure}
  \begin{figure}[H]
\includegraphics[page=11,scale=.95]{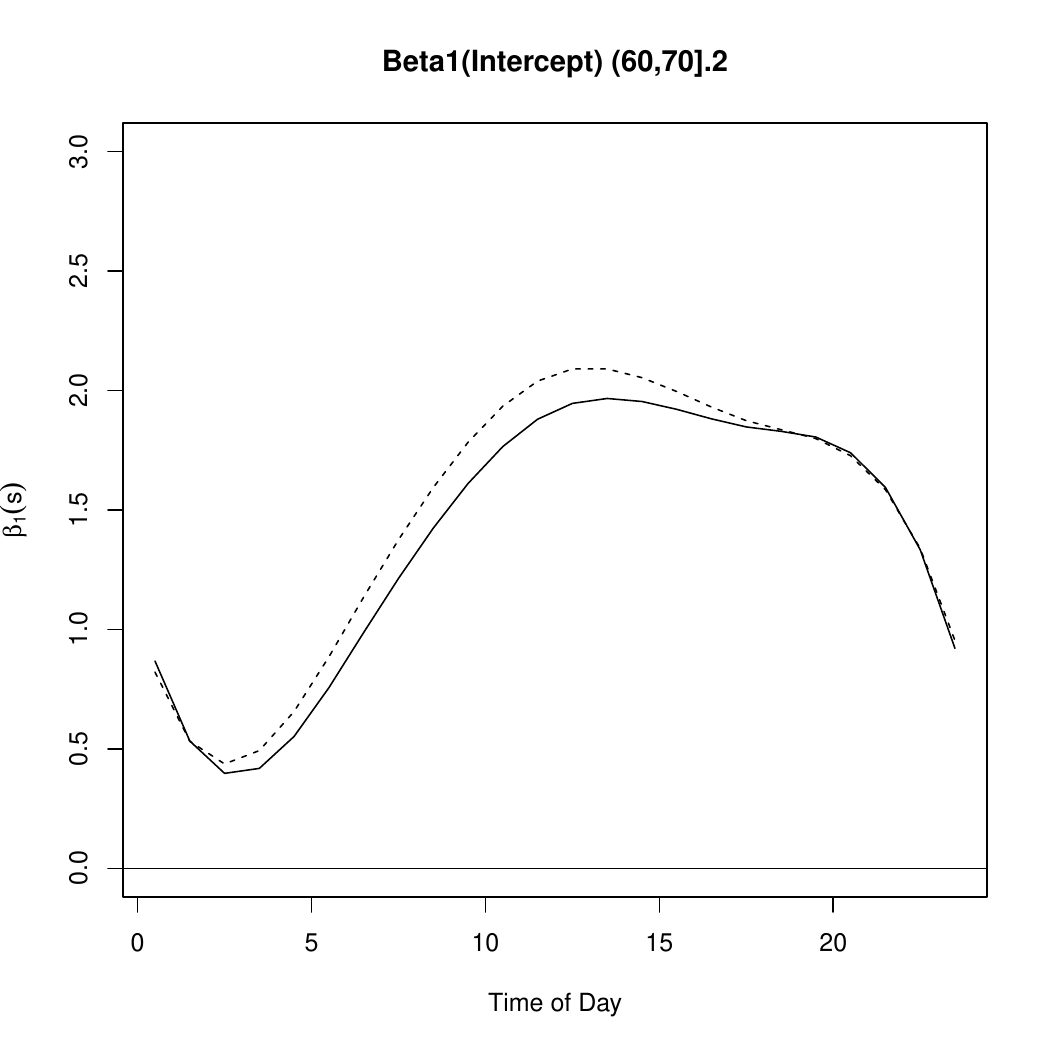}
  \caption{Predicted functional trajectories ($\hat{\beta}_1(s)+\hat{\beta}_{1i}(s)$) of diurnal MIMS for males from the MuFuMES method in the NHANES application. The other continuous covariates were held at their average values. The solid line corresponds to the fixed effect prediction $\hat{\beta}_1(s)$ and the dashed line corresponds to the mixed effects prediction $\hat{\beta}_1(s)+\hat{\beta}_{1i}(s)$. The age-by-race groups are indicated in the header.}
  \end{figure}
  \begin{figure}[H]
\includegraphics[page=12,scale=.95]{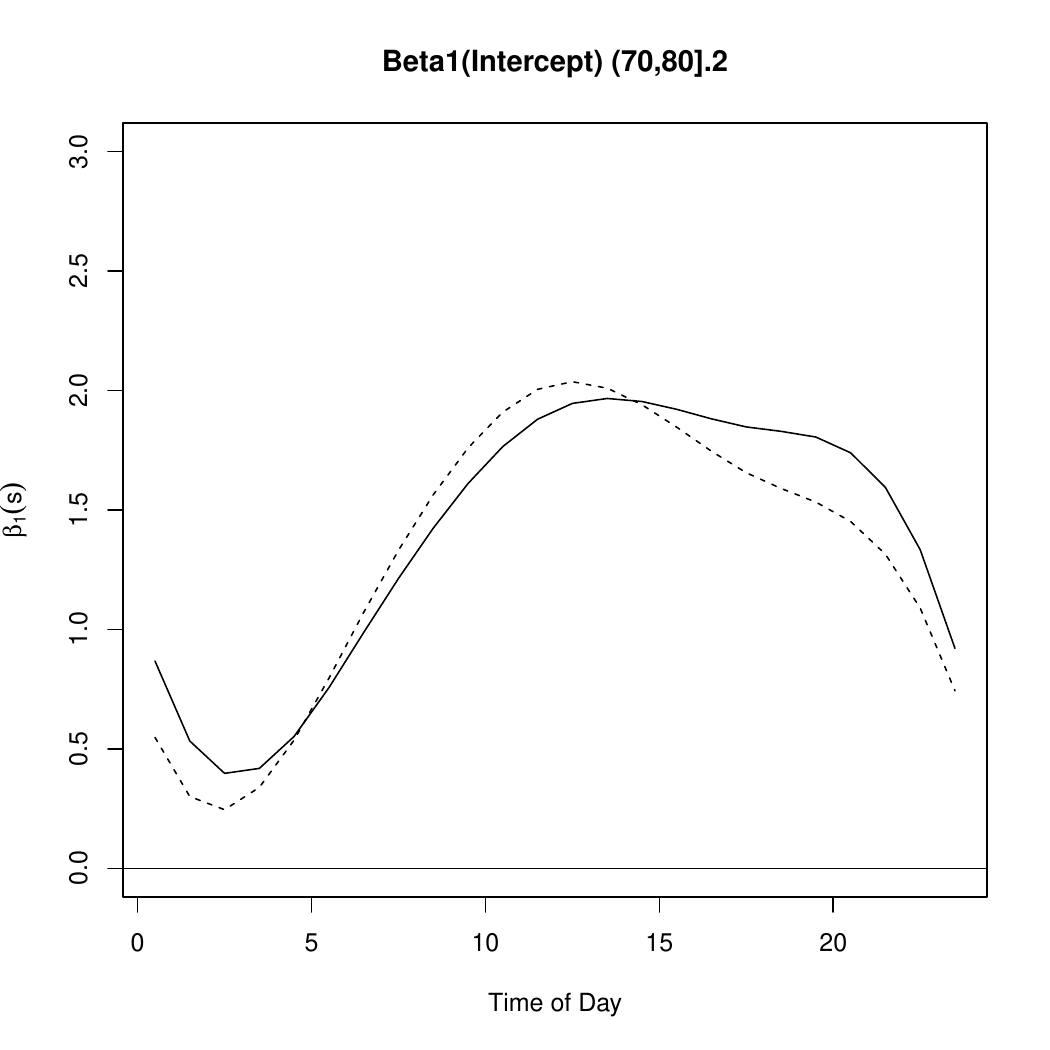}
  \caption{Predicted functional trajectories ($\hat{\beta}_1(s)+\hat{\beta}_{1i}(s)$) of diurnal MIMS for males from the MuFuMES method in the NHANES application. The other continuous covariates were held at their average values. The solid line corresponds to the fixed effect prediction $\hat{\beta}_1(s)$ and the dashed line corresponds to the mixed effects prediction $\hat{\beta}_1(s)+\hat{\beta}_{1i}(s)$. The age-by-race groups are indicated in the header.}
  \end{figure}
  \begin{figure}[H]
\includegraphics[page=13,scale=.95]{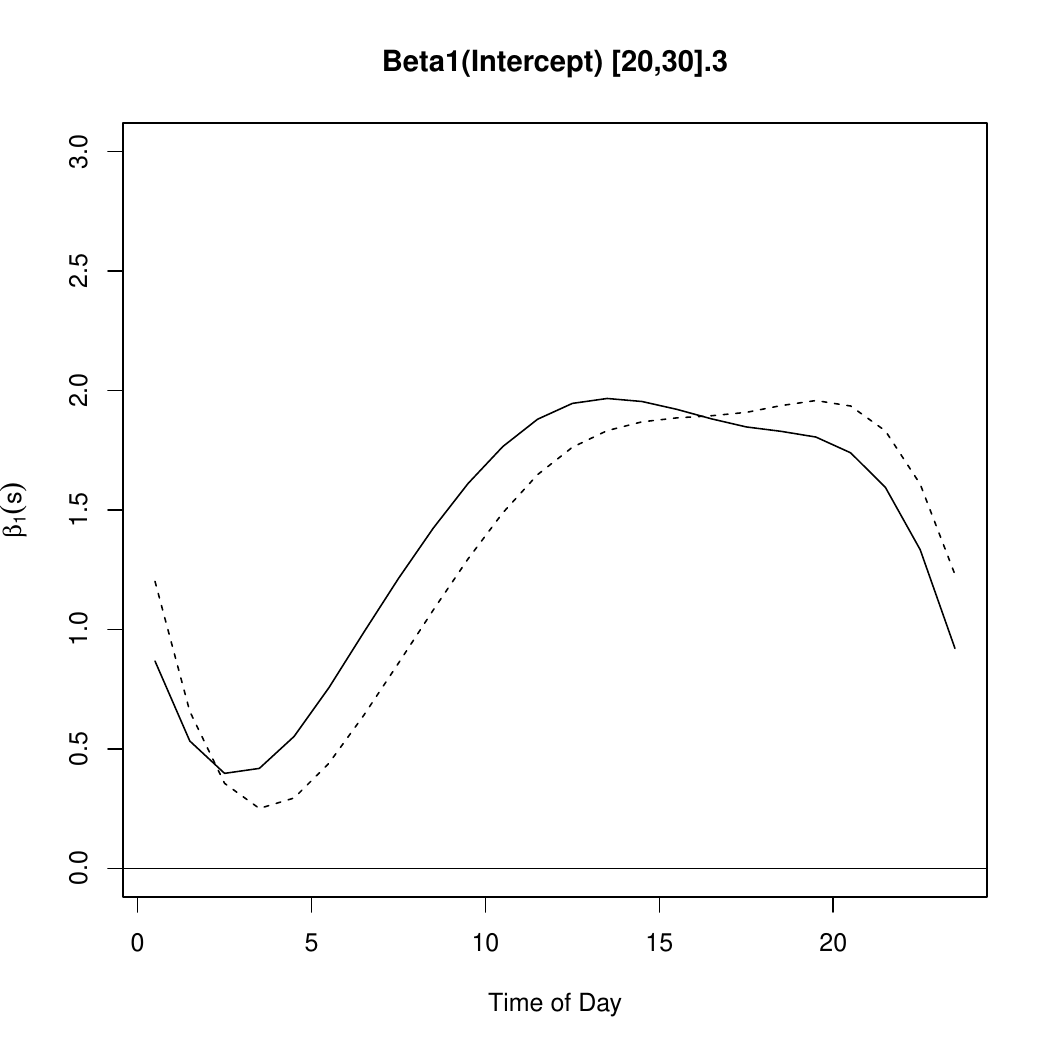}
  \caption{Predicted functional trajectories ($\hat{\beta}_1(s)+\hat{\beta}_{1i}(s)$) of diurnal MIMS for males from the MuFuMES method in the NHANES application. The other continuous covariates were held at their average values. The solid line corresponds to the fixed effect prediction $\hat{\beta}_1(s)$ and the dashed line corresponds to the mixed effects prediction $\hat{\beta}_1(s)+\hat{\beta}_{1i}(s)$. The age-by-race groups are indicated in the header.}
  \end{figure}
  \begin{figure}[H]
\includegraphics[page=14,scale=.95]{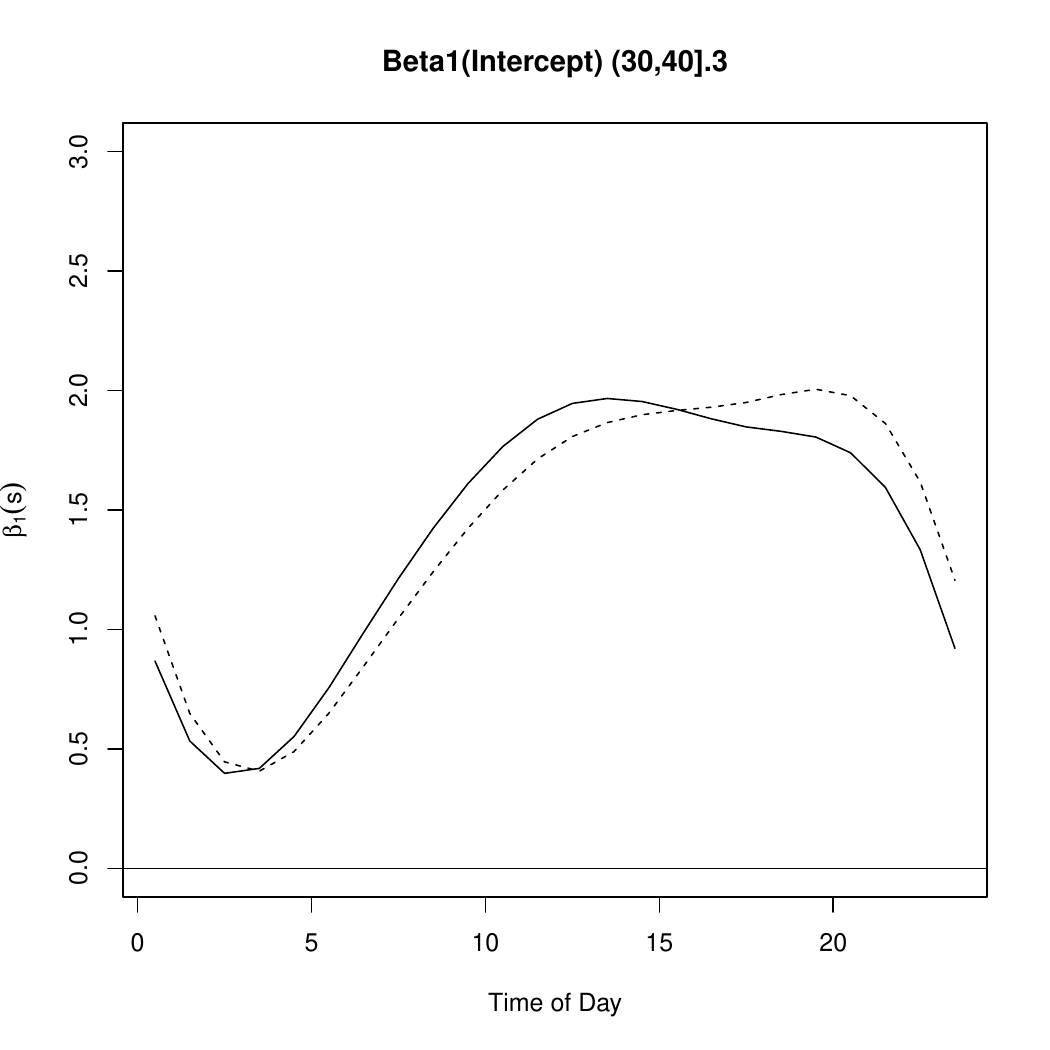}
  \caption{Predicted functional trajectories ($\hat{\beta}_1(s)+\hat{\beta}_{1i}(s)$) of diurnal MIMS for males from the MuFuMES method in the NHANES application. The other continuous covariates were held at their average values. The solid line corresponds to the fixed effect prediction $\hat{\beta}_1(s)$ and the dashed line corresponds to the mixed effects prediction $\hat{\beta}_1(s)+\hat{\beta}_{1i}(s)$. The age-by-race groups are indicated in the header.}
  \end{figure}
  \begin{figure}[H]
\includegraphics[page=15,scale=.95]{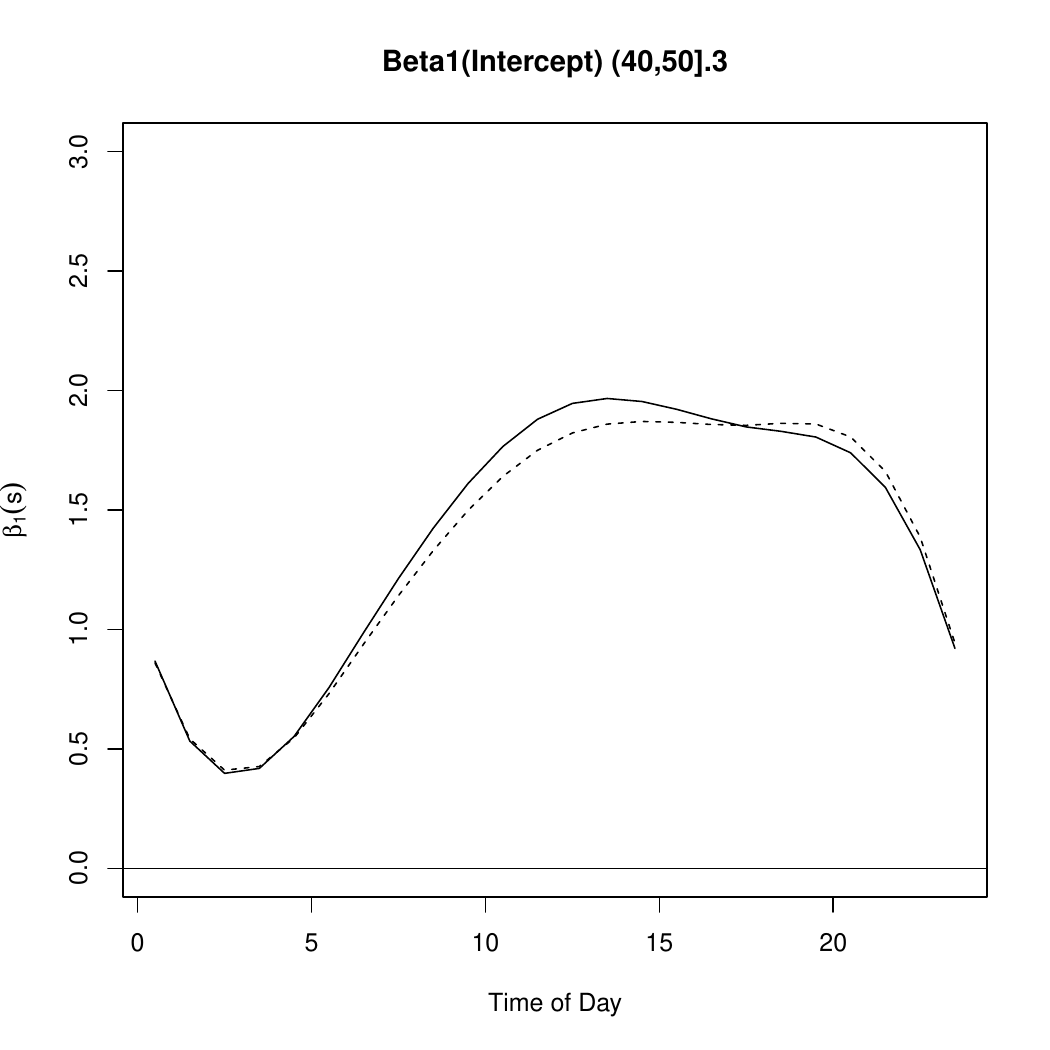}
  \caption{Predicted functional trajectories ($\hat{\beta}_1(s)+\hat{\beta}_{1i}(s)$) of diurnal MIMS for males from the MuFuMES method in the NHANES application. The other continuous covariates were held at their average values. The solid line corresponds to the fixed effect prediction $\hat{\beta}_1(s)$ and the dashed line corresponds to the mixed effects prediction $\hat{\beta}_1(s)+\hat{\beta}_{1i}(s)$. The age-by-race groups are indicated in the header.}
  \end{figure}
  \begin{figure}[H]
\includegraphics[page=16,scale=.95]{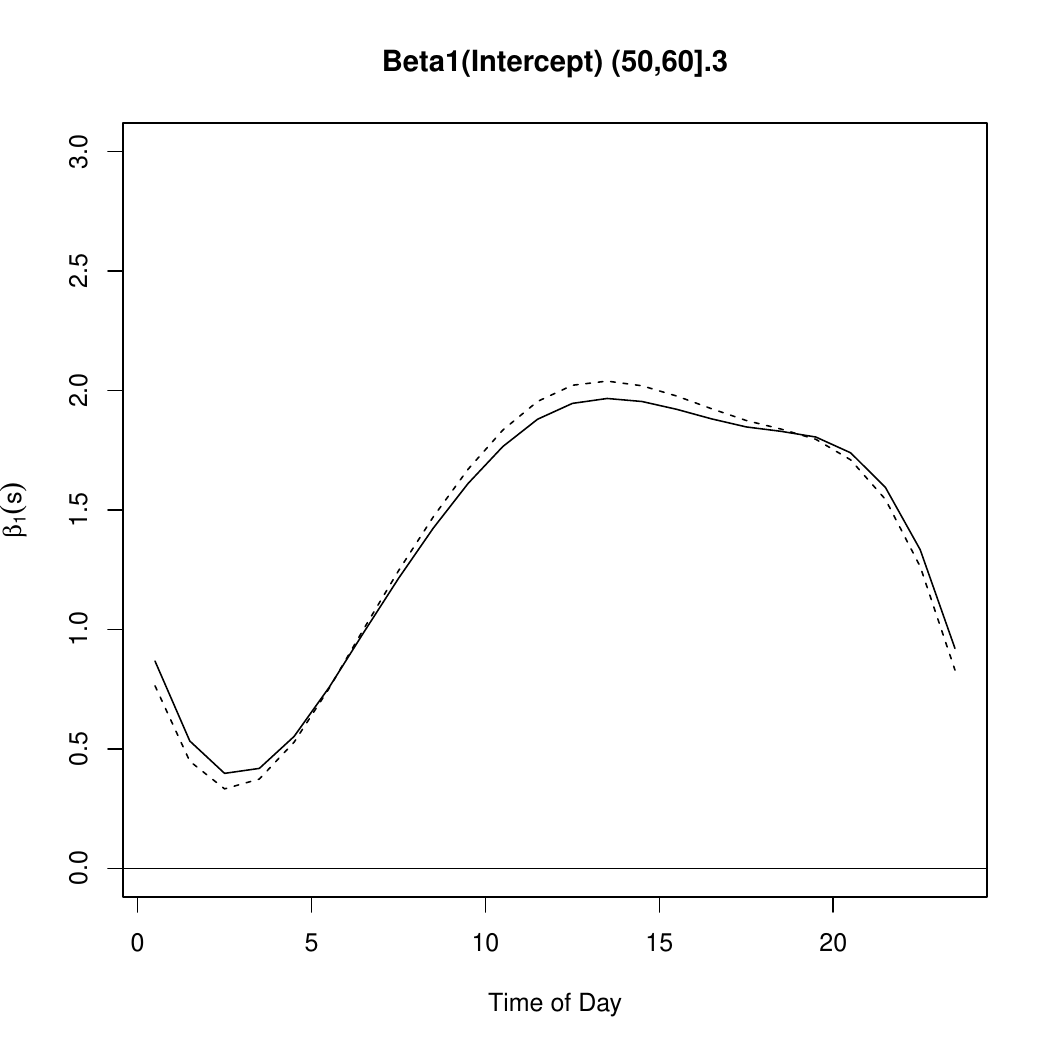}
  \caption{Predicted functional trajectories ($\hat{\beta}_1(s)+\hat{\beta}_{1i}(s)$) of diurnal MIMS for males from the MuFuMES method in the NHANES application. The other continuous covariates were held at their average values. The solid line corresponds to the fixed effect prediction $\hat{\beta}_1(s)$ and the dashed line corresponds to the mixed effects prediction $\hat{\beta}_1(s)+\hat{\beta}_{1i}(s)$. The age-by-race groups are indicated in the header.}
  \end{figure}
  \begin{figure}[H]
\includegraphics[page=17,scale=.95]{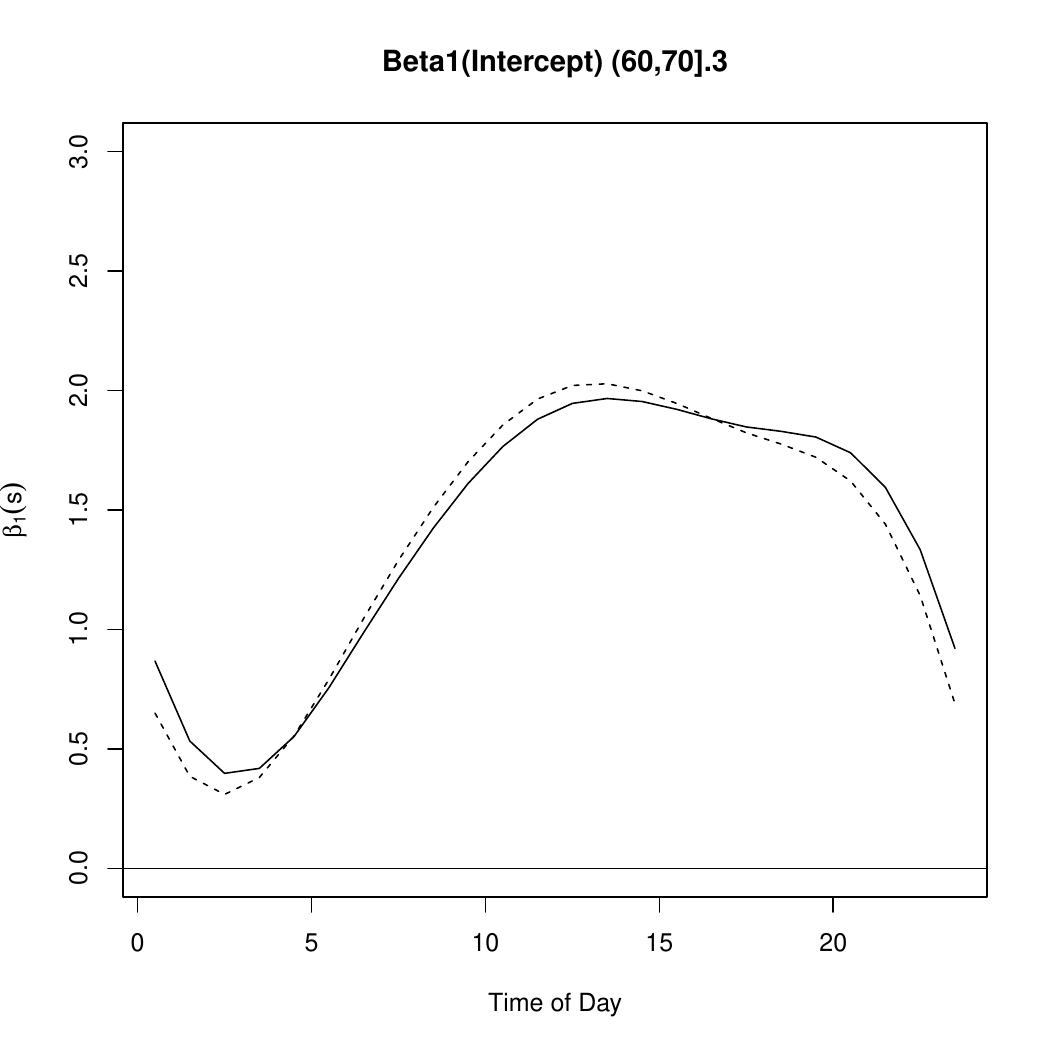}
  \caption{Predicted functional trajectories ($\hat{\beta}_1(s)+\hat{\beta}_{1i}(s)$) of diurnal MIMS for males from the MuFuMES method in the NHANES application. The other continuous covariates were held at their average values. The solid line corresponds to the fixed effect prediction $\hat{\beta}_1(s)$ and the dashed line corresponds to the mixed effects prediction $\hat{\beta}_1(s)+\hat{\beta}_{1i}(s)$. The age-by-race groups are indicated in the header.}
  \end{figure}
  \begin{figure}[H]
\includegraphics[page=18,scale=.95]{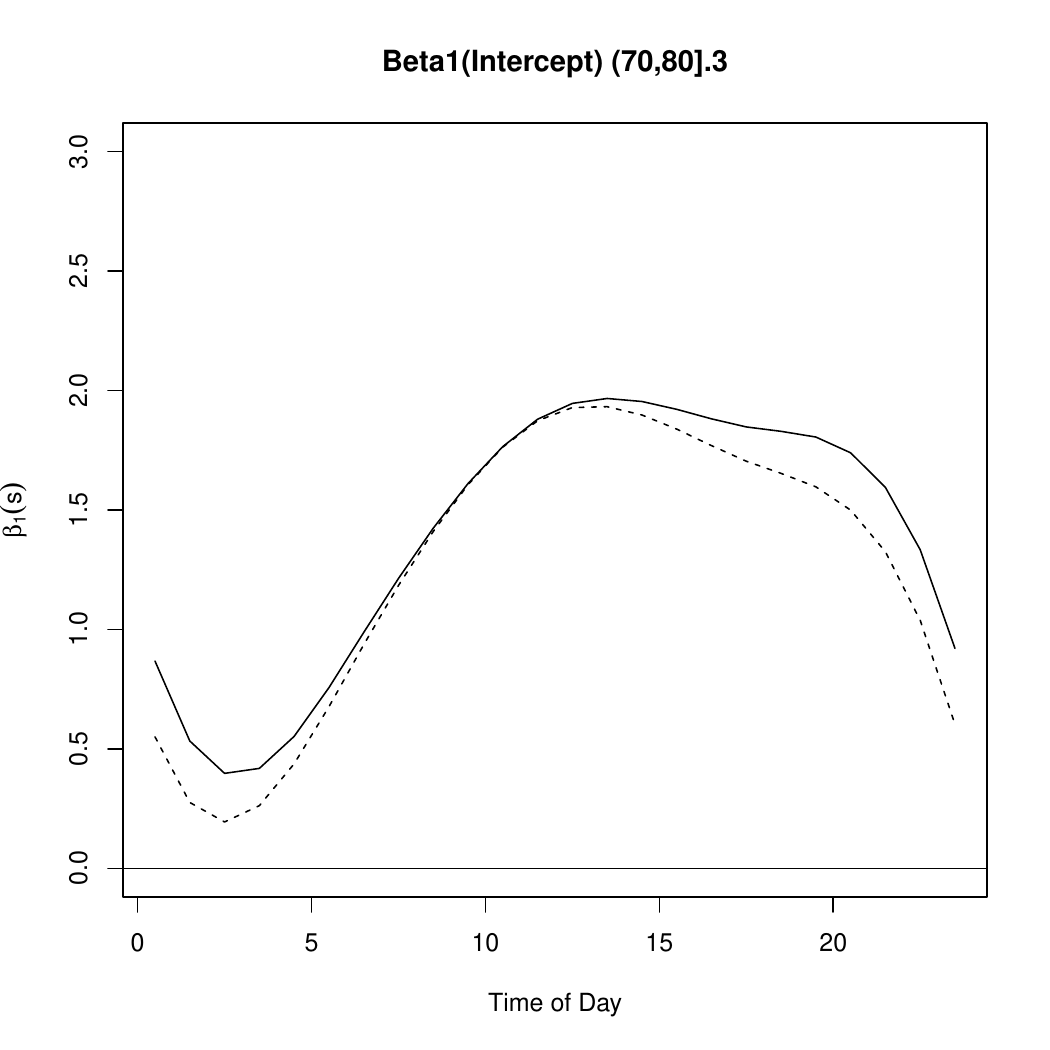}
  \caption{Predicted functional trajectories ($\hat{\beta}_1(s)+\hat{\beta}_{1i}(s)$) of diurnal MIMS for males from the MuFuMES method in the NHANES application. The other continuous covariates were held at their average values. The solid line corresponds to the fixed effect prediction $\hat{\beta}_1(s)$ and the dashed line corresponds to the mixed effects prediction $\hat{\beta}_1(s)+\hat{\beta}_{1i}(s)$. The age-by-race groups are indicated in the header.}
  \end{figure}
  \begin{figure}[H]
\includegraphics[page=19,scale=.95]{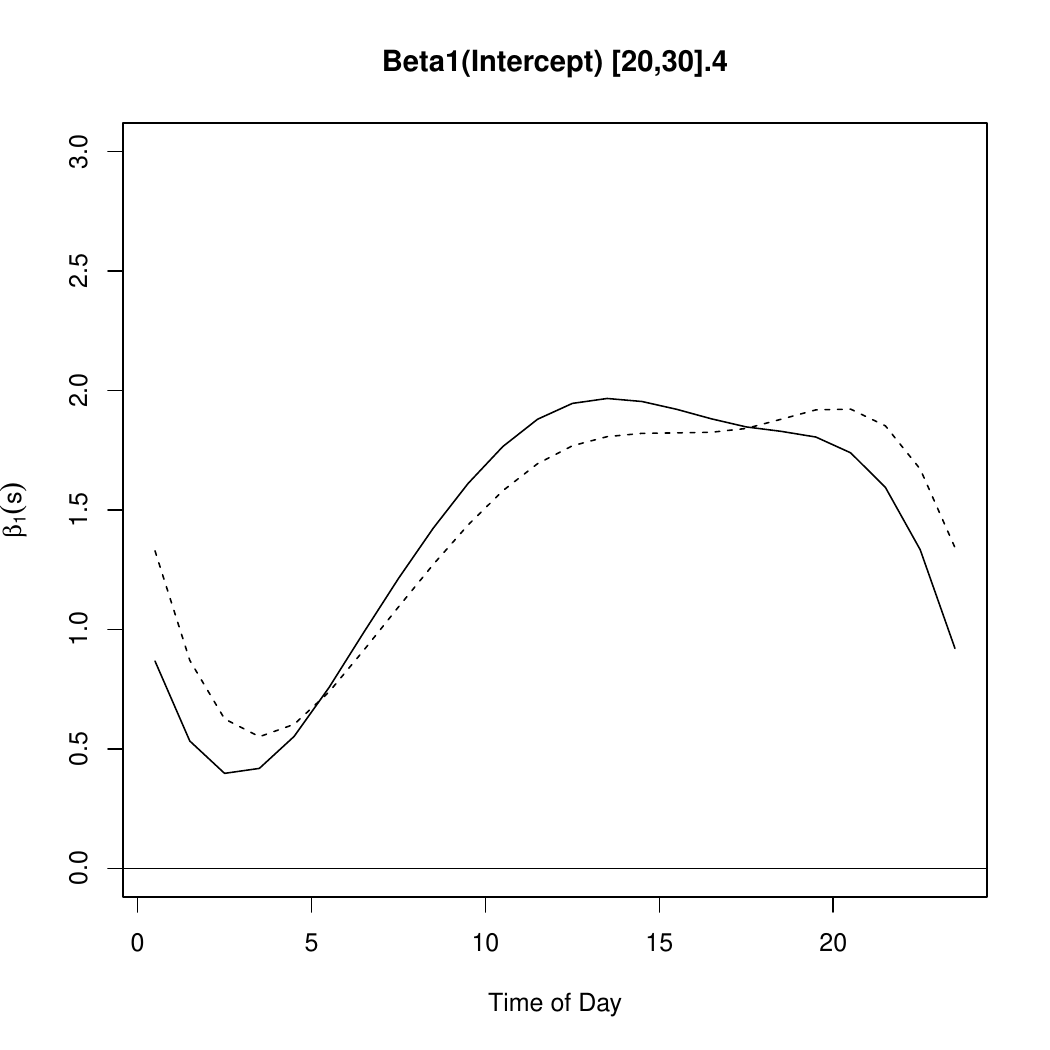}
  \caption{Predicted functional trajectories ($\hat{\beta}_1(s)+\hat{\beta}_{1i}(s)$) of diurnal MIMS for males from the MuFuMES method in the NHANES application. The other continuous covariates were held at their average values. The solid line corresponds to the fixed effect prediction $\hat{\beta}_1(s)$ and the dashed line corresponds to the mixed effects prediction $\hat{\beta}_1(s)+\hat{\beta}_{1i}(s)$. The age-by-race groups are indicated in the header.}
  \end{figure}
  \begin{figure}[H]
\includegraphics[page=20,scale=.95]{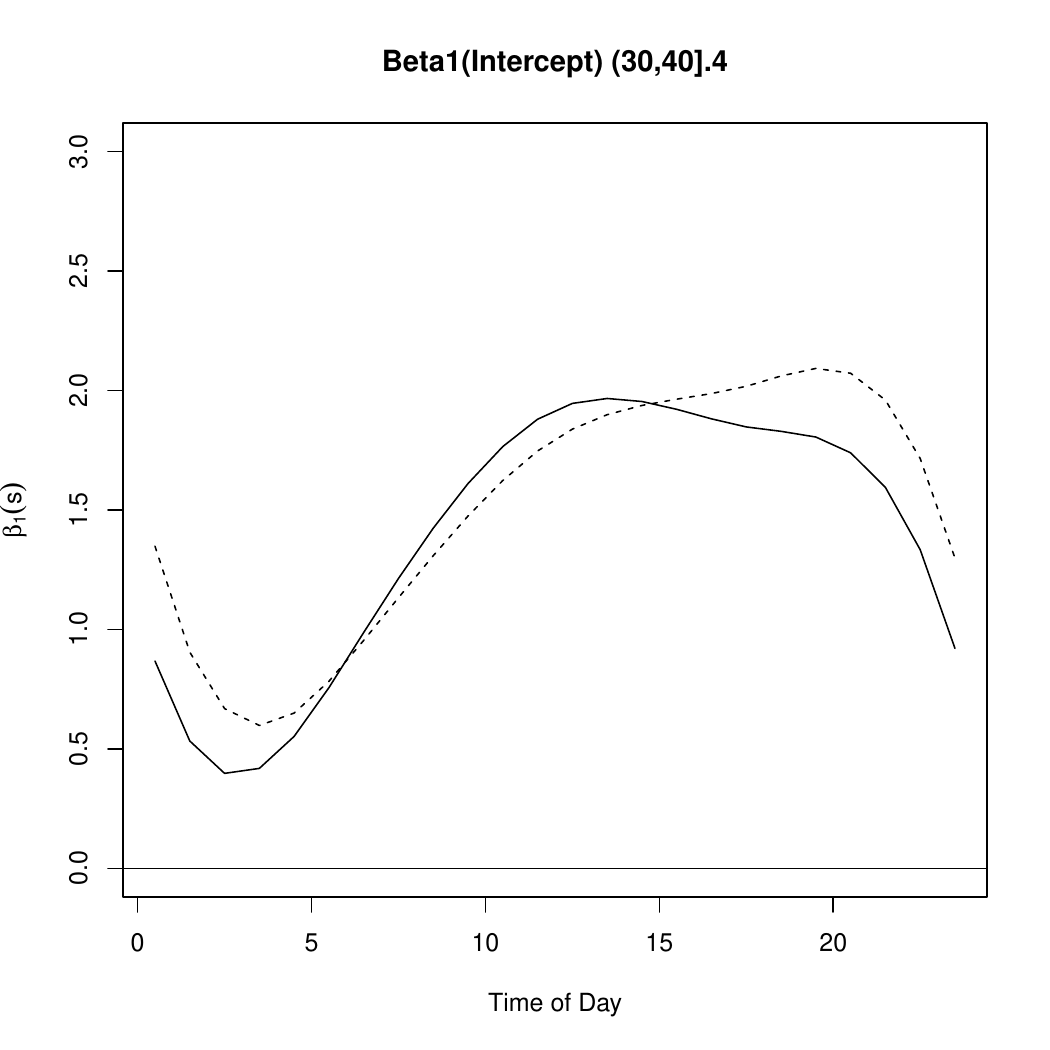}
  \caption{Predicted functional trajectories ($\hat{\beta}_1(s)+\hat{\beta}_{1i}(s)$) of diurnal MIMS for males from the MuFuMES method in the NHANES application. The other continuous covariates were held at their average values. The solid line corresponds to the fixed effect prediction $\hat{\beta}_1(s)$ and the dashed line corresponds to the mixed effects prediction $\hat{\beta}_1(s)+\hat{\beta}_{1i}(s)$. The age-by-race groups are indicated in the header.}
  \end{figure}
  \begin{figure}[H]
\includegraphics[page=21,scale=.95]{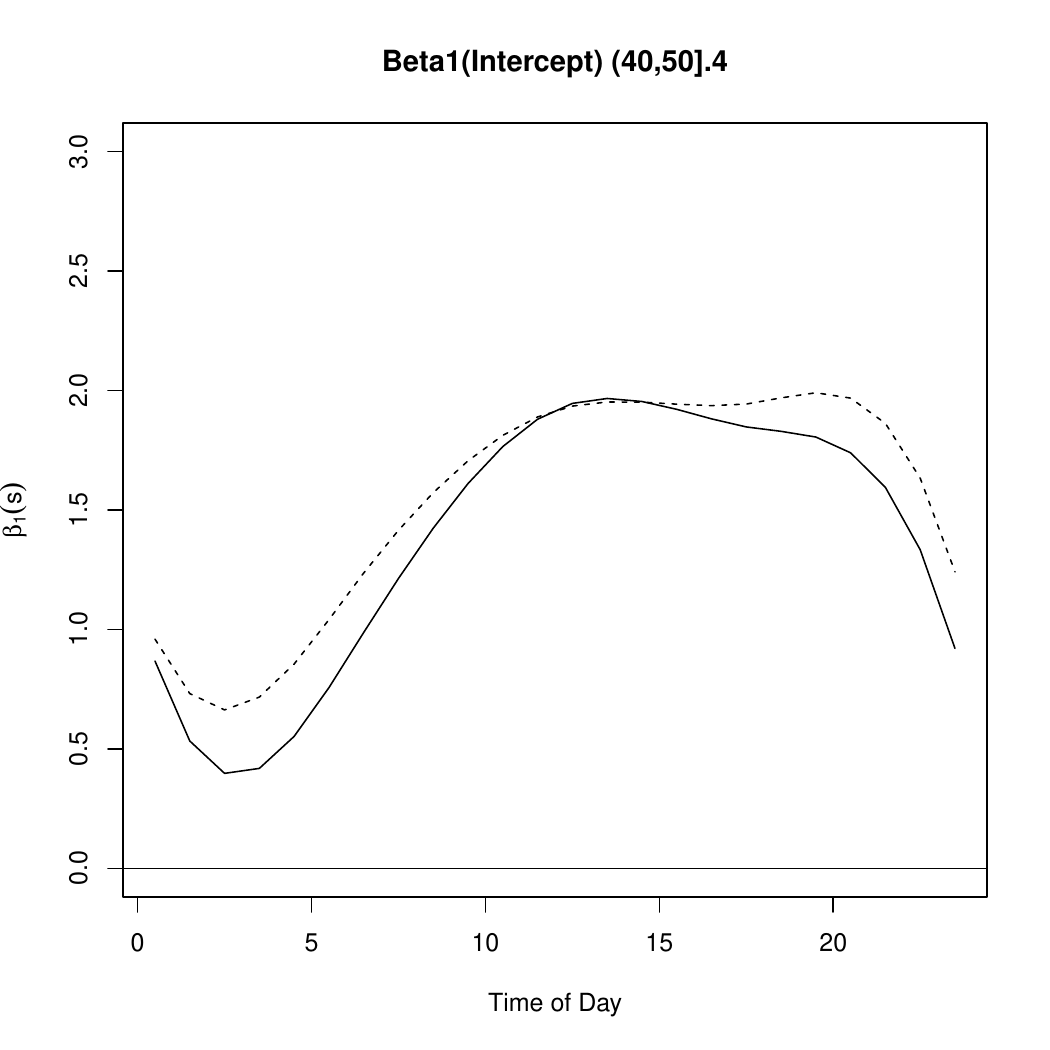}
  \caption{Predicted functional trajectories ($\hat{\beta}_1(s)+\hat{\beta}_{1i}(s)$) of diurnal MIMS for males from the MuFuMES method in the NHANES application. The other continuous covariates were held at their average values. The solid line corresponds to the fixed effect prediction $\hat{\beta}_1(s)$ and the dashed line corresponds to the mixed effects prediction $\hat{\beta}_1(s)+\hat{\beta}_{1i}(s)$. The age-by-race groups are indicated in the header.}
  \end{figure}
  \begin{figure}[H]
\includegraphics[page=22,scale=.95]{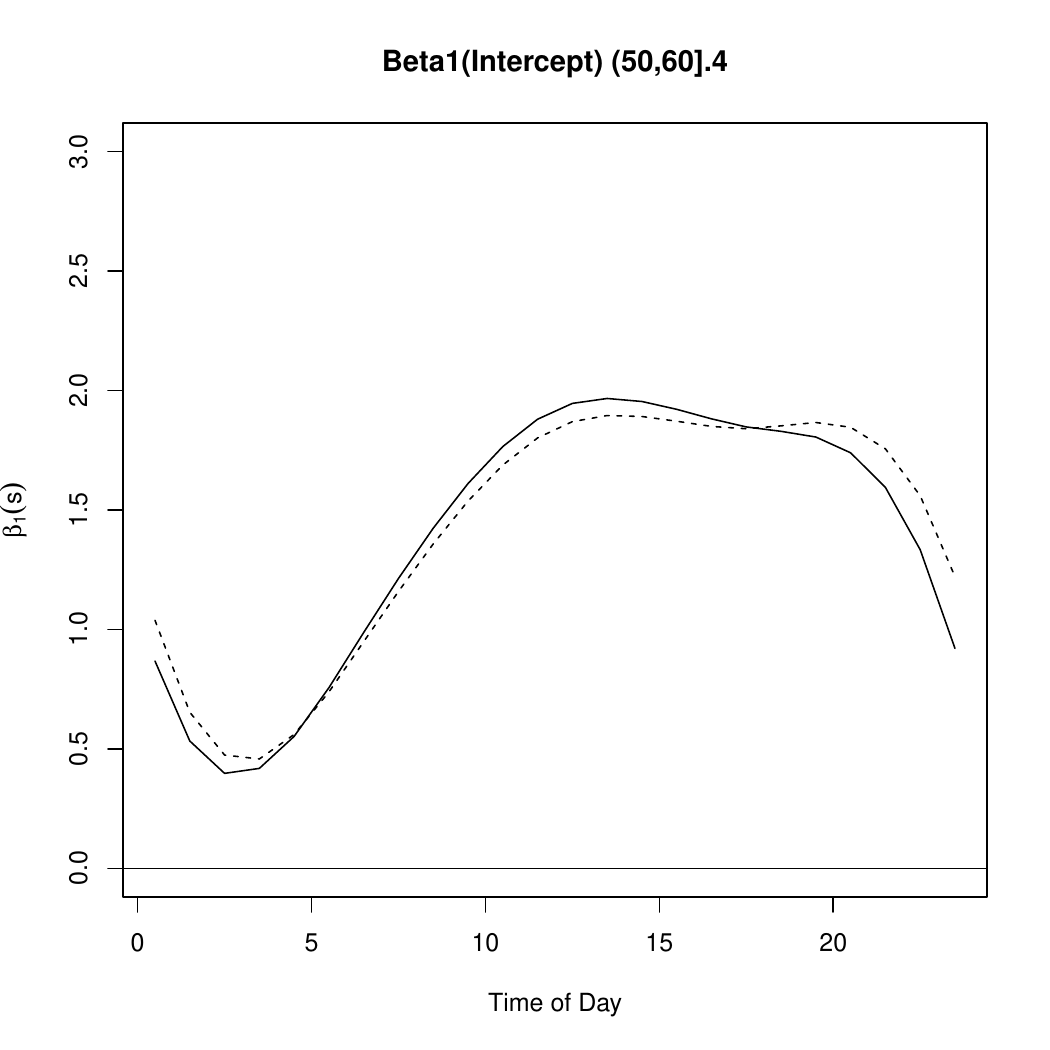}
  \caption{Predicted functional trajectories ($\hat{\beta}_1(s)+\hat{\beta}_{1i}(s)$) of diurnal MIMS for males from the MuFuMES method in the NHANES application. The other continuous covariates were held at their average values. The solid line corresponds to the fixed effect prediction $\hat{\beta}_1(s)$ and the dashed line corresponds to the mixed effects prediction $\hat{\beta}_1(s)+\hat{\beta}_{1i}(s)$. The age-by-race groups are indicated in the header.}
  \end{figure}
  \begin{figure}[H]
\includegraphics[page=23,scale=.95]{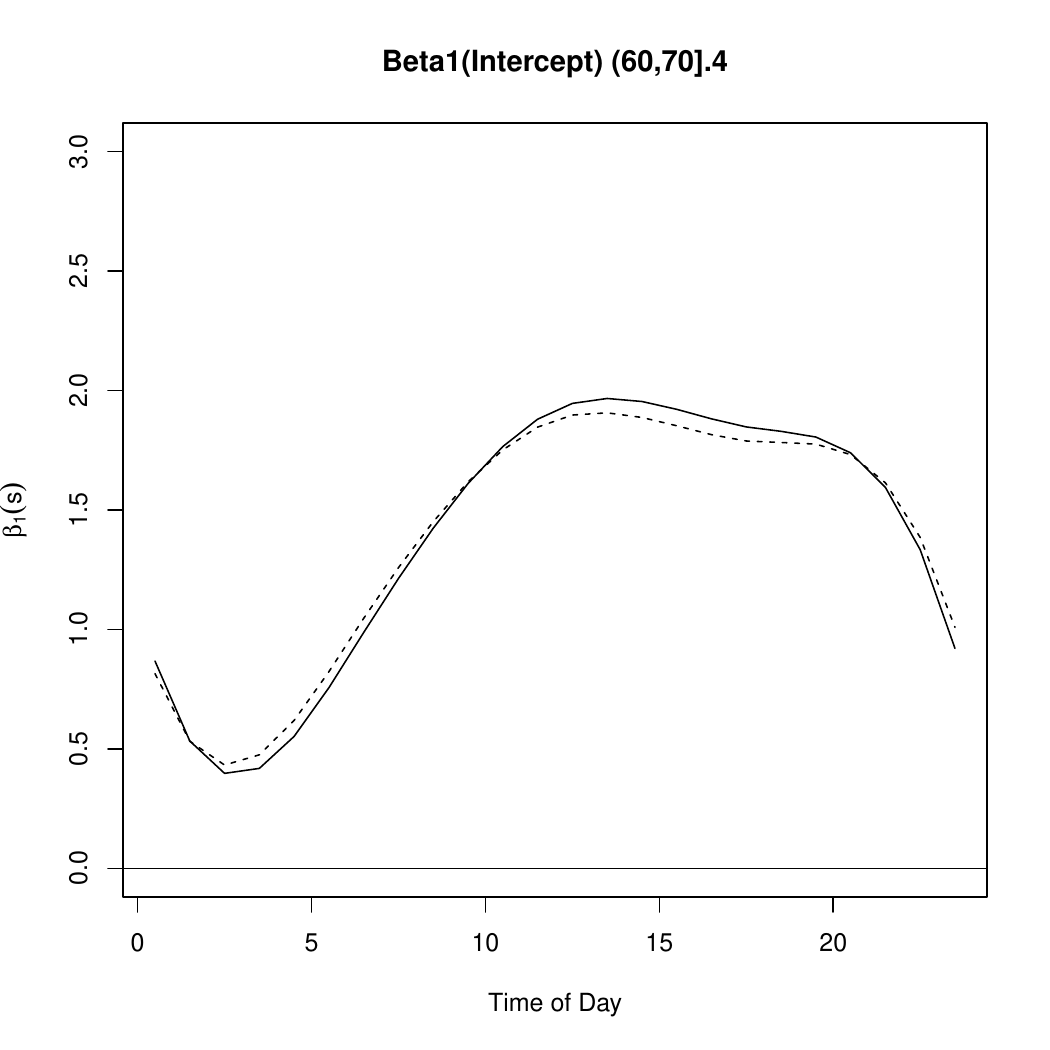}
  \caption{Predicted functional trajectories ($\hat{\beta}_1(s)+\hat{\beta}_{1i}(s)$) of diurnal MIMS for males from the MuFuMES method in the NHANES application. The other continuous covariates were held at their average values. The solid line corresponds to the fixed effect prediction $\hat{\beta}_1(s)$ and the dashed line corresponds to the mixed effects prediction $\hat{\beta}_1(s)+\hat{\beta}_{1i}(s)$. The age-by-race groups are indicated in the header.}
  \end{figure}
  \begin{figure}[H]
\includegraphics[page=24,scale=.95]{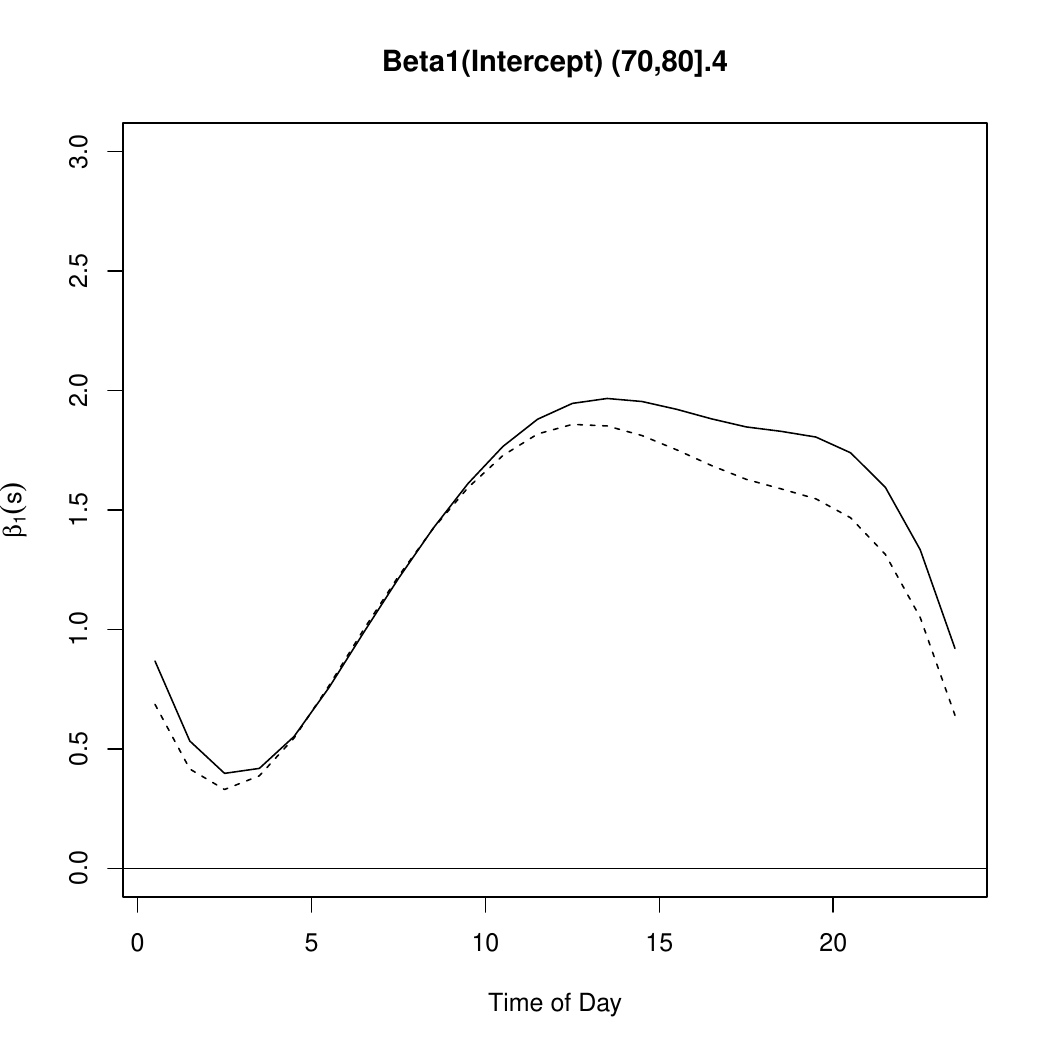}
  \caption{Predicted functional trajectories ($\hat{\beta}_1(s)+\hat{\beta}_{1i}(s)$) of diurnal MIMS for males from the MuFuMES method in the NHANES application. The other continuous covariates were held at their average values. The solid line corresponds to the fixed effect prediction $\hat{\beta}_1(s)$ and the dashed line corresponds to the mixed effects prediction $\hat{\beta}_1(s)+\hat{\beta}_{1i}(s)$. The age-by-race groups are indicated in the header.}
  \end{figure}
  \begin{figure}[H]
\includegraphics[page=25,scale=.95]{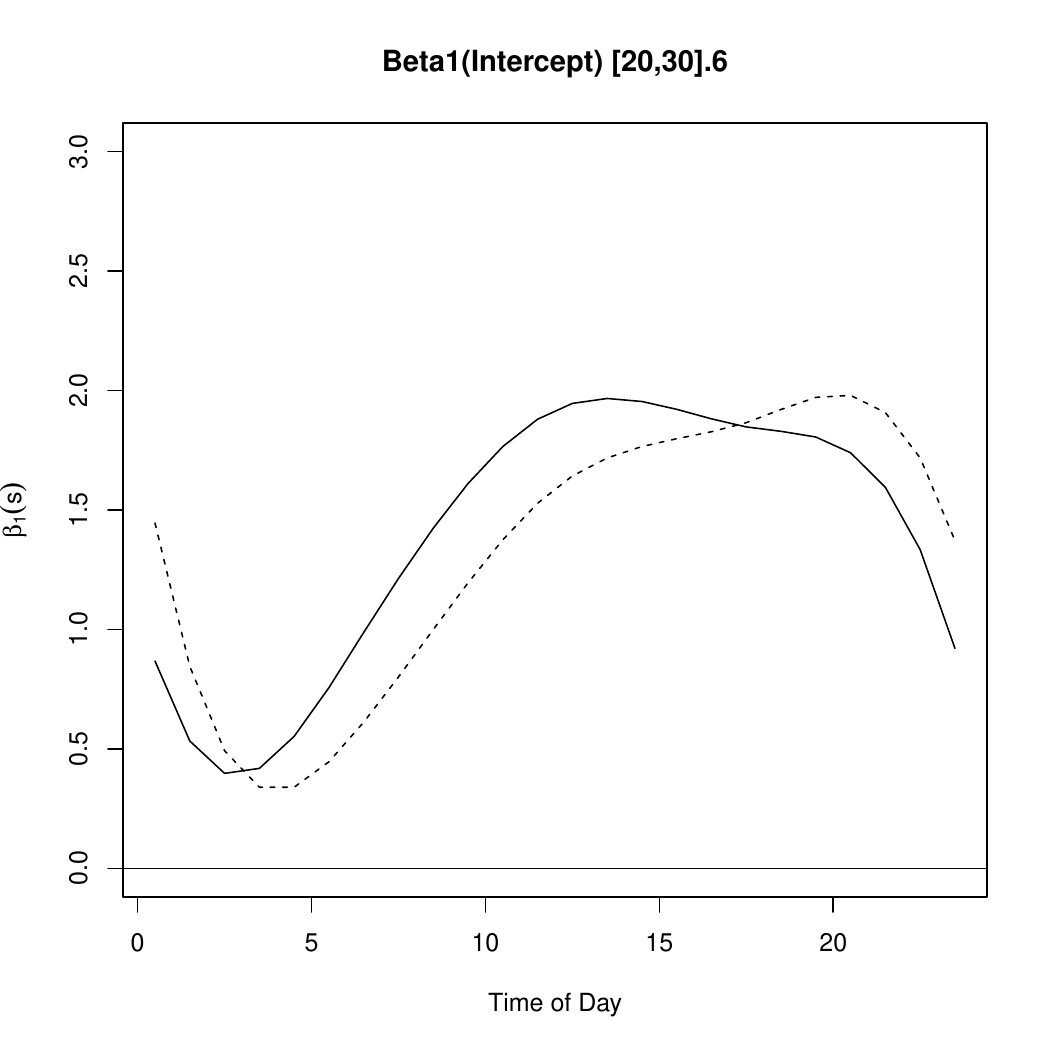}
  \caption{Predicted functional trajectories ($\hat{\beta}_1(s)+\hat{\beta}_{1i}(s)$) of diurnal MIMS for males from the MuFuMES method in the NHANES application. The other continuous covariates were held at their average values. The solid line corresponds to the fixed effect prediction $\hat{\beta}_1(s)$ and the dashed line corresponds to the mixed effects prediction $\hat{\beta}_1(s)+\hat{\beta}_{1i}(s)$. The age-by-race groups are indicated in the header.}
  \end{figure}
  \begin{figure}[H]
\includegraphics[page=26,scale=.95]{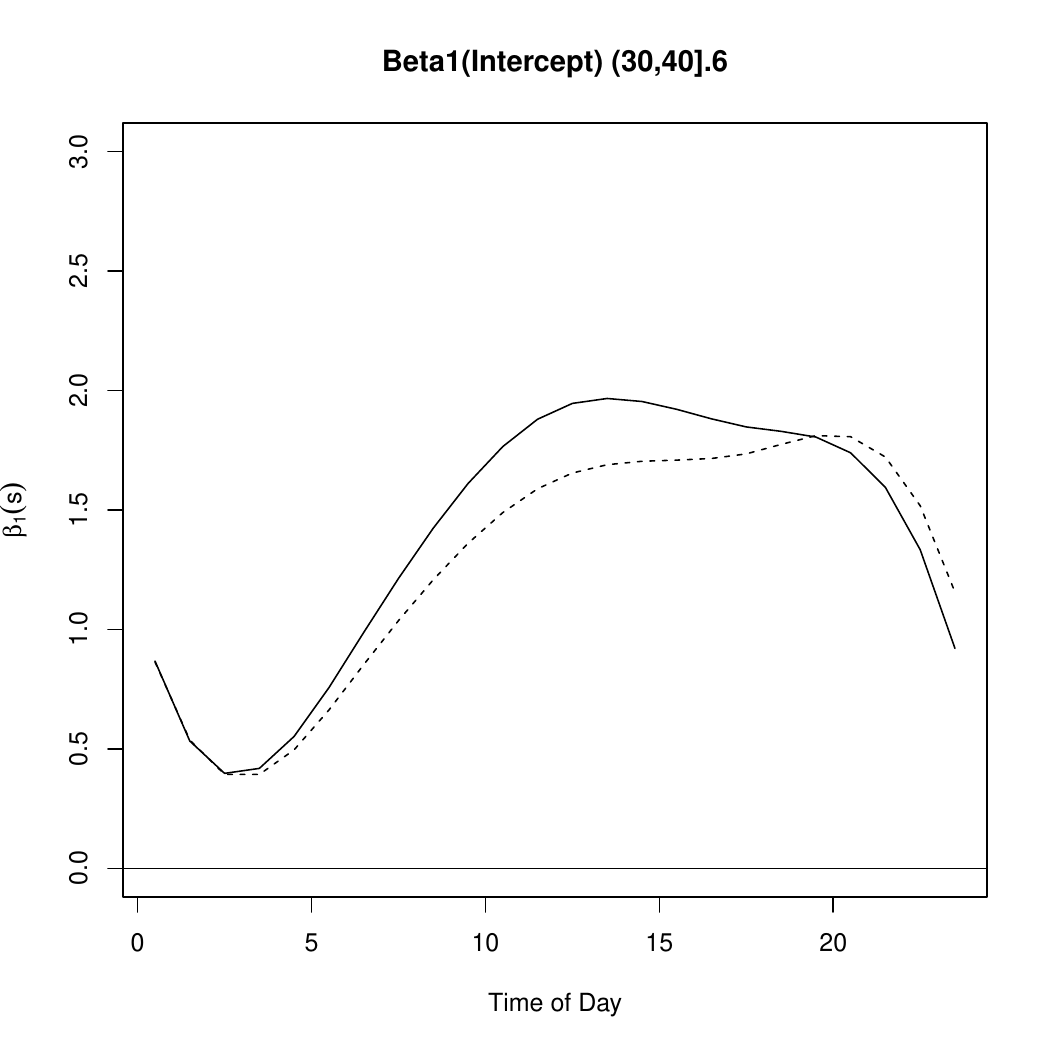}
  \caption{Predicted functional trajectories ($\hat{\beta}_1(s)+\hat{\beta}_{1i}(s)$) of diurnal MIMS for males from the MuFuMES method in the NHANES application. The other continuous covariates were held at their average values. The solid line corresponds to the fixed effect prediction $\hat{\beta}_1(s)$ and the dashed line corresponds to the mixed effects prediction $\hat{\beta}_1(s)+\hat{\beta}_{1i}(s)$. The age-by-race groups are indicated in the header.}
  \end{figure}
  \begin{figure}[H]
\includegraphics[page=27,scale=.95]{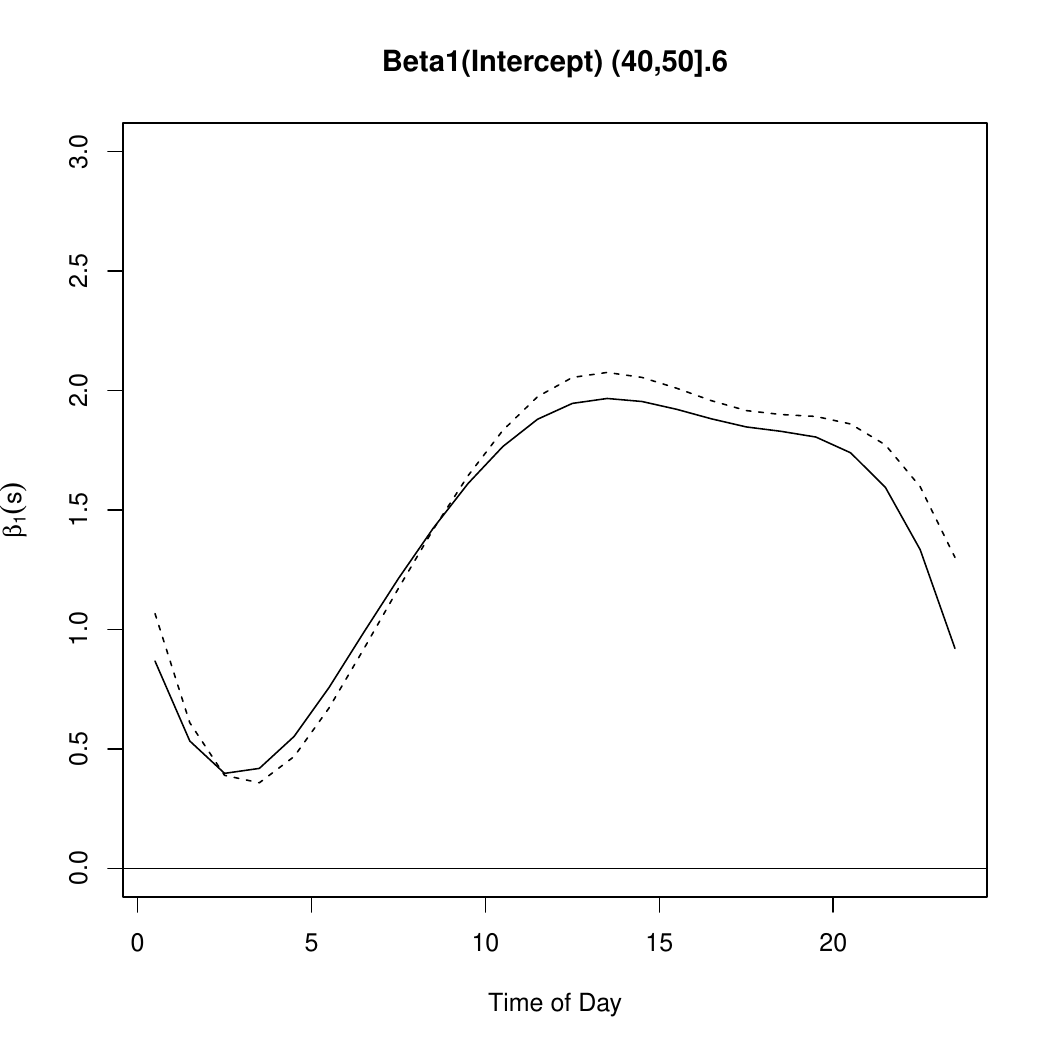}
  \caption{Predicted functional trajectories ($\hat{\beta}_1(s)+\hat{\beta}_{1i}(s)$) of diurnal MIMS for males from the MuFuMES method in the NHANES application. The other continuous covariates were held at their average values. The solid line corresponds to the fixed effect prediction $\hat{\beta}_1(s)$ and the dashed line corresponds to the mixed effects prediction $\hat{\beta}_1(s)+\hat{\beta}_{1i}(s)$. The age-by-race groups are indicated in the header.}
  \end{figure}
  \begin{figure}[H]
\includegraphics[page=28,scale=.95]{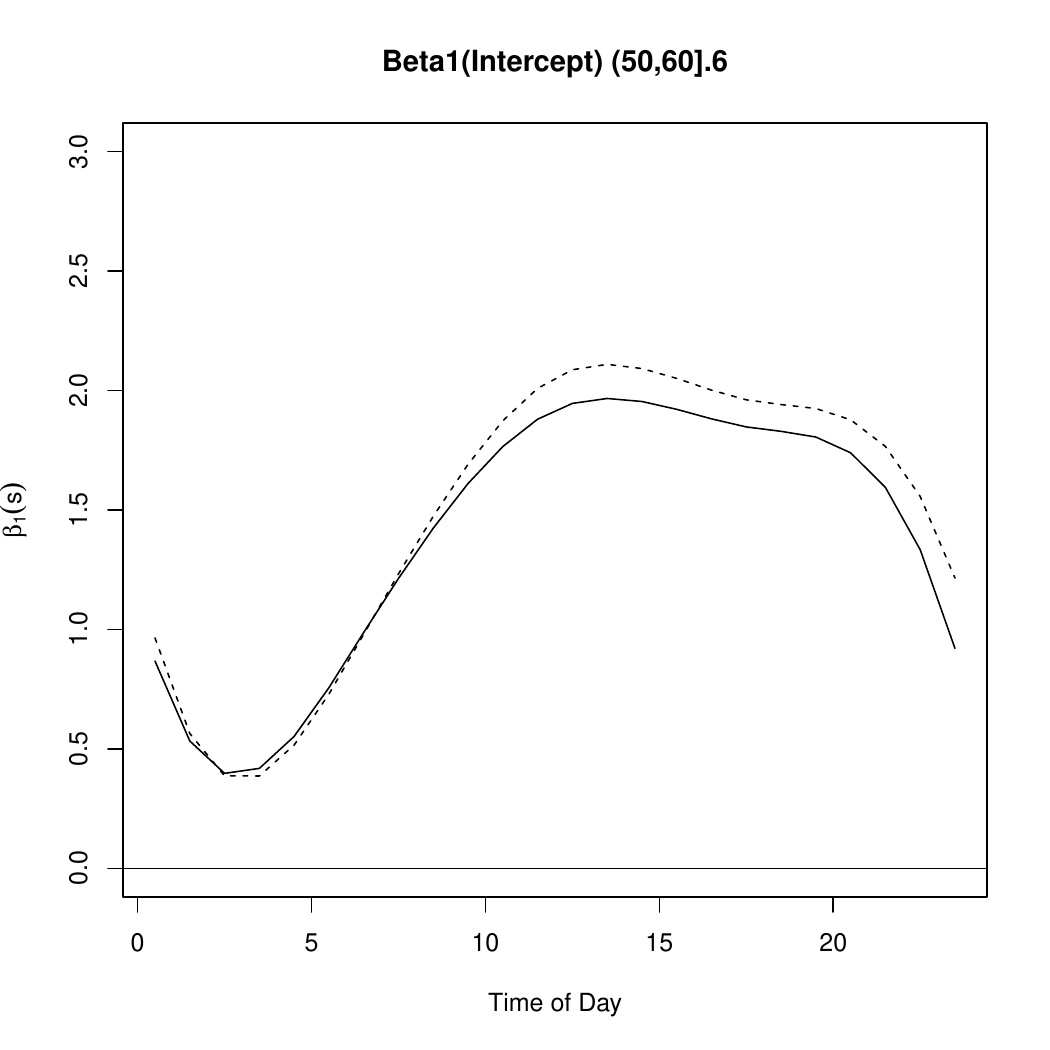}
  \caption{Predicted functional trajectories ($\hat{\beta}_1(s)+\hat{\beta}_{1i}(s)$) of diurnal MIMS for males from the MuFuMES method in the NHANES application. The other continuous covariates were held at their average values. The solid line corresponds to the fixed effect prediction $\hat{\beta}_1(s)$ and the dashed line corresponds to the mixed effects prediction $\hat{\beta}_1(s)+\hat{\beta}_{1i}(s)$. The age-by-race groups are indicated in the header.}
  \end{figure}
  \begin{figure}[H]
\includegraphics[page=29,scale=.95]{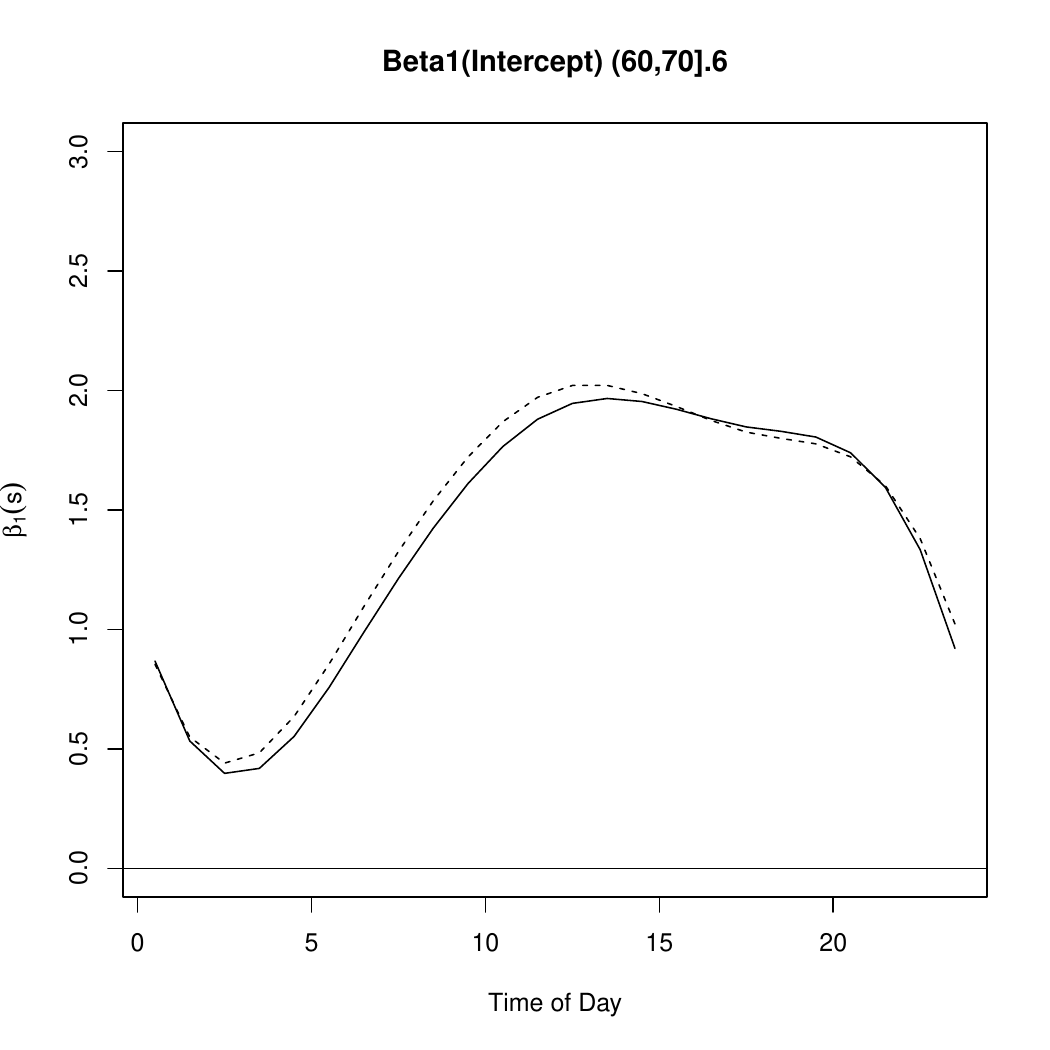}
  \caption{Predicted functional trajectories ($\hat{\beta}_1(s)+\hat{\beta}_{1i}(s)$) of diurnal MIMS for males from the MuFuMES method in the NHANES application. The other continuous covariates were held at their average values. The solid line corresponds to the fixed effect prediction $\hat{\beta}_1(s)$ and the dashed line corresponds to the mixed effects prediction $\hat{\beta}_1(s)+\hat{\beta}_{1i}(s)$. The age-by-race groups are indicated in the header.}
  \end{figure}
  \begin{figure}[H]
\includegraphics[page=30,scale=.95]{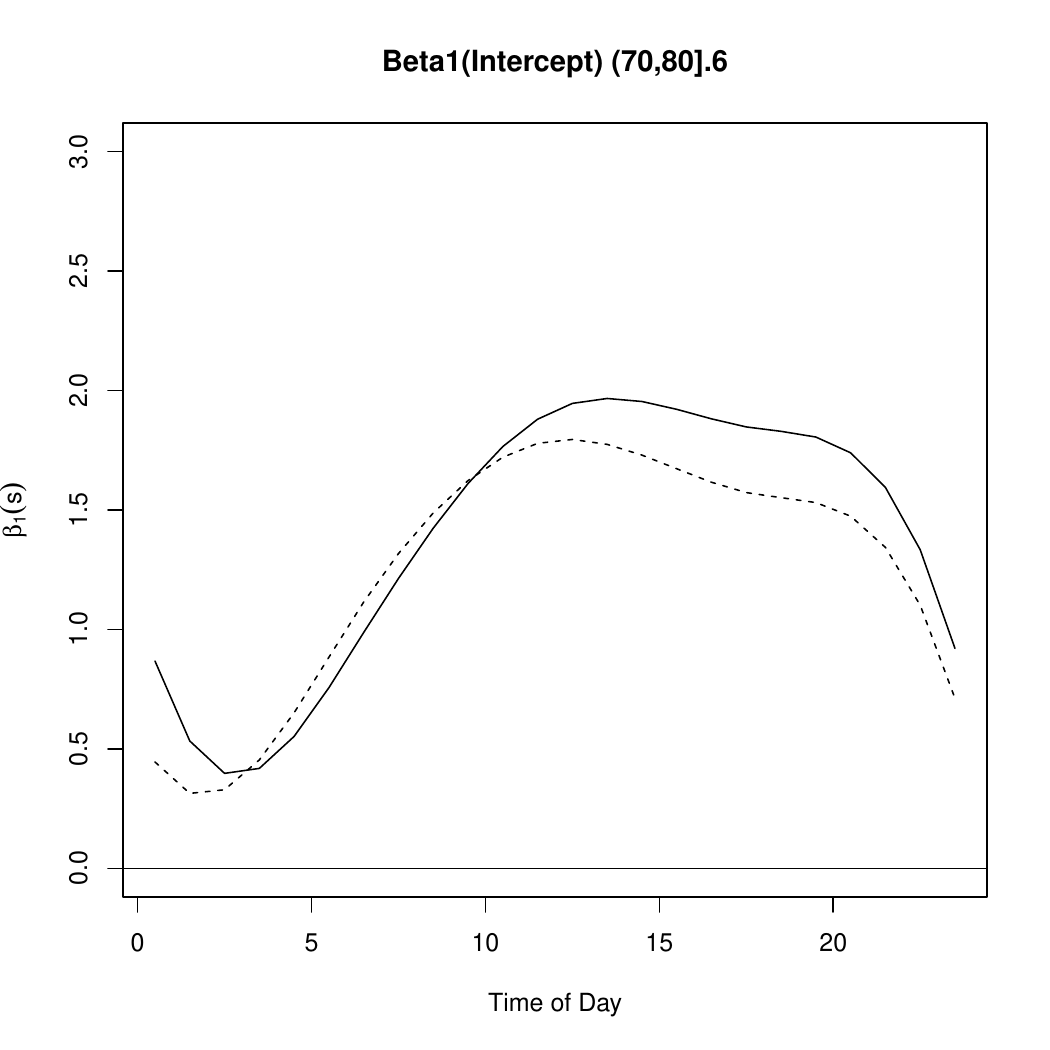}
  \caption{Predicted functional trajectories ($\hat{\beta}_1(s)+\hat{\beta}_{1i}(s)$) of diurnal MIMS for males from the MuFuMES method in the NHANES application. The other continuous covariates were held at their average values. The solid line corresponds to the fixed effect prediction $\hat{\beta}_1(s)$ and the dashed line corresponds to the mixed effects prediction $\hat{\beta}_1(s)+\hat{\beta}_{1i}(s)$. The age-by-race groups are indicated in the header.}
  \end{figure}
  \begin{figure}[H]
\includegraphics[page=31,scale=.95]{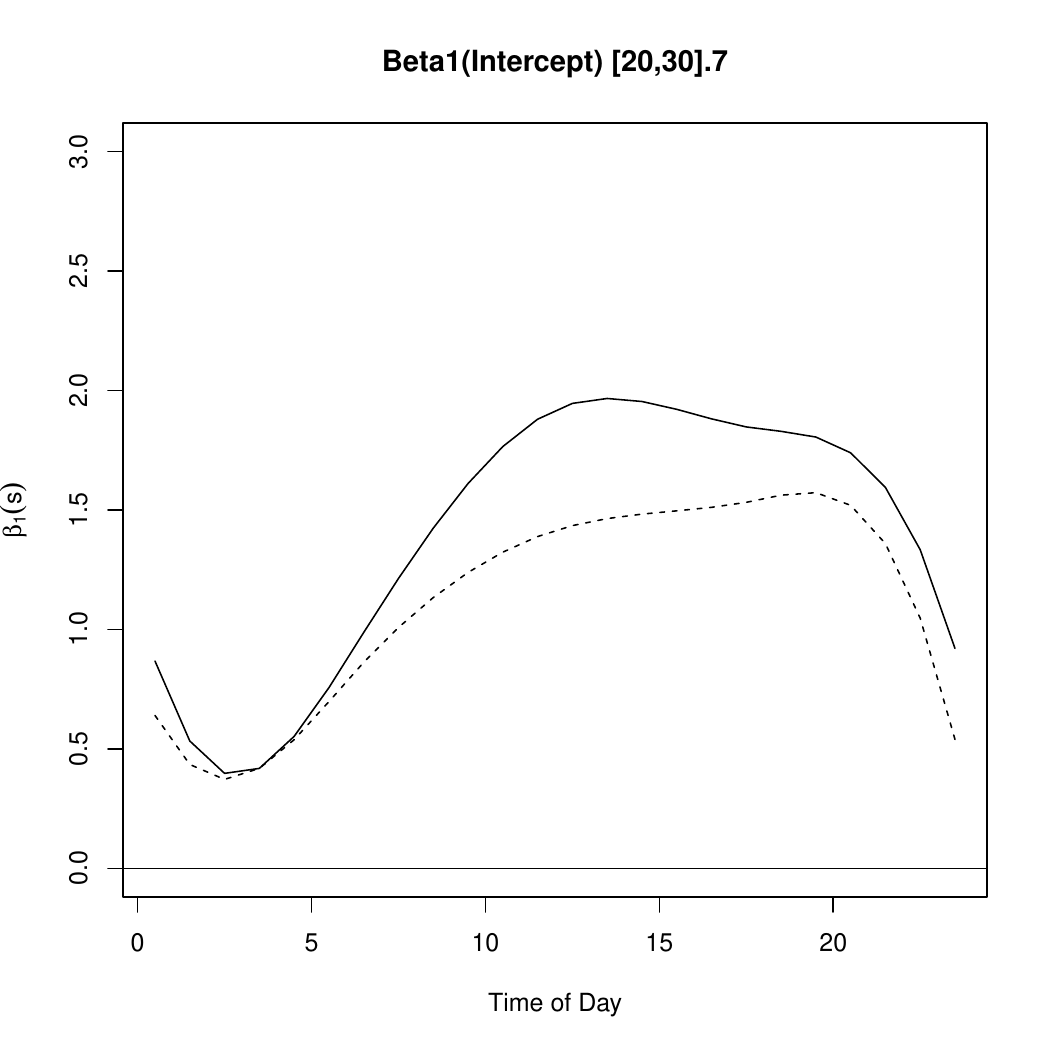}
  \caption{Predicted functional trajectories ($\hat{\beta}_1(s)+\hat{\beta}_{1i}(s)$) of diurnal MIMS for males from the MuFuMES method in the NHANES application. The other continuous covariates were held at their average values. The solid line corresponds to the fixed effect prediction $\hat{\beta}_1(s)$ and the dashed line corresponds to the mixed effects prediction $\hat{\beta}_1(s)+\hat{\beta}_{1i}(s)$. The age-by-race groups are indicated in the header.}
  \end{figure}
  \begin{figure}[H]
\includegraphics[page=32,scale=.95]{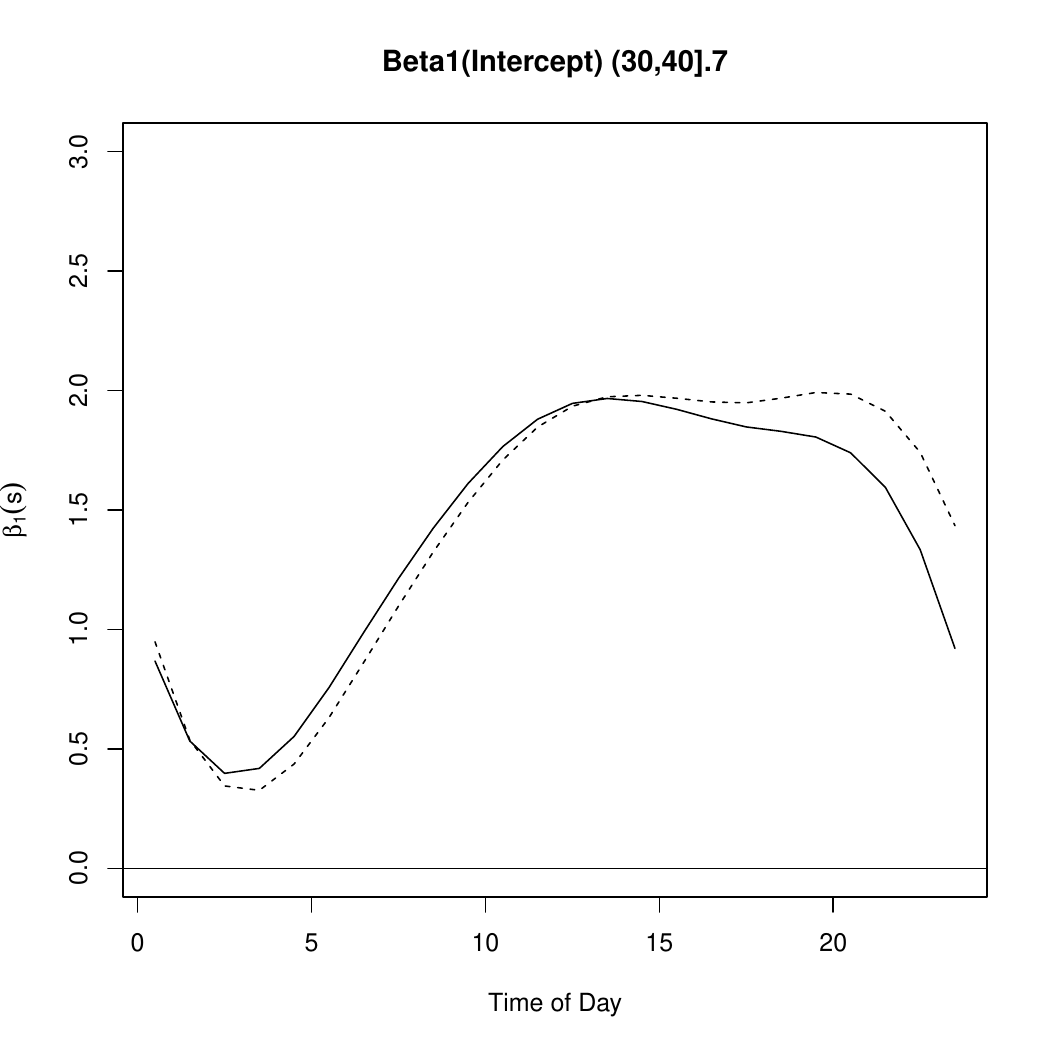}
  \caption{Predicted functional trajectories ($\hat{\beta}_1(s)+\hat{\beta}_{1i}(s)$) of diurnal MIMS for males from the MuFuMES method in the NHANES application. The other continuous covariates were held at their average values. The solid line corresponds to the fixed effect prediction $\hat{\beta}_1(s)$ and the dashed line corresponds to the mixed effects prediction $\hat{\beta}_1(s)+\hat{\beta}_{1i}(s)$. The age-by-race groups are indicated in the header.}
  \end{figure}
  \begin{figure}[H]
\includegraphics[page=33,scale=.95]{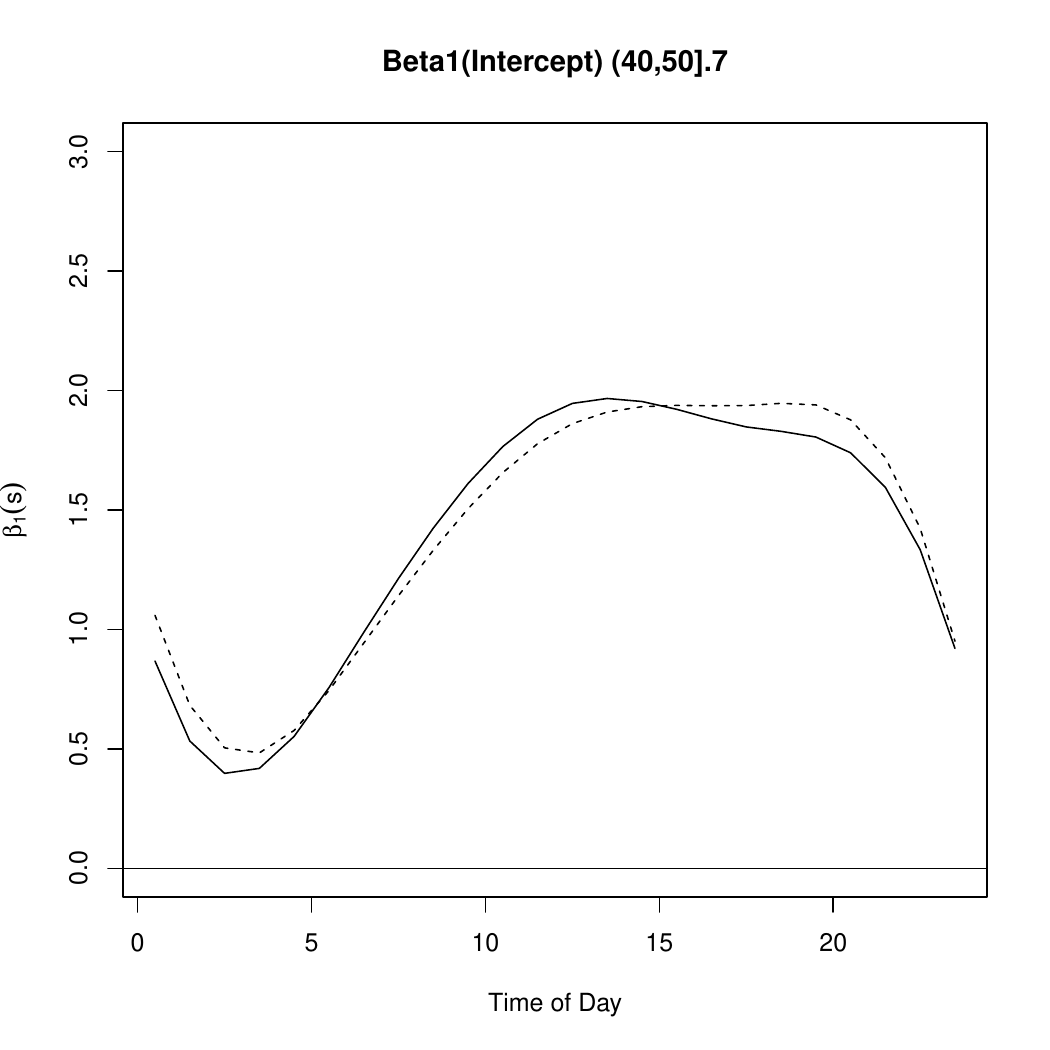}
  \caption{Predicted functional trajectories ($\hat{\beta}_1(s)+\hat{\beta}_{1i}(s)$) of diurnal MIMS for males from the MuFuMES method in the NHANES application. The other continuous covariates were held at their average values. The solid line corresponds to the fixed effect prediction $\hat{\beta}_1(s)$ and the dashed line corresponds to the mixed effects prediction $\hat{\beta}_1(s)+\hat{\beta}_{1i}(s)$. The age-by-race groups are indicated in the header.}
  \end{figure}
  \begin{figure}[H]
\includegraphics[page=34,scale=.95]{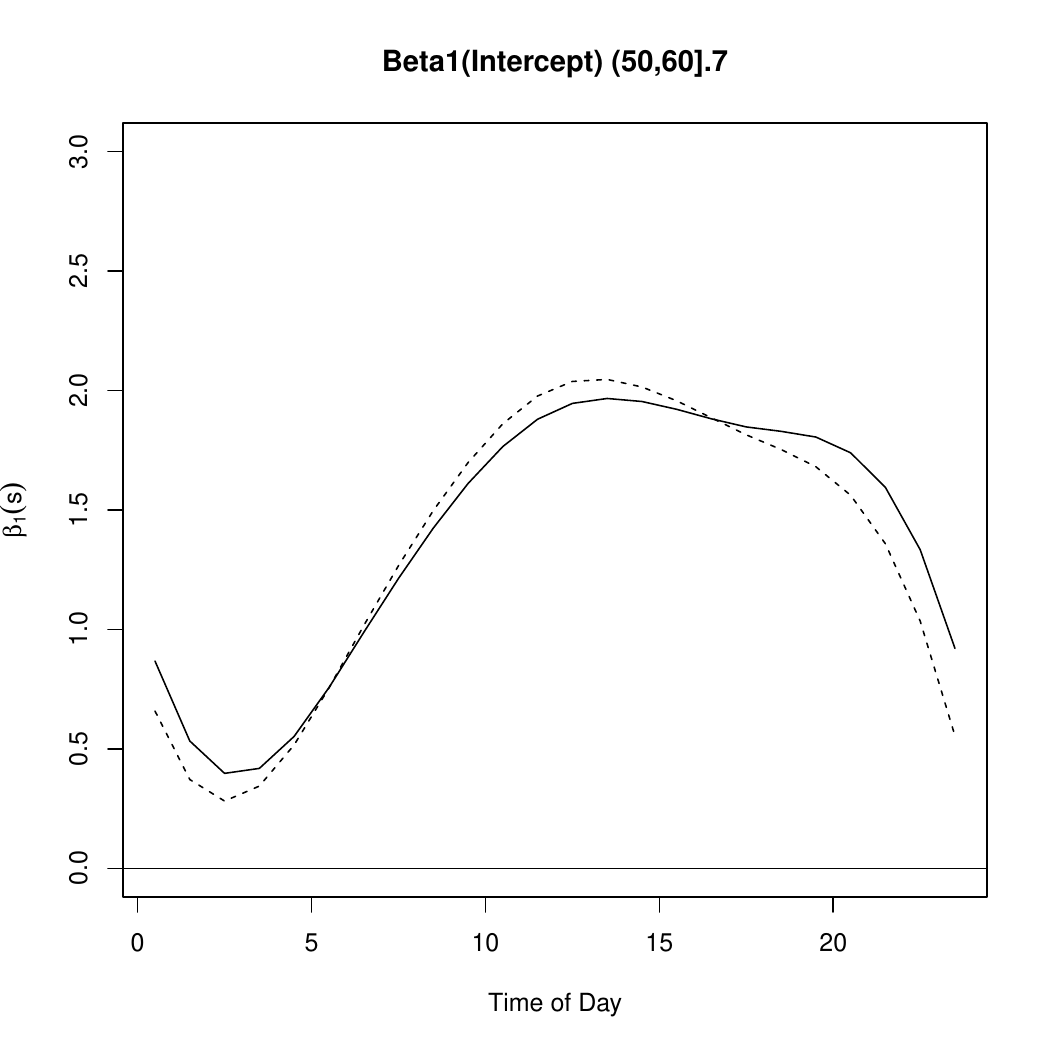}
  \caption{Predicted functional trajectories ($\hat{\beta}_1(s)+\hat{\beta}_{1i}(s)$) of diurnal MIMS for males from the MuFuMES method in the NHANES application. The other continuous covariates were held at their average values. The solid line corresponds to the fixed effect prediction $\hat{\beta}_1(s)$ and the dashed line corresponds to the mixed effects prediction $\hat{\beta}_1(s)+\hat{\beta}_{1i}(s)$. The age-by-race groups are indicated in the header.}
  \end{figure}
  \begin{figure}[H]
\includegraphics[page=35,scale=.95]{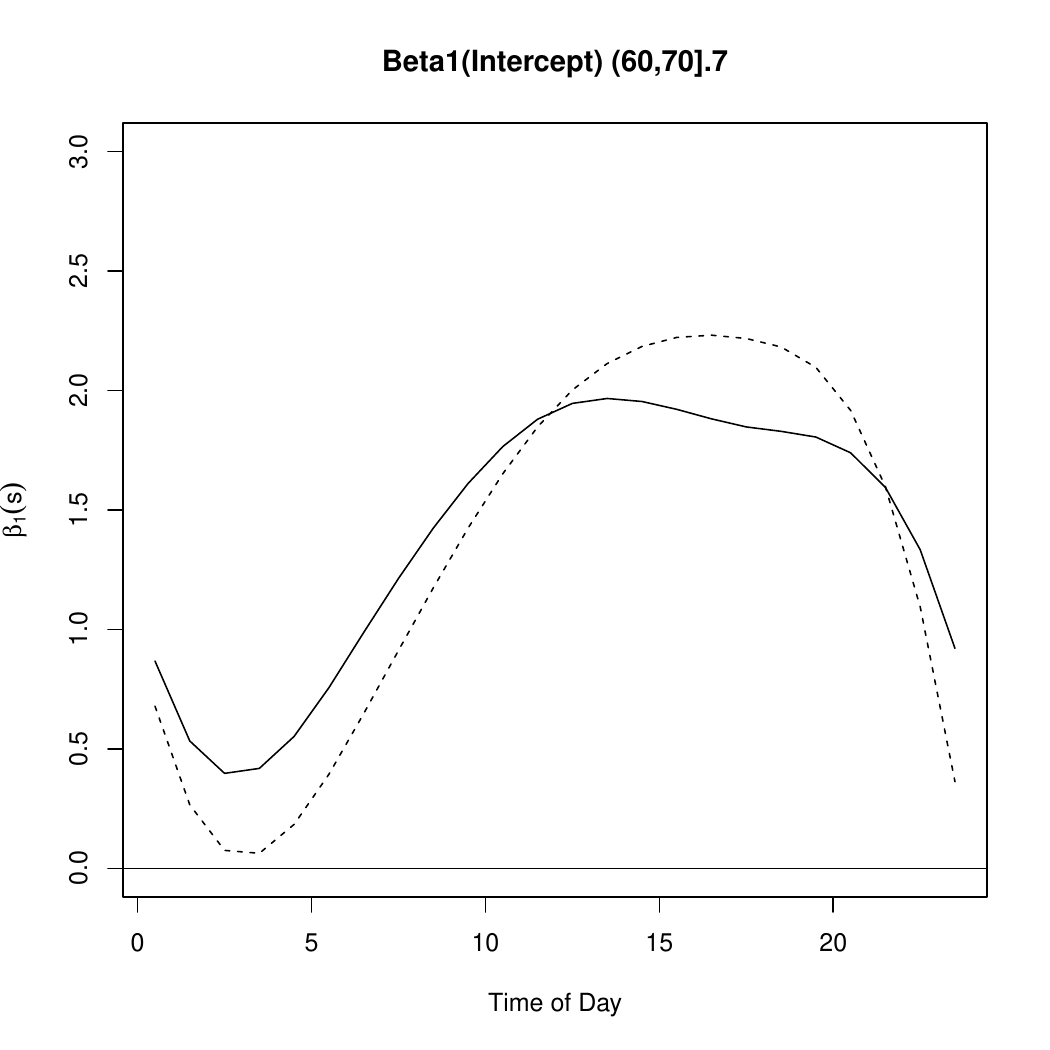}
  \caption{Predicted functional trajectories ($\hat{\beta}_1(s)+\hat{\beta}_{1i}(s)$) of diurnal MIMS for males from the MuFuMES method in the NHANES application. The other continuous covariates were held at their average values. The solid line corresponds to the fixed effect prediction $\hat{\beta}_1(s)$ and the dashed line corresponds to the mixed effects prediction $\hat{\beta}_1(s)+\hat{\beta}_{1i}(s)$. The age-by-race groups are indicated in the header.}
  \end{figure}
  \begin{figure}[H]
\includegraphics[page=36,scale=.95]{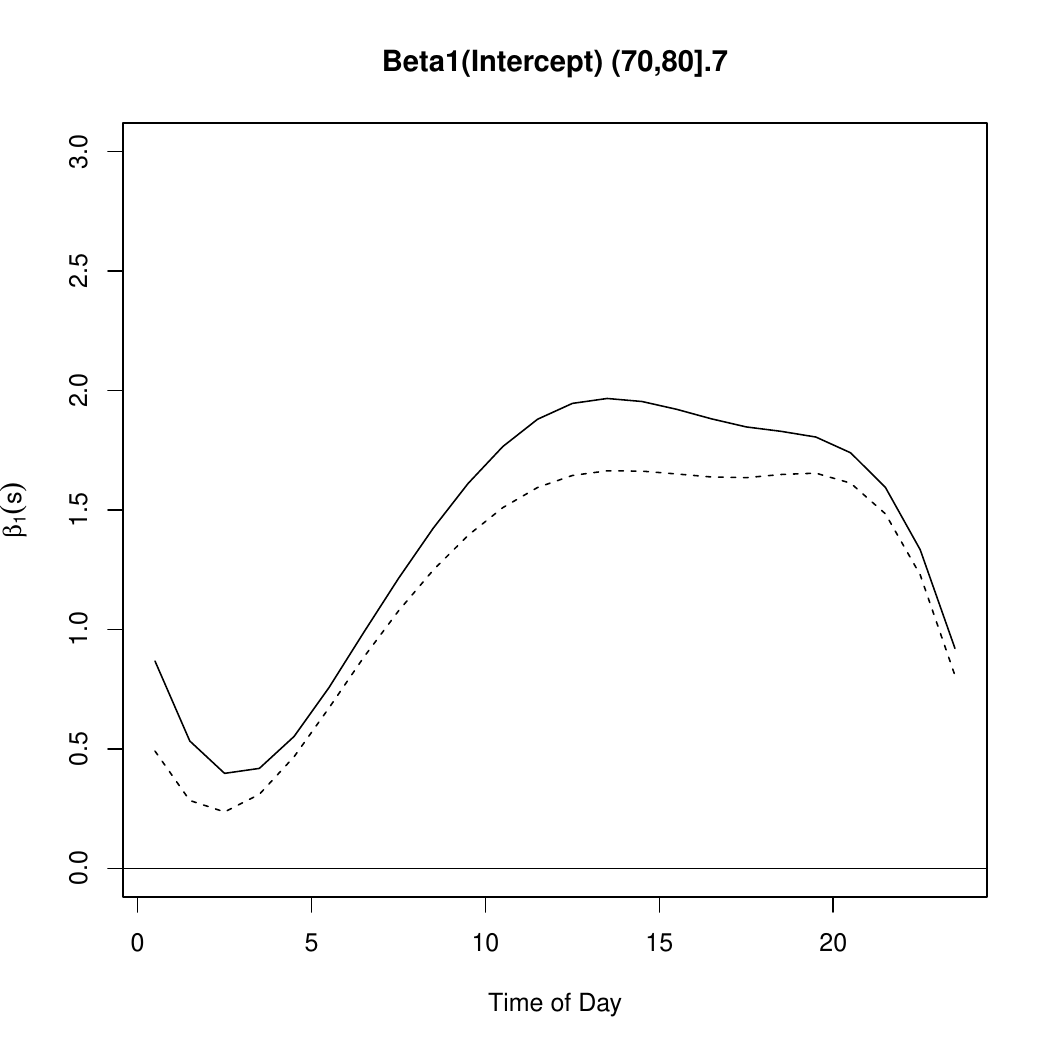}
  \caption{Predicted functional trajectories ($\hat{\beta}_1(s)+\hat{\beta}_{1i}(s)$) of diurnal MIMS for males from the MuFuMES method in the NHANES application. The other continuous covariates were held at their average values. The solid line corresponds to the fixed effect prediction $\hat{\beta}_1(s)$ and the dashed line corresponds to the mixed effects prediction $\hat{\beta}_1(s)+\hat{\beta}_{1i}(s)$. The age-by-race groups are indicated in the header.}
  \end{figure}


\bibliographystyle{Chicago}
\bibliography{refs}